\documentclass[fleqn,usenatbib]{mnras}
\usepackage{graphicx}
\usepackage{amsmath}
\usepackage{amssymb}
\usepackage{multicol}
\usepackage{bm}	
\usepackage{pdflscape}
\usepackage{comment}
\usepackage{subcaption}
\usepackage{tcolorbox}
\usepackage{mwe}
\usepackage{float}
\usepackage{caption}
\usepackage{amsmath}
\usepackage{mathtools}
\usepackage{soul}
\usepackage[mathscr]{eucal}
\usepackage{hyperref,tablefootnote,footnotehyper}
\usepackage[normalem]{ulem}
\hypersetup{colorlinks}
\usepackage[T1]{fontenc}
\usepackage{ae,aecompl}
\usepackage{newtxtext,newtxmath}
\usepackage{etoolbox}

\title[X-ray - radio interaction in Abell 1569 ]{\hspace{0.5cm}The complex intracluster medium of Abell~1569 and its interaction\\\hspace{5.8cm}with central radio galaxies}
\author[J. Tiwari $\&$ K.P. Singh]{\hspace{5.3cm} Juhi Tiwari $^{1}$\thanks{e-mail: tiwarijuhi92@gmail.com}
and Kulinder Pal Singh $^{1,2}$\thanks{e-mail: kpsinghx52@gmail.com}
\\\\
\hspace{2.7cm}$^{1}$Department of Physical Sciences, Indian Institute of Science Education and Research Mohali, Knowledge city,\\\hspace{5.1cm}Sector 81, Manauli, Sahibzada Ajit Singh Nagar, Punjab 140306, India\\
\hspace{2.7cm}${^2}$Tata Institute of Fundamental Research, 1 Homi Bhabha Road, Mumbai 400005, India
}
\begin{document}
\label{firstpage}
\pagerange{\pageref{firstpage}--\pageref{lastpage}}
\maketitle

\begin{abstract}
We present the first in-depth study of X-ray emission from a nearby ($z$$\sim$0.0784) galaxy cluster Abell 1569 using an archival \textit{Chandra} observation. A1569 consists of two unbound subclusters -- a northern subcluster (A1569N) hosting a double-lobed radio galaxy 1233+169 at its centre, and a southern subcluster (A1569S) harbouring a wide-angle-tailed (WAT) radio source 1233+168. X-ray emission from A1569N and A1569S extends to a radius $r$$\sim$248 kpc and $r$$\sim$370 kpc, respectively, indicating that the two gas clumps are group-scale systems. The two subclusters have low X-ray luminosities ($\sim$$10^{42-43}$ erg s$^{-1}$), average elemental abundances $\sim$$1/4$ Z$_{\odot}$, low average temperatures ($\sim$2 keV), and lack large ($r$$\gtrsim$40--50 kpc) cool cores associated with the intracluster gas. We detect a pair of cavities coincident with the radio lobes of 1233+169 in A1569N. The total mechanical power associated with the cavity pair is an order of magnitude larger than the X-ray radiative loss in the cavity-occupied region, providing corroborating evidence for cavity-induced heating of the intragroup gas in A1569N. A1569S exhibits possible evidence for a small-scale cluster-subcluster merger, as indicated by its high central entropy, and the presence of local gas elongation and a density discontinuity in between the bent radio tails of 1233+168. The discontinuity is indicative of a weak merger shock with Mach Number, $M$$\sim$1.7. The most plausible geometry for the ongoing interaction is a head-on merger occurring between A1569S and a subcluster falling in from the west along the line bisecting the WAT tails.   
\end{abstract}

\begin{keywords}
X-rays: galaxies: clusters -- galaxies: clusters: intracluster medium -- galaxies: clusters: general -- galaxies: clusters: individual (A1569) -- galaxies: groups: individual (A1569N, A1569S) -- radio continuum: galaxies
\end{keywords}

\section{Introduction} \label{intro}
Optical and X-ray observations of galaxy clusters reveal that they are dynamically evolving systems containing a significant degree of substructure \citep{fabian1992,oergerle2001}. The galaxy number density maps in the optical \citep{geller1982,flin2006} and X-ray surface brightness (SB) images \citep{gomez1997_9clusters,jones1999,schuecker2001} of clusters often show multiple peaks corresponding to individual groups of galaxies. Within the hierarchical structure formation scenario, mergers of these smaller galaxy groups and subclusters lead to the assembly of larger cluster units. \citep{voit2005}. 

The central galaxies of clusters and groups more often than not host active galactic nuclei (AGN) that emit radio synchrotron emission \citep{mcnamara2007,fabian2012}. Radio emission from these galaxies generally extends well beyond the optical boundaries, sometimes out to hundreds of kiloparsec \citep{miller2001}, and hence it is expected that the radio-emitting regions interact with the hot X-ray emitting intracluster/intragroup medium (ICM/IGM\footnote{The terms ICM and IGM are used synonymously in this text.}). Substructure in the form of deficits in the ICM X-ray emission \citep{reynolds2005,sanders2009} is commonly observed as a result of the interaction of the jets and lobes of the radio galaxy with the surrounding cluster gas. These deficits dubbed `cavities' are regions where the radio plasma has displaced the X-ray emitting gas, creating a low-density bubble which rises buoyantly and expands, distributing energy to the surrounding ICM (e.g., \cite{churazov2002}). A number of X-ray observations show cavities in galaxy clusters and groups, coincident with extended emission from the central radio galaxy \citep{mcnamara2000,blanton2001,chon2012}. The sizes of the cavities created by the radio lobes vary widely from a few kpc to tens and hundreds of kpc in diameter \citep{birzan2004,blanton2010}. Radio galaxies in clusters are also known to heat the surrounding gas. Although the details of how the radio jets heat the ICM are still not very clear, jet-driven cavity heating \citep{churazov2001} and shock heating \citep{fabian2003,nulsen2005} are the two popularly proposed mechanisms.

Substructure in the hot ICM also arises as a result of mergers and often shows up as an asymmetry and/or local distortions in the X-ray SB distribution \citep{burns1996,roettiger1996,schuecker2001,sarazin2002}. Mergers between clusters produce moderately supersonic shocks (Mach number: $M \lesssim$ 3) \citep{sarazin2002} that compress and heat the intracluster gas \citep{markevitch2007}. Since the X-ray flux is proportional to the square of gas density, the shock resultant density discontinuities are visible as sharp edges in the SB maps of clusters \citep{sanders2016}.  Detection of shock fronts in X-rays is a key observational tool for studying mergers in galaxy clusters since these can be used to estimate the merger-driven gas bulk flow velocities in the plane of the sky and constrain the merger geometry. Sharp edges in cluster X-ray images can also arise when the rapidly moving dense gas core of an infalling subcluster survives a major merger and forms a prominent contact discontinuity or boundary with the ambient hot gas of the host cluster (e.g., the discontinuity observed in Abell 3667 -- \citet{vikhlinin2001}). These contact discontinuities are better known as cold fronts. Cold fronts can also be induced in the cool cores of relaxed galaxy clusters by minor merger events. Such mergers displace the cold central intracluster gas from the centre of the cluster potential and subsequently cause it to slosh around the potential minimum (e.g., the sloshing spiral feature in Abell 2029 -- \citet{paterno2013}).

Evidence for interaction between radio galaxies and the host cluster gas also comes from observations of bent-tail radio sources -- a class of radio galaxies in which the jets and tails are significantly bent or distorted from the expected straight line trajectory \citep{ryle1968,owen1976}. The characteristic shape of these sources is attributed to the weather of the surrounding ICM, and hence, these sources are useful tracers of galaxy clusters/groups \citep{gNv2009}. Relative motion ($\gtrsim$ 1000 km s$^{-1}$) between the extended radio galaxy and the dense ICM, resulting in significant ram pressure on the radio-emitting material of the jets, is a popularly proposed physical mechanism to explain the observed jet bending. The relative motion can result from the galaxy moving at a high peculiar velocity through the dense gas -- as is believed to be the case for narrow-angle-tailed (NAT) galaxies (e.g., \citet{odea1985,venkatesan1994}), or by the gas itself moving across the galaxy (e.g., bulk flows in the ICM induced by cluster-subcluster mergers) -- in case of wide-angle-tailed (WAT) sources (e.g., \citet{pinkney1994,douglass2011}).          

Abell~1569 (hereafter A1569) is a low-redshift ($z=0.0784$) galaxy cluster with richness class 0 \citep{abell1989}. Its optical centre is at R.A.(J2000) = 12$^h$36$^m$18$^s$ and Dec.(J2000) = +16\textdegree{}35$'$00$''$ and it has $\sim$56 member galaxies within a projected radius of 23.5 arcmin measured from this centre \citep{gomez1997_a1569}. A1569 has been studied in great detail at optical wavelength by \citet{gomez1997_a1569} using data from the \textit{First Palomar Sky Survey} (\textit{POSS I}) and the 2.3-m \textit{Bok telescope} at the \textit{University of Arizona Steward Observatory}. The authors detected optical substructure in the cluster with the help of one, two, and three-dimensional statistical tests. They detected the presence of two unbound subclusters in A1569, one towards the northwest and the other in the southeast direction. We refer to the former as A1569N and the latter as A1569S in the following text. The subclumps were found to be segregated both spatially and kinematically. The optical centre of A1569N was determined as R.A.(J2000) = 12$^h$36$^m$13$^s$.07 and Dec.(J2000) = +16\textdegree{}36$'$21$''$.09, while that of A1569S as R.A.(J2000) = 12$^h$36$^m$15$^s$.65 and Dec.(J2000) = +16\textdegree{}31$'$48$''$.27. The estimated number of member galaxies is 25 for A1569N and 29 for A1569S. A1569N is found to have a mean velocity of 23782 km s$^{-1}$ with velocity dispersion of 622 km s$^{-1}$, while these values are 20740 km s$^{-1}$ and 433 km s$^{-1}$ respectively, for A1569S.

A1569 was first studied by \citet{abramo83} as part of an X-ray survey of clusters of galaxies using data from the \textit{Einstein Observatory}. The authors did not report the presence of substructure in the cluster. Substructure in X-ray emission from A1569 was first detected by \citet{gomez1997_9clusters} using the \textit{Roentgen Satellite} (\textit{ROSAT}) position sensitive proportional counter (PSPC) data with an exposure time of 2.85 ks. Their low-resolution (15\arcsec $\times$ 15\arcsec pixel size) PSPC image (fig. 1d of their paper) showed the presence of two X-ray clumps -- A1569N and A1569S, coinciding with the optically detected subclusters in A1569 (fig. 3f of \cite{gomez1997_a1569}).
Analysis of X-ray spectrum of A1569S averaged over a circular region of 4.9 arcmin radius centred on its X-ray peak by \citet{gomez1997_9clusters} found the cluster gas in A1569S to possess an average temperature of $1.4^{+0.9}_{-0.3}$ keV. The fit was performed in the energy range 0.5--2.05 keV, and due to low number of counts ($210\pm{16}$), the elemental abundance of the ICM could not be constrained and was fixed at 0.3 Z$_{\odot}$ during the fit. The X-ray luminosity of A1569S in the energy band 0.5--2.05 keV was estimated to be $1.7\pm{0.3} \times 10^{43} \text{erg s}^{-1}$. An average spectral analysis of A1569N was not performed due to sparse X-ray data. Furthermore, a spectral analysis of different regions within A1569S was not carried out by \citet{gomez1997_9clusters}. The authors calculated a low mass inflow rate of 0.4 M$_{\odot}$ yr$^{-1}$ into the centre of A1569S indicating the absence of a significant cluster-wide cooling flow.

A particularly interesting feature of A1569 is that both its subclusters host extended radio galaxies at their centres. The double-lobed radio source, 1233+169, is hosted by A1569N, whereas A1569S harbours a WAT radio galaxy, 1233+168 \citep{owenledlow1997,gomez1997_a1569}. Any interaction between 1233+169 and the surrounding gas in A1569N has never been explored before since the \textit{ROSAT} study by \citet{gomez1997_a1569} concentrated on A1569S. The authors found that relative velocities of radio sources with respect to the ICM should be in the range 400--2500 km s$^{-1}$ in order to explain the observed bent-tail radio morphology. Radio source 1233+168 was found to exhibit a small peculiar velocity of 215 km s$^{-1}$ which was not sufficient to explain the bending of the WAT according to ram-pressure models. Additionally, the X-ray SB and optical spatial galaxy distributions of A1569S were found to exhibit similar extended morphologies. Direct evidence of merger activity in A1569S was not found due to the low resolution of the PSPC data, but based on the points mentioned above, \citet{gomez1997_a1569} postulated that a cluster-subcluster merger could provide a reasonable explanation for the observed optical and X-ray properties of A1569S and the WAT morphology.

Here, we present a study of the hot X-ray emitting gas within the two subclusters of A1569 using the superior spatial and energy resolution ($\sim$0.5 arcsec; $\sim$120 eV) of the Advanced CCD Imaging Spectrometer (ACIS) aboard the \textit{Chandra} X-ray Observatory (CXO) over the \textit{ROSAT} PSPC ($\sim$25 arcsec, $\sim$550 eV). We investigate the thermodynamic properties of the ICM of A1569 by performing a detailed spectral analysis within the two substructures, A1569N and A1569S. Our work emphasizes on the interaction of the central radio galaxies with the surrounding cluster gas. We take the study of \citet{gomez1997_a1569} a step further to search for cavities and merger signatures in the ICM of A1569N and A1569S. These features enable us to understand the effect of radio mode feedback operating in the subclusters and the influence of the gas environment on the radio galaxies.

The paper is organized as follows. The X-ray and radio observations used here and details of data reduction are provided in Section 2. The imaging analysis is presented in Section 3 followed by the X-ray spectral analysis and radial profiles of the gas properties in Section 4. Estimates of the X-ray luminosity, gas mass, and the total mass of A1569N and A1569S are given in Section 5. Section 6 describes the energetics of the detected cavities and the X-ray SB discontinuity detection. A discussion based on the results is presented in Section 7. Finally, we present our conclusions in Section 8. We use H$_0$ = 67.4 km s$^{-1}$ Mpc$^{-1}$ , $\Omega_{m}$ = 0.315 and $\Omega_{\Lambda}$= 0.685 based on the findings of \citet{planck2020}, in all our calculations. At the redshift $z = 0.0793$ of A1569N, 1 arcsec $=$ 1.55 kpc, while at redshift $z = 0.0691$ corresponding to A1569S, 1 arcsec $=$ 1.37 kpc \citep{wright2006}.

\begin{figure*}
\begin{tabular}{cc}
\includegraphics[width=0.47\linewidth,height=7.3cm]{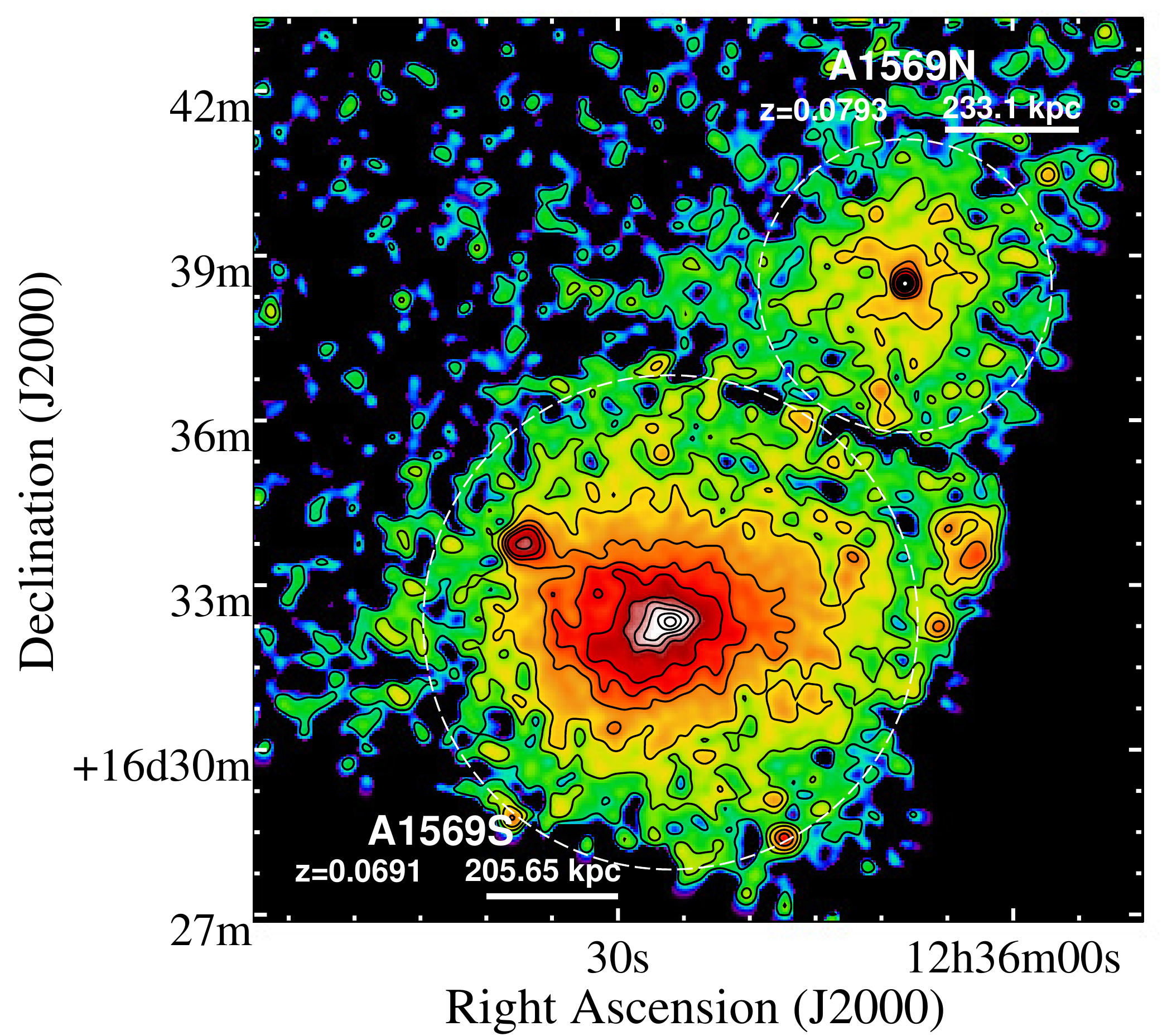}&\includegraphics[width=0.46\linewidth,height=7.3cm]{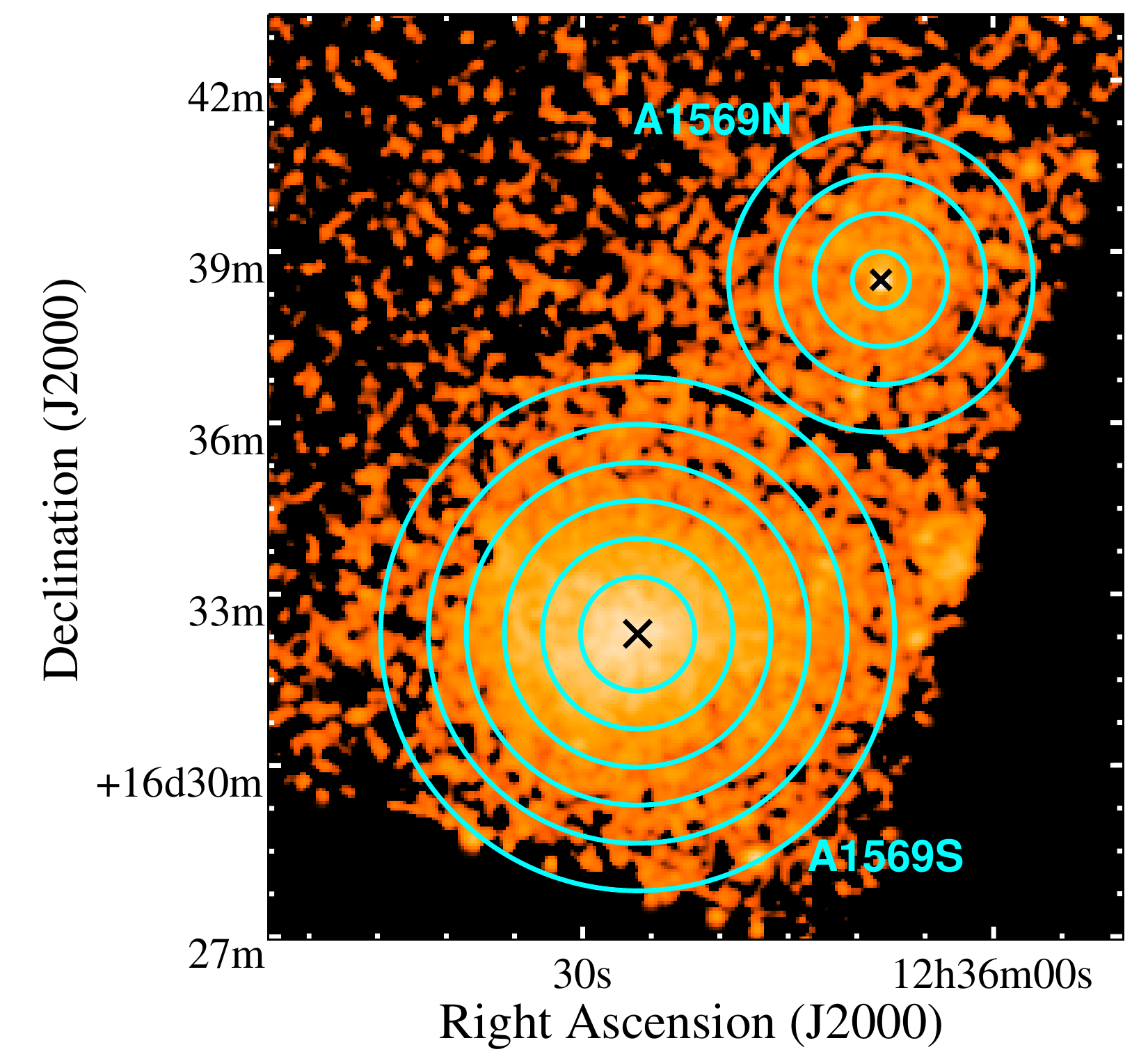}\\
\hspace{1.4cm}(a)&\hspace{1.5cm}(b)
\end{tabular}
\begin{tabular}{cc}
\includegraphics[width=0.46\linewidth,height=6.2cm]{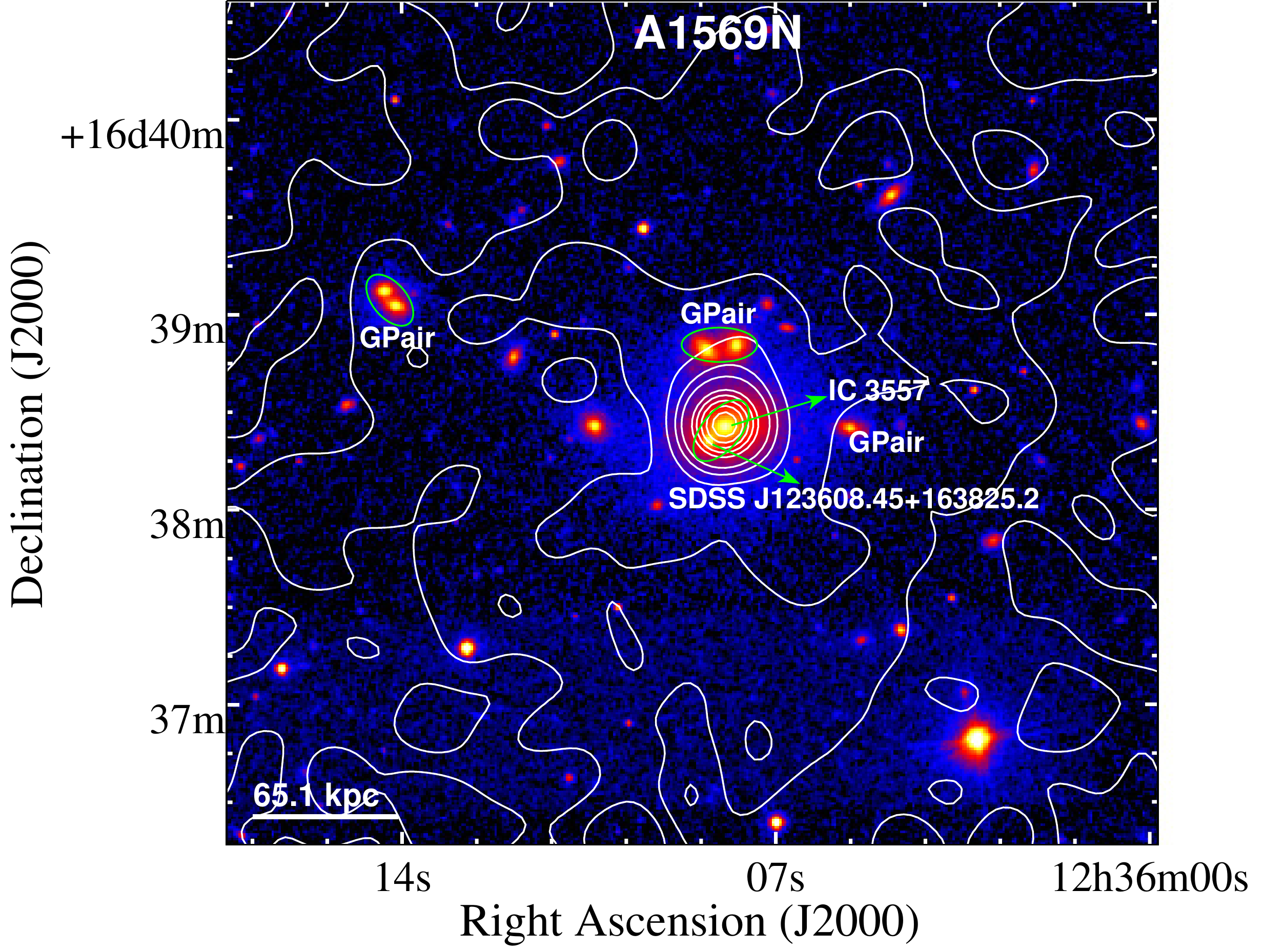}&\includegraphics[width=0.46\linewidth,height=6.2cm]{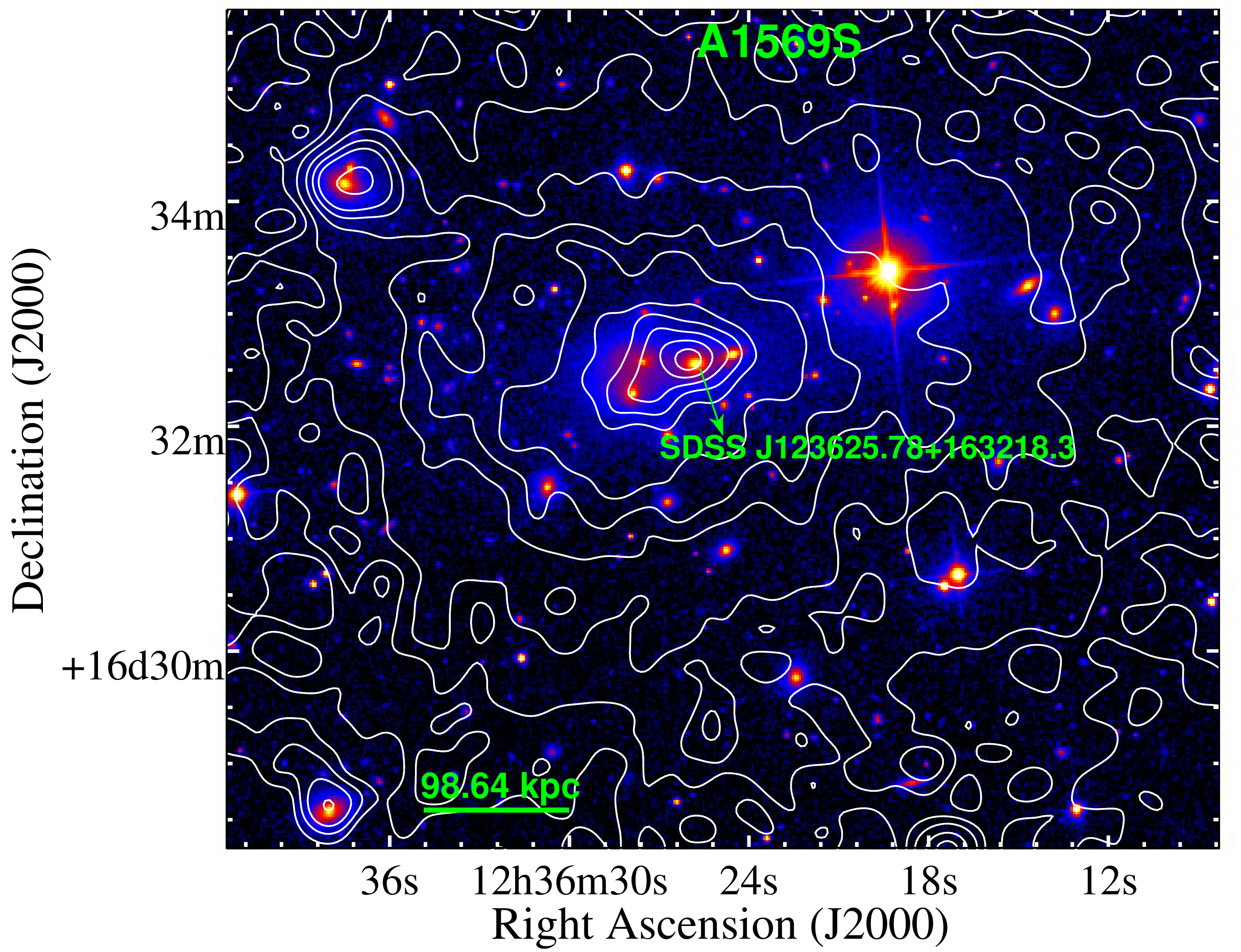}\\
\hspace{0.8cm}(c)&\hspace{1.5cm}(d)
\end{tabular}
\caption{\textbf{(a)} The 0.4--4.0 keV exposure-corrected \textit{Chandra} image of A1569 obtained after elimination of point sources and subtraction of particle background. The image has a pixel size of 0.492 arcsec and was smoothed using a gaussian kernel with $\sigma$ = 15 pixels. Overlaid logarithmic contours range from 1.5 x 10$^{-10}$ to 1.0 x 10$^{-8}$ photons cm$^{-2}$ s$^{-1}$ pixel$^{-1}$. \textbf{(b)} Regions used for the azimuthally averaged spectral analysis of A1569N and A1569S. The \textit{black cross} symbols mark the positions of the X-ray peaks. \textbf{(c)} and \textbf{(d)} Optical r-band images of the central regions of A1569N and A1569S, respectively, from the \textit{Sloan Digital Sky Survey} with overlaid X-ray contours from the \textit{Chandra} image shown in panel (a). The brightest cluster galaxy in each subcluster is labelled and few galaxy pairs (Gpair) in A1569N identified from SIMBAD are highlighted in green ellipses.}\label{morphology}
\end{figure*}

\begin{figure*}\setlength{\columnsep}{5pt}
\begin{multicols}{3}
\subcaptionbox{}{\includegraphics[width=\linewidth,height=5cm]{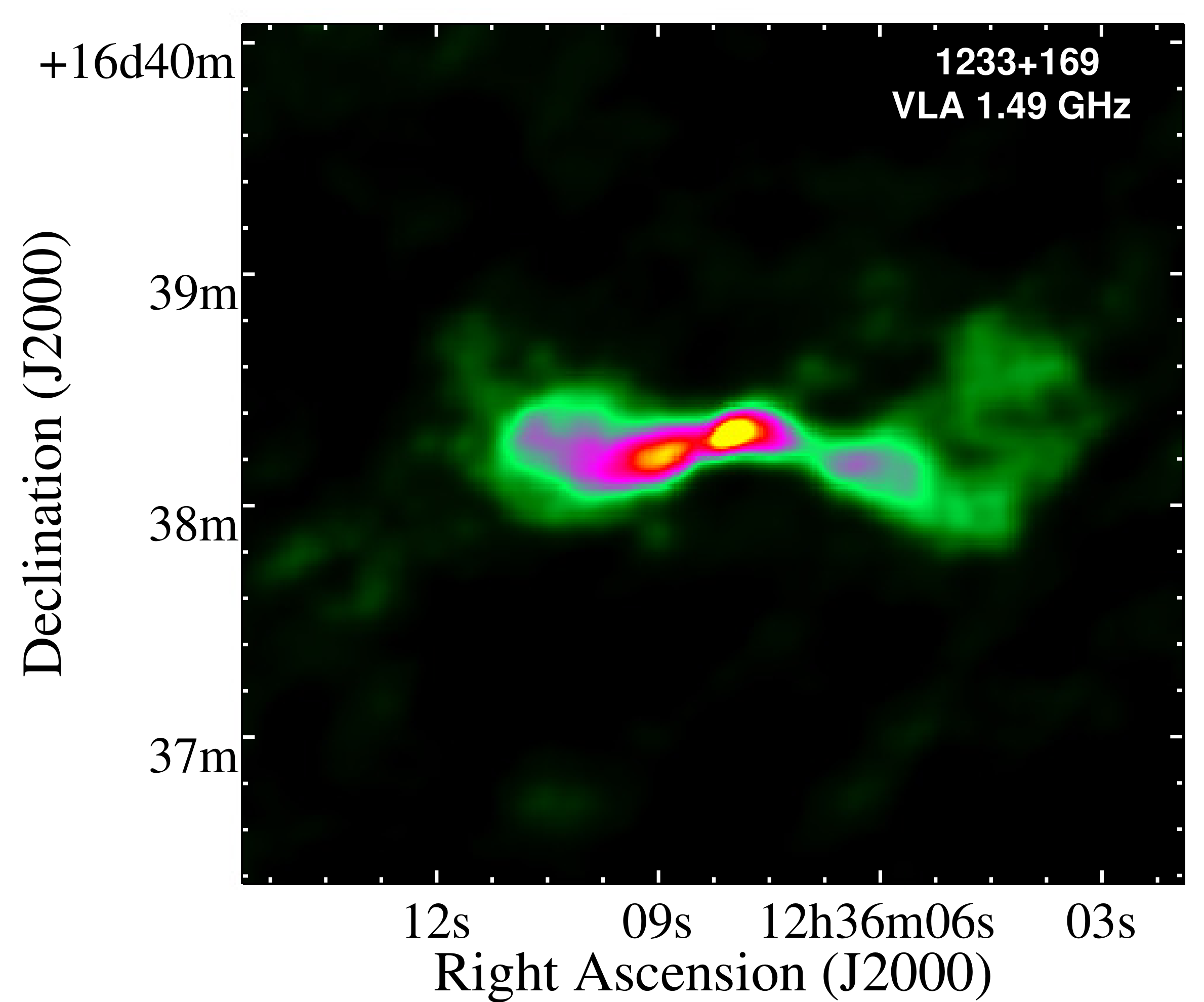}}\par
\subcaptionbox{}{\includegraphics[width=\linewidth,height=5cm]{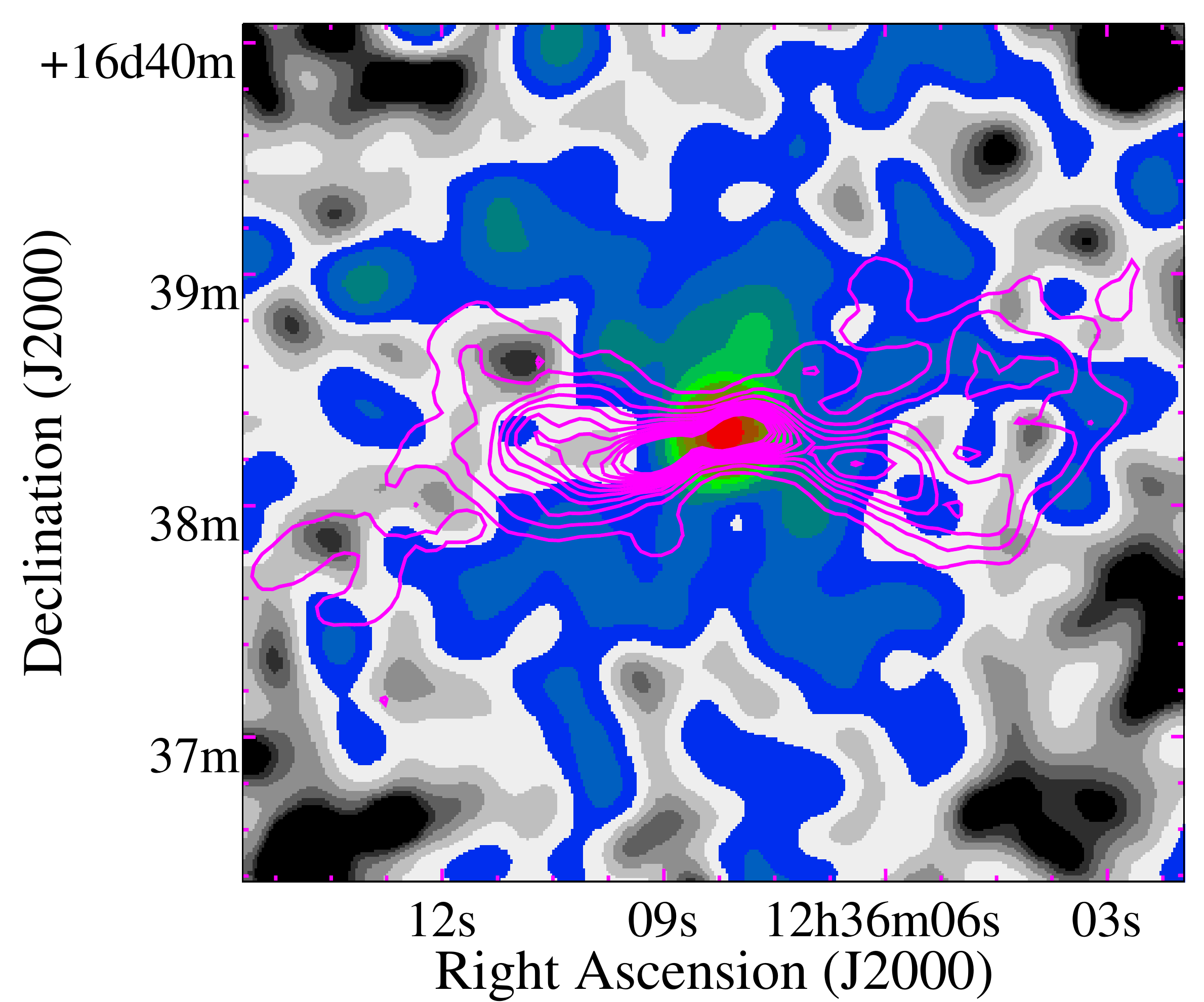}}\par
\subcaptionbox{}{\includegraphics[width=\linewidth,height=5cm]{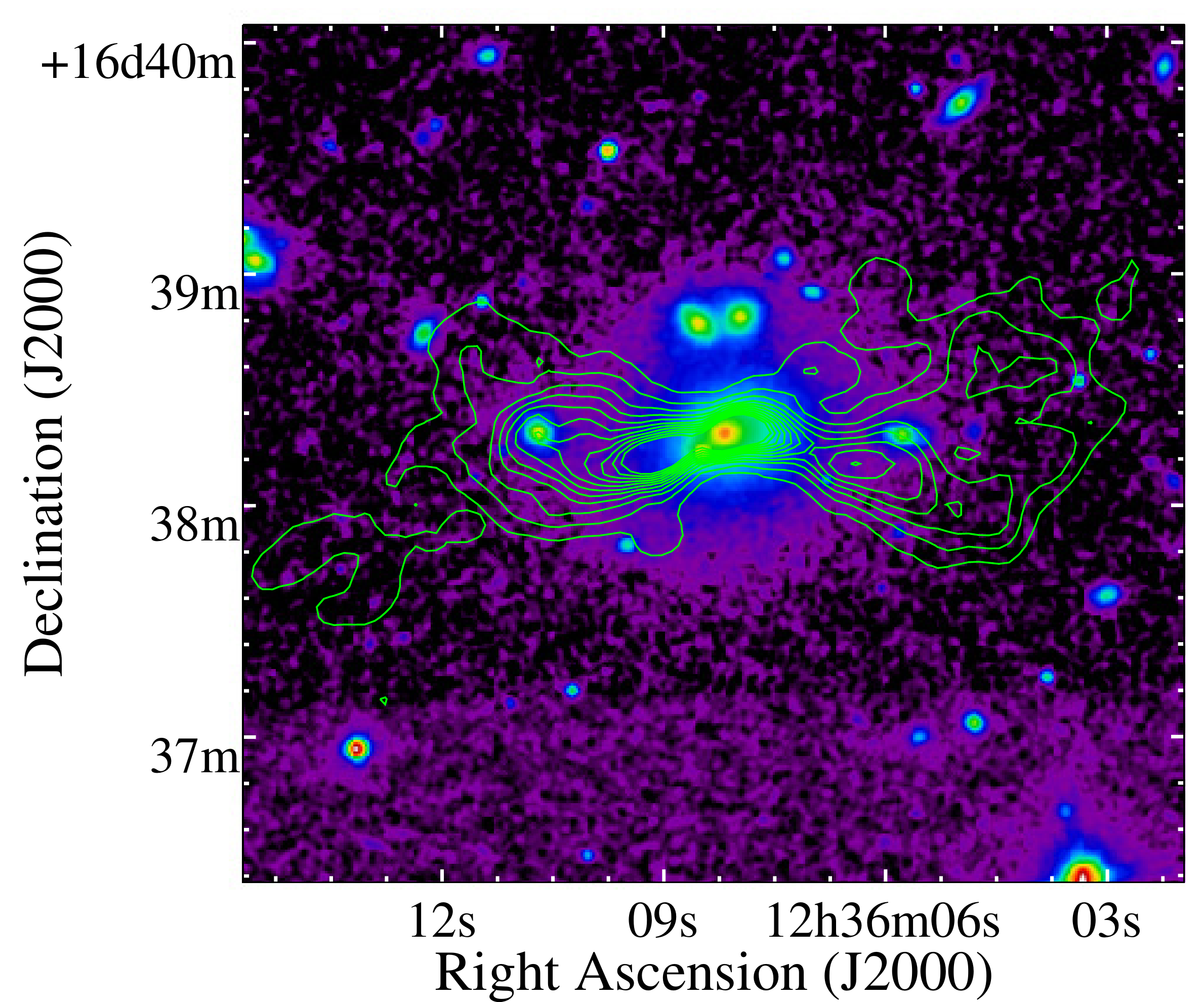}}\par
\end{multicols}
\begin{multicols}{3}
\subcaptionbox{}{\includegraphics[width=\linewidth,height=5.3cm]{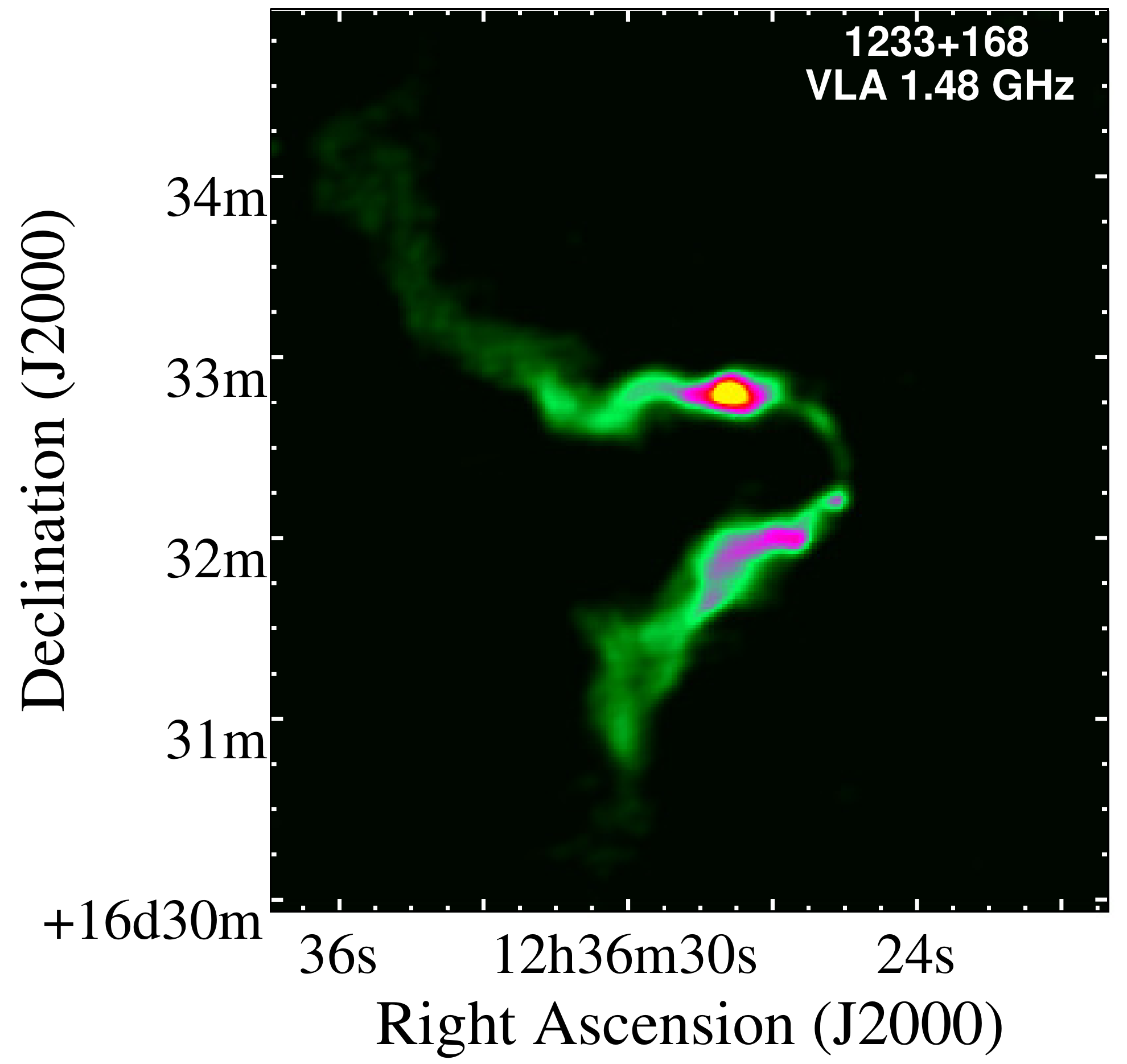}}\par
\subcaptionbox{}{\includegraphics[width=\linewidth,height=5.3cm]{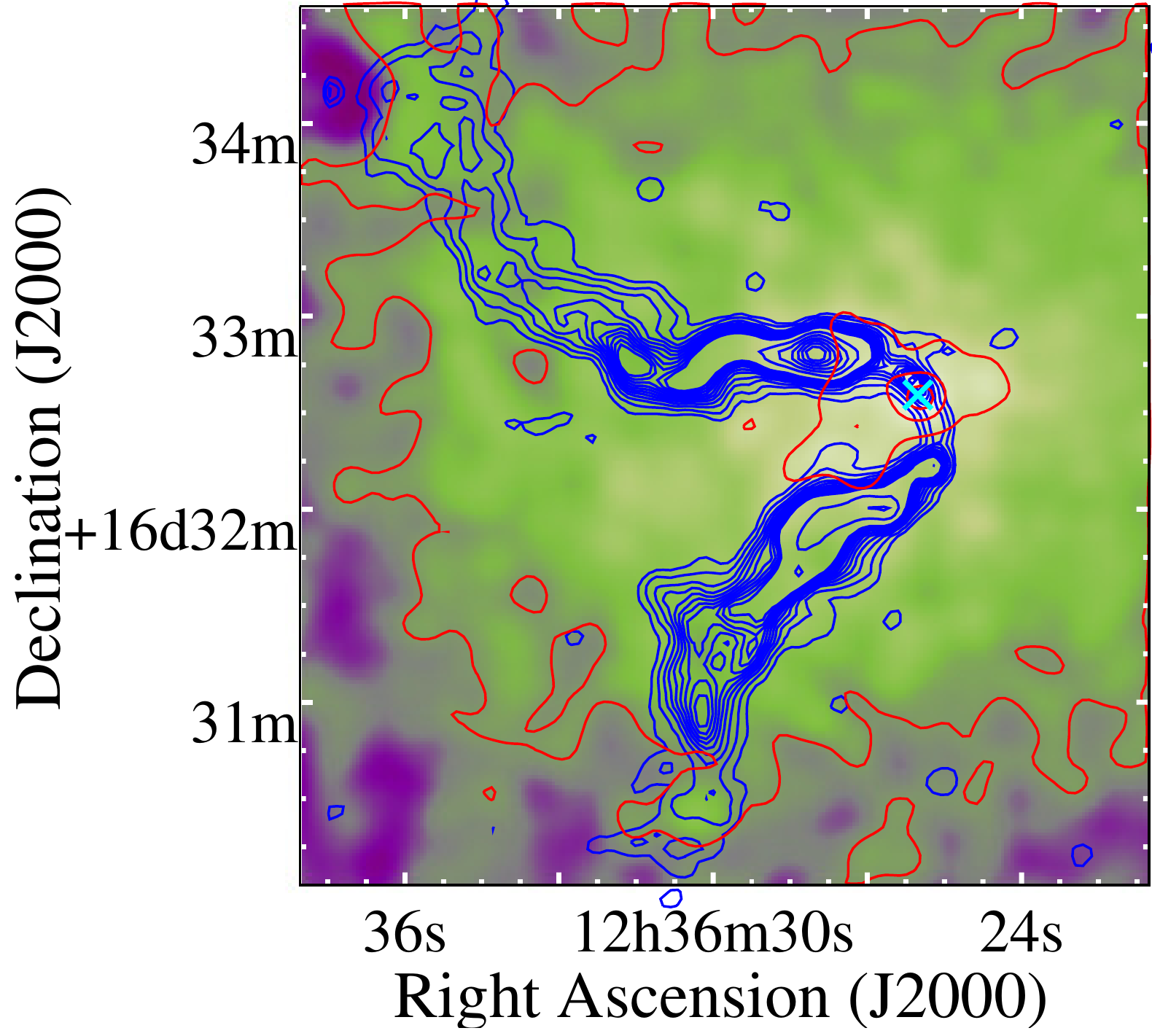}}\par
\subcaptionbox{}{\includegraphics[width=\linewidth,height=5.3cm]{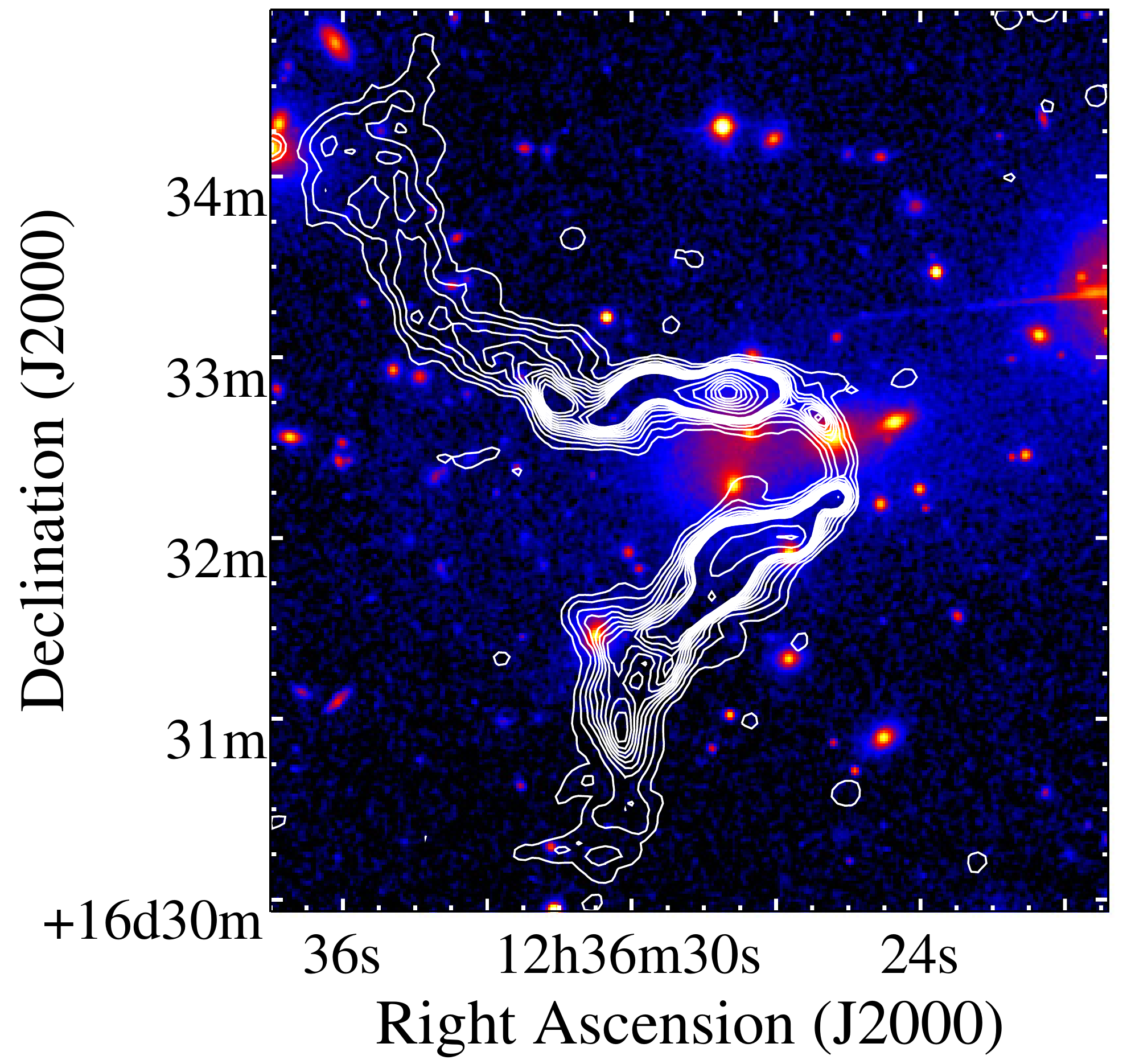}}\par
\end{multicols}
\caption{\textbf{\textit{left:}} The combined B and C-configuration VLA radio image of galaxy 1233+169 (1.49 GHz) in A1569N (\textit{top}) and 1233+168 (1.48 GHz) in A1569S (\textit{bottom}). \textbf{\textit{middle:}} \textit{Chandra} zoomed-in images (pixel size: 0.492 arcsec; images smoothed using a gaussian with $\sigma$ = 8 pixels) around the positions of 1233+169 in A1569N (\textit{top}) and 1233+168 in A1569S (\textit{bottom}), with the VLA radio contours of the galaxies superimposed. In the \textit{bottom} panel, X-ray surface brightness contours are also shown in \textit{red} color, and the \textit{cross} symbol in \textit{cyan} color marks the optical position of the BCG 1233+168. The 1.49 GHz contour levels of 1233+169 are 0.4, 1.4, 2.7, 4.1, 5.5, 6.9, 8.2, 9.6, 11.0, 12.3, and 13.7 mJy beam$^{-1}$ (beam size: $5.00 \times4.23$ arcsec$^2$). The 1.48 GHz contour levels of 1233+168 are 0.3, 0.7, 1.4, 2.1, 2.8, 3.5, 4.2, 4.9, 5.6, 6.3, 7.0, 14.0, 21.0, 28.0, 35.0, 42.0, and 49.0 mJy beam$^{-1}$ (beam size: $5.00 \times 4.23$ arcsec$^2$). \textbf{\textit{right:}} Optical r-band images of the central regions of A1569N (\textit{top}) and A1569S (\textit{bottom}), respectively, with the VLA radio contours of the central galaxies overlaid.}
\label{radio_overlays}
\end{figure*}

\section{Observations and Data Reduction}\label{obs} 
\subsection{X-ray}\label{xraydata}
We have used a single \textit{Chandra} ACIS-I X-ray observation of A1569 in this work. The observational data were obtained from the HEASARC\footnote{High Energy Astrophysics Science Archive Research Center\\  \url{https://heasarc.gsfc.nasa.gov/}} archive and have the following specifications:
\\Observation ID: 6100\\
Date of observation: 2005 April 07\\ 
Pointing RA (J2000): 12$^h$36$^m$36$^s$.59\\
Pointing Dec (J2000): +16\textdegree{}36$'$50$''$.00\\
Data mode: Very Faint\\
Useful exposure time: 41.75 ks

\vspace{0.3cm}X-ray data were reduced, screened and analysed using Chandra Interactive Analysis of Observations (CIAO) software version 4.12 and Calibration Database (CALDB) version 4.9.1 as described in detail in \S2.2 of \citet{hercules2021}. The description therein also includes procedures used for point source subtraction, image generation, background subtraction and spectral extraction. All spectra were appropriately grouped using the FTOOL \textit{grppha} and analysed using XSPEC version 12.11.0 \citep{xspec1996}. In order to constrain the X-ray background (XRB) parameters in spectral fitting, we obtained the \textit{ROSAT} All Sky Survey (\textit{RASS}) diffuse background spectrum and PSPC response\footnote{These are publicly available via the HEASARC X-ray background tool \url{https://heasarc.gsfc.nasa.gov/cgi-bin/Tools/xraybg/xraybg.pl}}, within a circular region of radius 0\textdegree{}.6 (centred on galactic coordinates \textit{lII}$=$284\textdegree{}.0; \textit{bII}$=$79\textdegree{}.7) in the vicinity of A1569. This region has neutral and total hydrogen column density values similar to those in the A1569 region.

\subsection{Radio}\label{radiodata}
We have used archival \textit{Very Large Array} (VLA) L-band observations of the two radio galaxies in A1569 in this work. The galaxy 1233+169 in A1569N and 1233+168 in A1569S, were observed by the VLA in B-configuration on January 12, 1992 (project code: AL252; total integration time: $\sim$16.83 minutes for both 1233+169 and 1233+168), and in C-configuration on January 29, 1991 (project code: AO104; total integration time: $\sim$7 minutes for 1233+169 and $\sim$6.5 minutes for 1233+168). The individual datasets were reduced and calibrated using the Astronomical Image Processing System (AIPS) software. The calibrated B-configuration and C-configuration datasets were then combined using the AIPS task DBCON to generate multi-configuration images of the two sources.

\subsection{Optical}\label{optical}
We have used the \textit{Sloan Digital Sky Survey} (SDSS) data release 12 (DR12) imaging data to generate an optical mosaic of A1569. Fifty-two \textit{r-band} images were retrieved from the DR12 Science Archive Server in a field of 30 arcmin centred at RA = 189\textdegree{}.11 and Dec = +16\textdegree{}.54. These images were mosaicked into a single image using SWarp version 2.38.0.

\begin{table*}
 \begin{center}
  \caption{Results of the two-dimensional $\beta$-model fitting to the \textit{Chandra} X-ray image of A1569N and A1569S. The errors are quoted at 90 per cent confidence level.}
  \setlength{\tabcolsep}{0pt}
  \begin{tabular}{ccccccc}
    \hline\hline
&\hspace{3.1cm}A1569N&&\hspace{3.2cm}A1569S&&&\\
&&&&&&\\
Parameter&Isotropic&Elliptical&Isotropic&Elliptical\\
\hline
Core radius $r_{c}$ (kpc)&$101.3_{-28.4}^{+38.1}$&$114.5_{-31.7}^{+44.8}$&$40.9_{-4.4}^{+4.7}$&$51.0_{-5.3}^{+5.7}$\\[0.15cm]
Centre position R.A. (J2000)&$12^{h}36^{m}08^{s}.8^{+0.5}_{-0.5}$&$12^{h}36^{m}08^{s}.9^{+0.5}_{-0.5}$&$12^{h}36^{m}26^{s}.6^{+0.1}_{-0.1}$&$12^{h}36^{m}26^{s}.6^{+0.2}_{-0.2}$\\[0.15cm]
Centre position Dec. (J2000)&$+16^\circ 38\arcmin 21\arcsec.41\pm{8\arcsec.0}$&$+16^\circ 38\arcmin 20\arcsec.51_{-9\arcsec.2}^{+8\arcsec.6}$&$+16^\circ 32\arcmin 17\arcsec.9\pm1\arcsec.7$&$ +16^\circ 32\arcmin 18\arcsec.4\pm{1\arcsec.5}$\\[0.15cm]
Ellipticity $ellip$&--&$0.22\pm{0.09}$&--&$0.27\pm{0.02}$\\[0.15cm]
$\theta$ (degree)&--&$49.2_{-14.8}^{+8.5}$&--&$8.6^{+2.9}_{-2.8}$\\[0.15cm]
Amplitude ($10^{-9}$ photon cm$^{-2}$ s$^{-1}$ pixel$^{-1}$)&$1.67\pm{0.09}$&$1.67_{-0.01}^{+0.02}$&$10.8\pm{0.08}$&$10.7_{-0.07}^{+0.08}$\\[0.15cm]
Index $\beta$&$0.30_{-0.01}^{+0.03}$&$0.28_{-0.02}^{+0.03}$&$0.35\pm{0.01}$&$0.36\pm{0.01}$\\[0.15cm]
Reduced C-stat&1.03&1.03&1.10&1.09\\
\hline
&Background Estimate&&&\\
\hline
Constant ($10^{-9}$ photon cm$^{-2}$ s$^{-1}$ pixel$^{-1}$) &$0.54\pm{0.01}$&&&\\[0.15cm]
Reduced C-stat&0.88&&&\\
\hline\hline
  \end{tabular}
   \label{2dbetatable}
  \end{center}
 \end{table*}

\section{Imaging analysis}\label{imaging}
\subsection{X-ray, optical and radio images}
\autoref{morphology}a shows the point-source-subtracted, exposure-corrected, and particle-background-subtracted \textit{Chandra} X-ray image of A1569 in the energy range 0.4$-$4.0 keV. Two gas subclumps, one towards the north and the other towards the south, are seen very clearly. 
As mentioned in \S1, we refer to the northern subclump as A1569N, and the southern subclump as A1569S. Based on this \textit{Chandra} image of A1569, the X-ray peak of A1569N is located at RA (J2000)$=$12$^{h}$36$^{m}$08$^{s}$.14 and Dec (J2000)$=$+16\textdegree{}38\arcmin29\arcsec.19, and that of A1569S is at RA (J2000)$=$12$^{h}$36$^{m}$25$^{s}$.97 and Dec (J2000)$=$+16\textdegree{}32\arcmin19\arcsec.40. A1569S appears to be elongated in the E-W direction while the gas in A1569N seems to be extended in the NW-SE direction. The large-scale morphology of the two subclusters is explored in more detail in the following subsection. 

\autoref{morphology}c and \autoref{morphology}d show the optical \textit{r-band} images of the central regions of A1569N and A1569S from the SDSS. The clustering of galaxies around the X-ray peaks of the subclumps is visible. The presence of two distinct galaxy groups within A1569 coinciding with the locations of the X-ray gas clumps was reported in the detailed optical analysis carried out by \citet{gomez1997_a1569}. The X-ray peak of A1569N coincides with the elliptical galaxy IC 3557 (SIMBAD\footnote{Set of Identifications, Measurements and Bibliography for Astronomical Data: \url{http://simbad.u-strasbg.fr/}}) shown in Fig.\ref{morphology}c. IC 3557 along with SDSS J123608.45+163825.2 (Fig.\ref{morphology}c) constitutes a pair of galaxies. IC 3557 is the brighter member of the pair and is classified as a brightest cluster galaxy (BCG) by \citet{szabo2011}. The X-ray peak of A1569S coincides with the BCG SDSS J123625.78+163218.3 (Fig.\ref{morphology}d).

\begin{figure}
\centering
\includegraphics[width=0.68\linewidth,height=5.5cm,angle=-1.2]{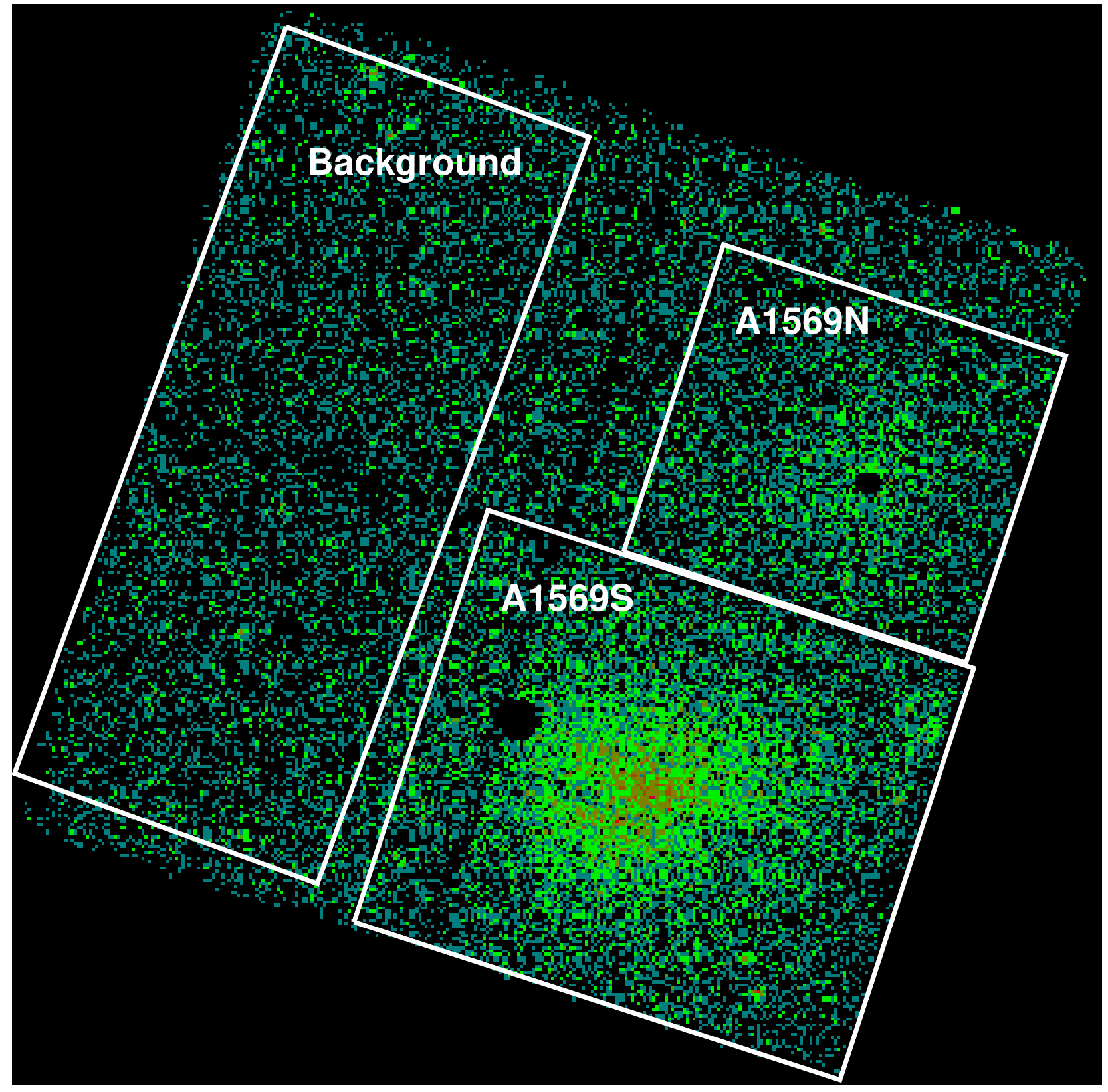}
\caption{Regions used for estimating the background, and subcluster properties of A1569N and A1569S by fitting a two-dimensional $\beta$-model to the \textit{Chandra} X-ray image. The image has a pixel size of 3.936 arcsec (8 times the original ACIS pixel size).}
\label{2dbeta}
\end{figure}

\autoref{radio_overlays}a and \autoref{radio_overlays}d display the VLA combined B and C-configuration L-band radio images of the galaxies 1233+169 and 1233+168, respectively. 1233+169 in A1569N extends to $\sim$104 kpc on a side, while the extended features of 1233+168 ($\sim$230 kpc) span almost the entire eastern side of A1569S. The total flux density of 1233+168 at 1.48 GHz is 1360 mJy while that of 1233+169 is 620 mJy. The total radio luminosity of 1233+168 at 1.48 GHz is $1.66 \times 10^{25}$ W Hz$^{-1}$ while that of 1233+169 is $1.02 \times 10^{25}$ W Hz$^{-1}$. High-resolution radio images of these two galaxies have been presented earlier by \citet{owenledlow1997}. However, their interaction with the surrounding intracluster gas has never been explored before. With the aim of investigating this prospect, we searched for X-ray structure in the vicinity of the extended features of the two radio galaxies. In \autoref{radio_overlays}b and \autoref{radio_overlays}e, we present the \textit{Chandra} zoomed-in images of the central regions of A1569N and A1569S respectively, with the radio contours of 1233+169 and 1233+168 overlaid. Fig. \ref{radio_overlays}b shows that the X-ray gas distribution in the region around 1233+169 is not azimuthally symmetrical. There appears to be an X-ray deficit in the regions where the extended features of 1233+169 lie, thus, indicating the presence of X-ray substructure adjacent to the radio galaxy in A1569N. Fig. \ref{radio_overlays}e, however, does not show obvious X-ray structure around the tails of 1233+168 in A1569S. In the central $\sim$70 kpc region of A1569S, however, the gas distribution clearly deviates from azimuthal symmetry and appears to be elongated. The X-ray extension lies between the tails of 1233+168 and is well beyond its optical light distribution. \autoref{radio_overlays}c and \autoref{radio_overlays}f show that the centres of both 1233+169 and 1233+168 coincide with the BCGs of the two subclusters.
 
\begin{figure*}\setlength{\columnsep}{5pt}
\centering
\begin{multicols}{2}
\subcaptionbox{}{\includegraphics[width=0.8\linewidth,height=5cm]{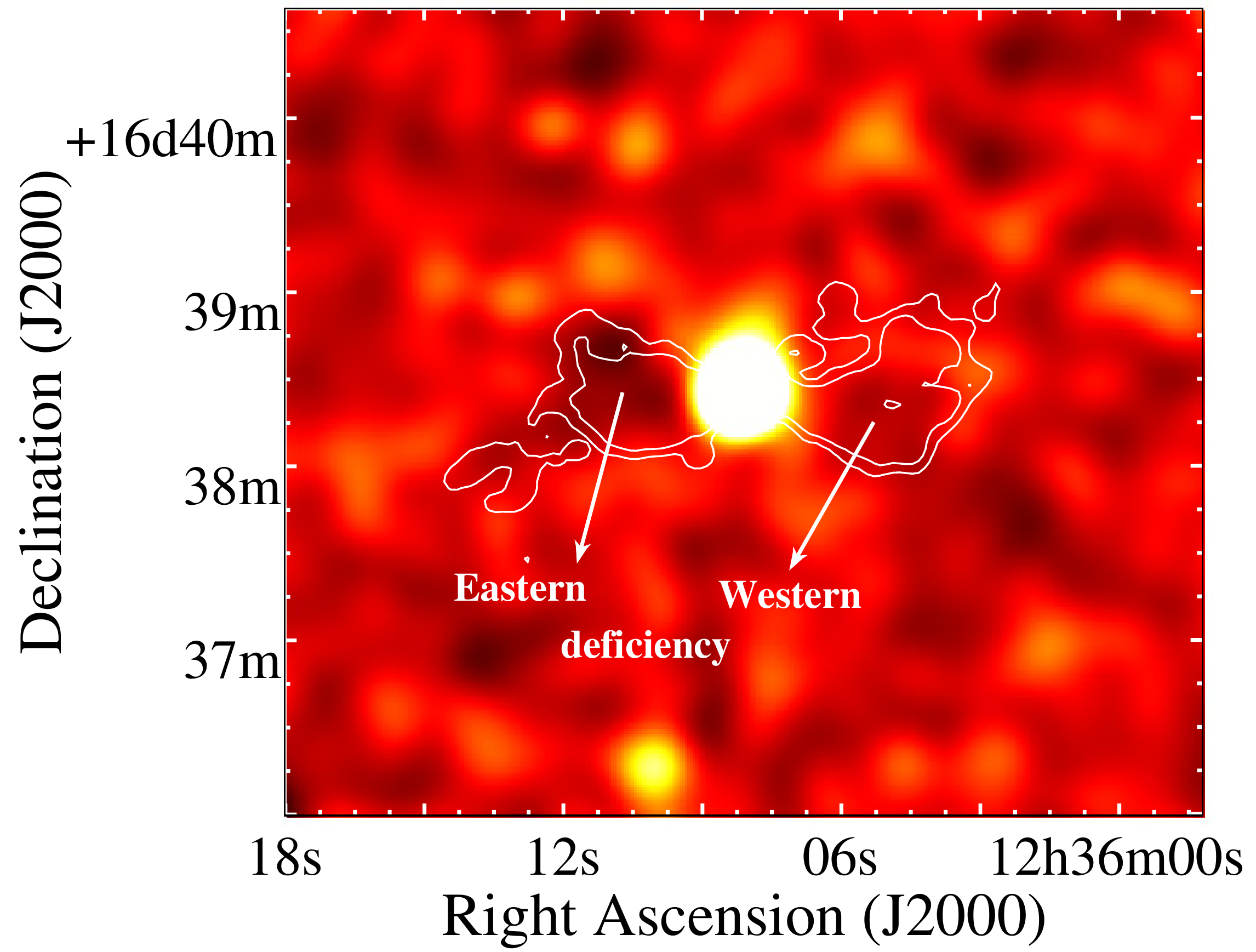}}\par
\subcaptionbox{}{\includegraphics[width=0.88\linewidth,height=5cm]{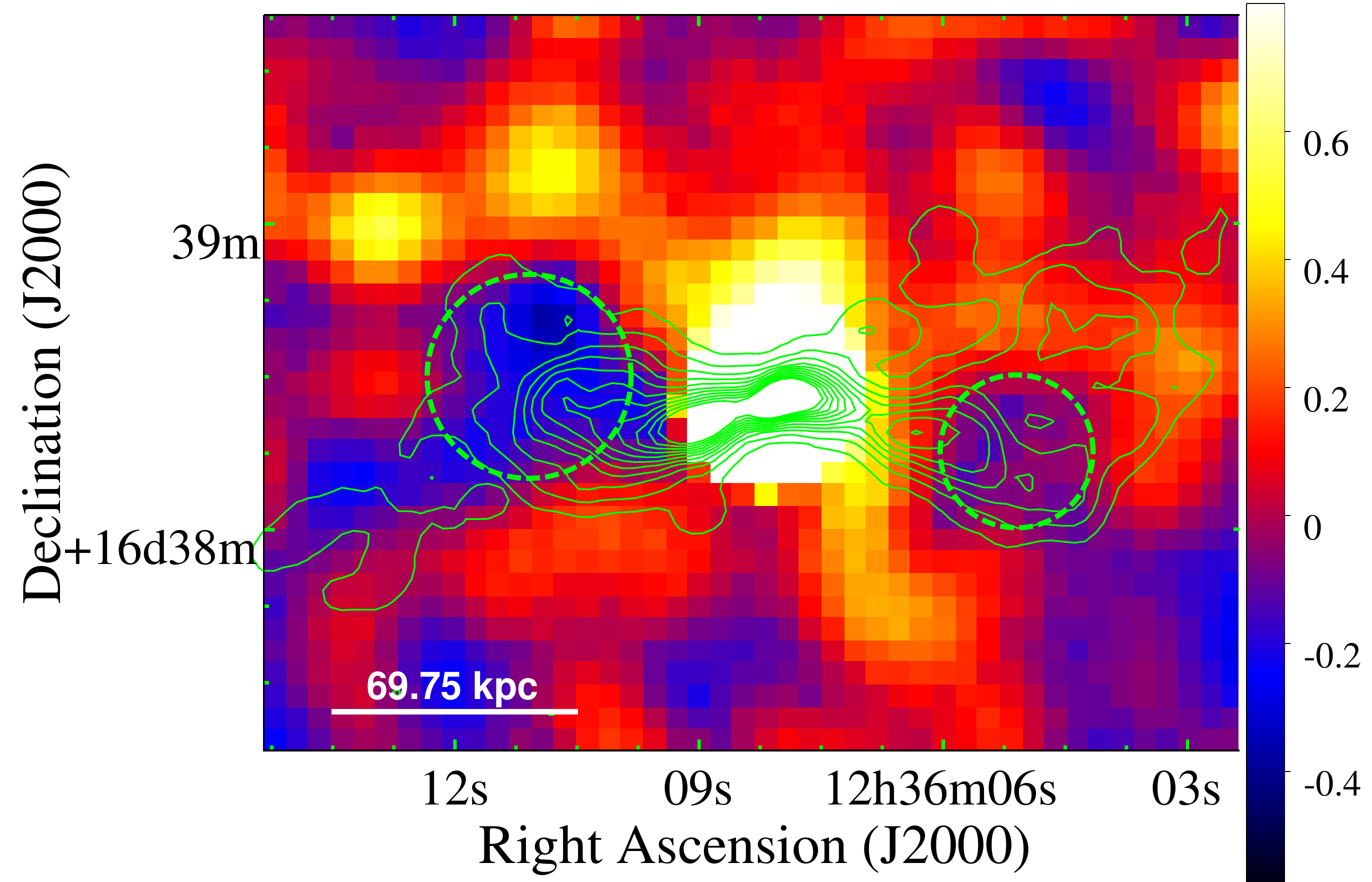}}\par
\end{multicols}
\begin{multicols}{3}
\subcaptionbox{}{\includegraphics[width=0.92\linewidth,height=5cm]{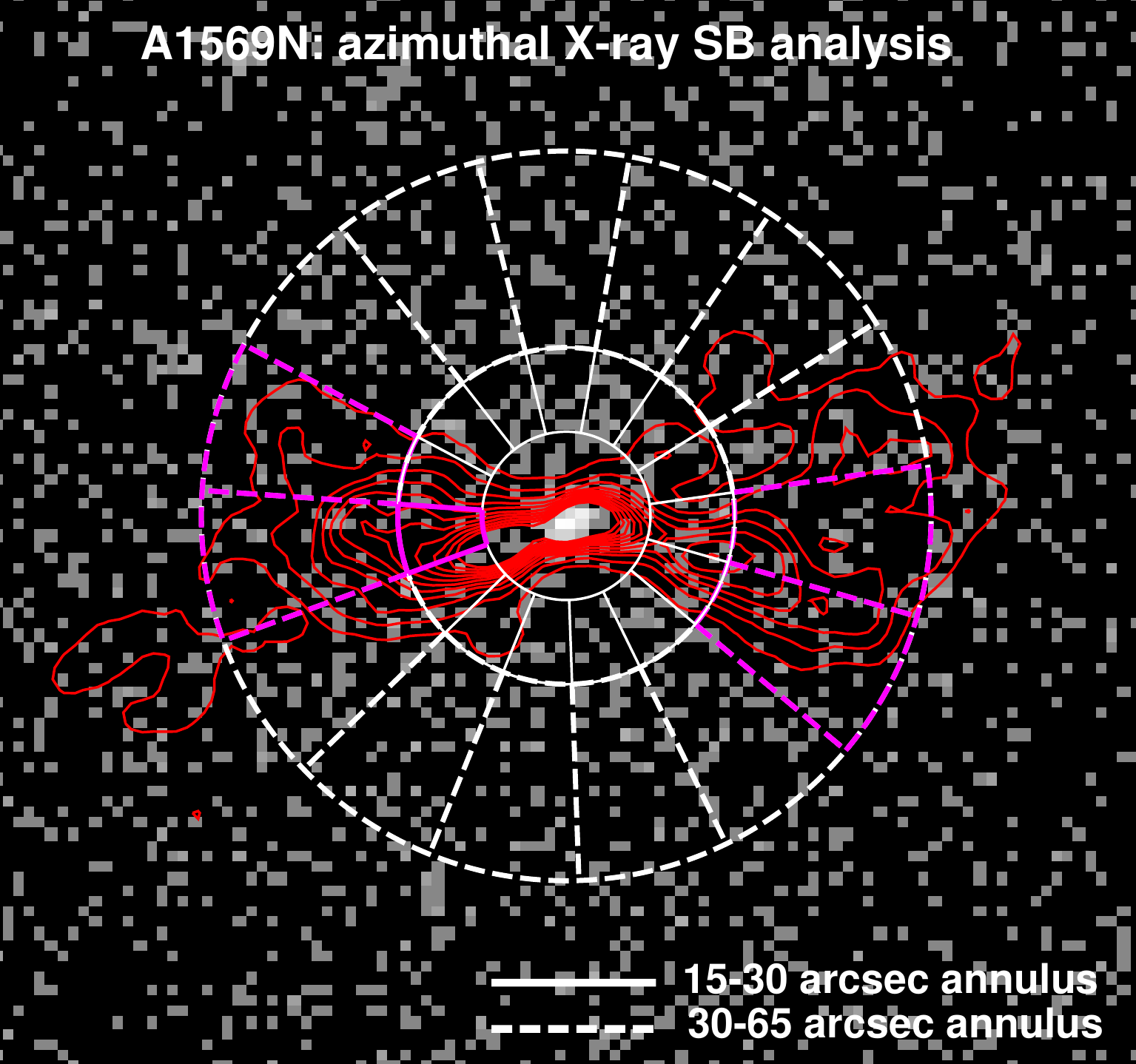}}\par
\subcaptionbox{}{\includegraphics[width=1.05\linewidth,height=4.2cm]{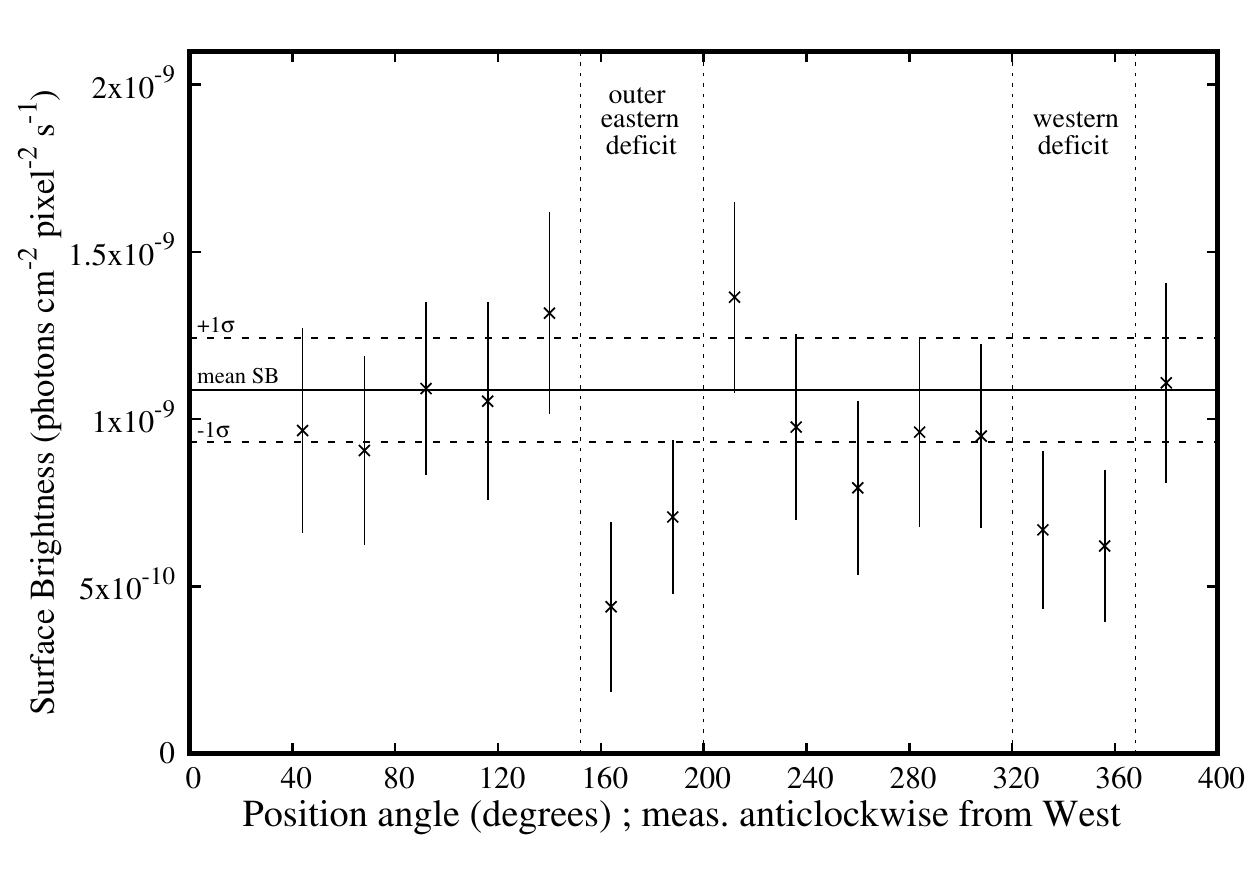}}\par
\subcaptionbox{}{\includegraphics[width=\linewidth,height=4cm]{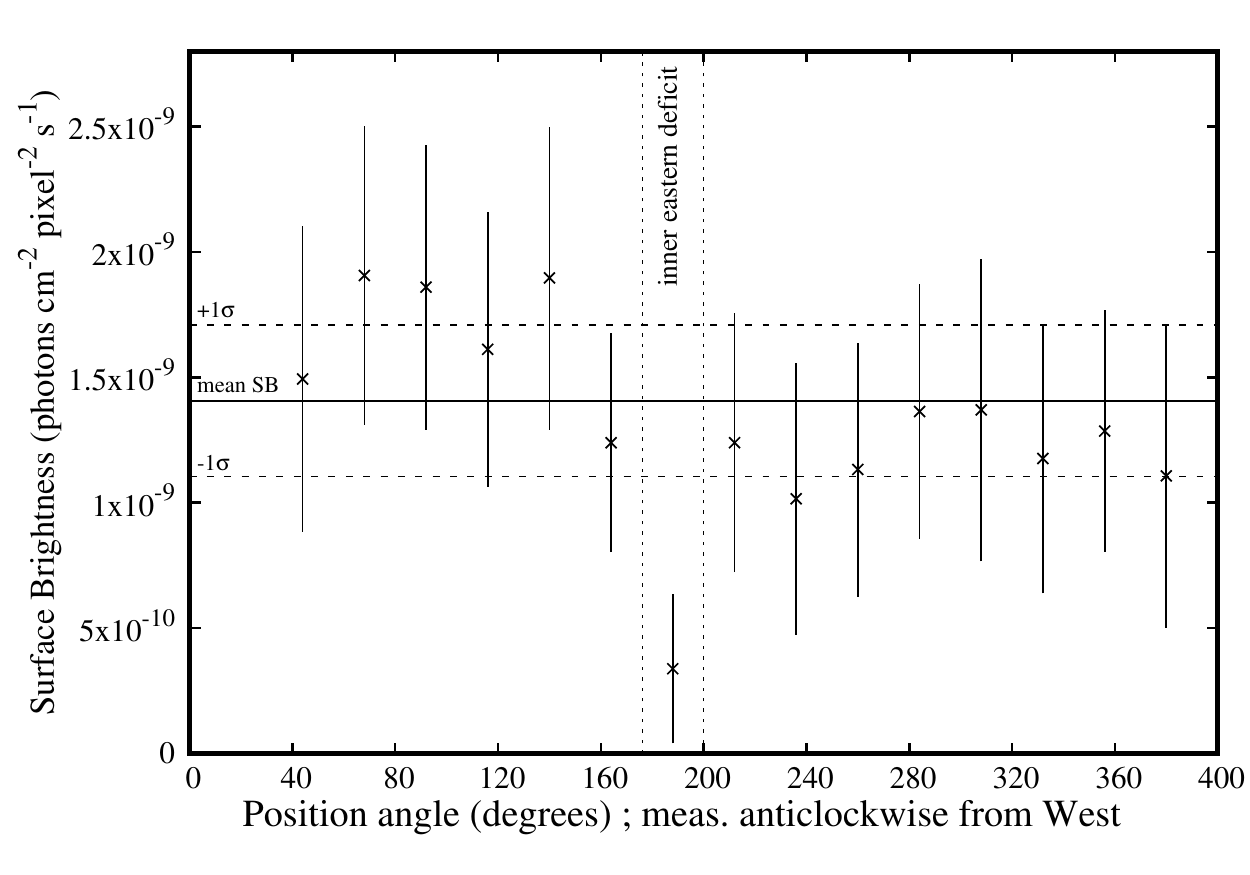}}\par
\end{multicols}
\caption{\textbf{(a)} Unsharp-masked image of the central region of A1569N with the outermost radio SB contours of galaxy 1233+169 superimposed in \textit{white} color. X-ray deficits spatially coincident with the extended features of 1233+169 are seen towards the east and west of the central peak. \textbf{(b)} Residual image of the central region of A1569N obtained after subtracting the best-fitting elliptical 2D-$\beta$ model from the data image. The radio contours of 1233+169 are overlaid in \textit{green}. The dashed \textit{green} circles represent the gas deficient regions. \textbf{(c)} The outer (30--65 arcsec) and inner (15--30 arcsec) annular regions used for the azimuthal SB analysis are outlined in \textit{white} dashed and \textit{white} solid line styles, respectively. Both annular regions are divided into fifteen 24\textdegree{} sectors. Radio SB contours of 1233+169 are overlaid in \textit{red} color. The eastern and western sectors within which a significant drop in the SB is observed are highlighted in \textit{magenta} colour. \textbf{(d)} 0.5--4.0 keV azimuthal SB profile obtained for the outer annulus (30--65 arcsec) using the fifteen 24\textdegree{} sectors. The solid line marks the mean SB calculated after excluding the 4 lowest data points. Dashed lines show the $1\sigma$ uncertainties on the mean. \textbf{(e)} 0.5--4.0 keV azimuthal SB profile obtained for the inner annulus (15--30 arcsec) using the fifteen 24\textdegree{} sectors. The solid line marks the mean SB calculated after excluding the lowest data point. Dashed lines show the 1$\sigma$ uncertainties on the mean; The X-ray deficits highlighted in (d) and (e) correspond to the \textit{magenta} sectors marked in (c).}
\label{cavity}
\end{figure*}

\subsection{Two-dimensional image fitting}\label{2Dfit_text}
The 0.4-4.0 keV \textit{Chandra} image was fitted with a 2D-$\beta$ model to extract the general characteristics of the large-scale X-ray morphology of each subcluster. CIAO's modelling and fitting application SHERPA was used for this purpose. \autoref{2dbeta} shows the image regions used for estimating the background and subcluster properties. Total counts are 7725 for the background, and 7705 and 17552 respectively, for the A1569N and A1569S source regions. We used a binning factor of 8, such that each pixel in the image has a size of 3.936 arcsec.

First, an estimate of the background surface brightness (SB) was obtained by modelling the background region with the constant 2D model (\textit{const2d}). Each source region was then fitted with the model \textit{beta2d+const2d}, where \textit{beta2d} is the isotropic 2D-$\beta$ model (model parameters $ellip$ and $\theta$ frozen to 0.0). The background value was fixed when fitting the source regions. An exposure map generated using the task \textit{fluximage} (\S 2.1) was supplied during the fit. The fitting results are listed in \autoref{2dbetatable}. We note that in case of A1569N, a circular region of radius 15 arcsec corresponding to the core of the central galaxy was excluded while modelling the subcluster source emission, since including the central region did not result in physically reasonable parameter values. We also tried fitting the model \textit{gauss2d+beta2d} (the former representing emission from the central galaxy and the latter modelling the cluster gas emission) to A1569N, but this did not result in a good fit. 

The source regions were also modelled with the elliptical 2D-$\beta$ model by thawing the paramaters $ellip$ and $\theta$. The results are provided in Table \ref{2dbetatable}. The position of the subcluster centre thus obtained, the core radius and $\beta$ value for both A1569N and A1569S are consistent with the results of the isotropic 2D-$\beta$ model. The ellipticity ($ellip$) and position angle ($\theta$) parameters are $0.27\pm{0.02}$ and $8^\circ.6_{-2.8}^{+2.9}$ respectively, for A1569S. Thus, we not only confirm but also robustly quantify the E-W elongation of A1569S previously reported by \citet{gomez1997_a1569}. The ellipticity and position angle values are $0.22\pm{0.09}$ and $49^\circ.2^{+8.5}_{-14.8}$ respectively, for A1569N. These values indicate that A1569N is elongated in the NW-SE direction. The $\beta$ values obtained for the two subclusters are comparable to those found in groups of galaxies \citep{helsdon2000,mulchaey2000}. 

\subsection{Search for X-ray deficits around 1233+169}
Motivated by the presence of X-ray substructure in the intracluster gas around the radio galaxy 1233+169 in A1569N (Fig. \ref{radio_overlays}b), we tested for the existence of potential cavities in the ICM of A1569N. An unsharp-masked image of the central region of A1569N was created by subtracting a heavily smoothed ($\sim$221 arcsec) X-ray image from a lightly smoothed ($\sim$36 arcsec) image. The resulting image with the radio contours of 1233+169 superimposed is shown in \autoref{cavity}a. The image shows X-ray deficits spatially coincident with the extended features of the radio galaxy 1233+169. The eastern deficit is stronger compared to the western deficit. We note that several smoothing scales were tried to generate the unsharp-masked image and the observed X-ray deficiencies were best visible with the scales mentioned above, likely because the smaller smoothing scale ($\sim$36 arcsec) is similar to the size of the extended radio features. The X-ray deficits seen in Fig. \ref{cavity}a, albeit mild in appearance due to limited photon statistics, are indicative of the presence of potential cavities carved out in the ICM of A1569N by the radio lobes of 1233+169. We note that the unsharp masking technique is highly sensitive to noise and the smoothing scales used. This, along with the low photon statistics, may have led to the over-sharpening of artefacts (chip gaps, bad pixels, unrecorded pixels) in Fig. \ref{cavity}a.

Additionally, we subtracted the best-fitting elliptical 2D-$\beta$ model (\autoref{2dbetatable}) from the data image of A1569N. The resulting residual image is shown in \autoref{cavity}b. Signs of X-ray deficits towards the east and west of the central X-ray peak are more distinctly observed in this image. The gas deficient regions overlap with the radio lobes of 1233+169 and are circled in \textit{green} color in Fig. \ref{cavity}b. These regions appear to be surrounded by excess X-ray emission, in the form of bright arms$/$rims, likely resulting from the displacement and compression of the ICM formerly present in the potential cavities by the radio lobes of 1233+169. We note that in the 2D-$\beta$ modelling of the gas emission from A1569N, the central galaxy was excluded (\S \ref{2Dfit_text}). The central region, therefore, shows up as an excess of X-ray emission in the residual image.  

To test the statistical significance of the X-ray deficits corresponding to the location of the extended radio features of 1233+169, we compared the X-ray surface brightness in the gas deficient regions with the brightness of other ICM regions at similar radii. We measured the $0.5-4.0$ keV background-subtracted and exposure-corrected azimuthal SB (with point sources removed) using an annulus with inner and outer radii of 30 and 65 arcsec, respectively,  divided into sectors with annular width of 24\textdegree{}. The choice of sectors with annular width smaller than 24\textdegree{} made it difficult to assess the significance of the probable cavities, owing to very few photons enclosed. The annular region was centred on the X-ray peak of A1569N and is shown in dashed \textit{white} sectors in \autoref{cavity}c. The SB profile was constructed using the point-source-subtracted and unsmoothed image of A1569N having a pixel size of 0.492 arcsec, and is shown in \autoref{cavity}d. The negative features seen in Fig. \ref{cavity}d correlate with the X-ray SB deficit towards the east and the west of the central peak of A1569N, and overlap with the position of the radio lobes of 1233+169. These sectors within which the X-ray deficits are observed, are highlighted in \textit{magenta} color in Fig. \ref{cavity}c. 
We also performed a similar azimuthal SB analysis in a smaller annular region covering the inner part of the potential eastern cavity. This region has inner and outer radii of 15 and 30 arcsec, and is marked in solid \textit{white} line style in Fig. \ref{cavity}c. The resulting azimuthal X-ray SB profile is shown in \autoref{cavity}e. We further estimated the significance of the X-ray deficits, spatially coincident with the radio features of 1233+169, by calculating the mean number of counts in the regions used for the azimuthal SB analysis. We excluded the sectors within which the negative SB features lie in calculating this mean value. The exposure-corrected and particle-background-subtracted mean value for the inner annulus (15--30 arcsec) is $16.7\pm{4.1}$ counts, whereas this value is $29.7\pm{5.4}$ counts for the outer annulus (30--65 arcsec). The value in the inner eastern deficit (Fig. \ref{cavity}e) is only a mere 3.1 counts, giving a 3.3$\sigma$ significance. The lowest value in the outer eastern deficit (Fig. \ref{cavity}d) is 13.4 counts and that in the western deficit (Fig. \ref{cavity}d) is 16.3 counts, giving significance of 3$\sigma$ and 2.5$\sigma$ respectively.

\begin{figure*}
\begin{multicols}{2}
\subcaptionbox{}{\includegraphics[width=0.6\linewidth,height=8cm,angle=-90]{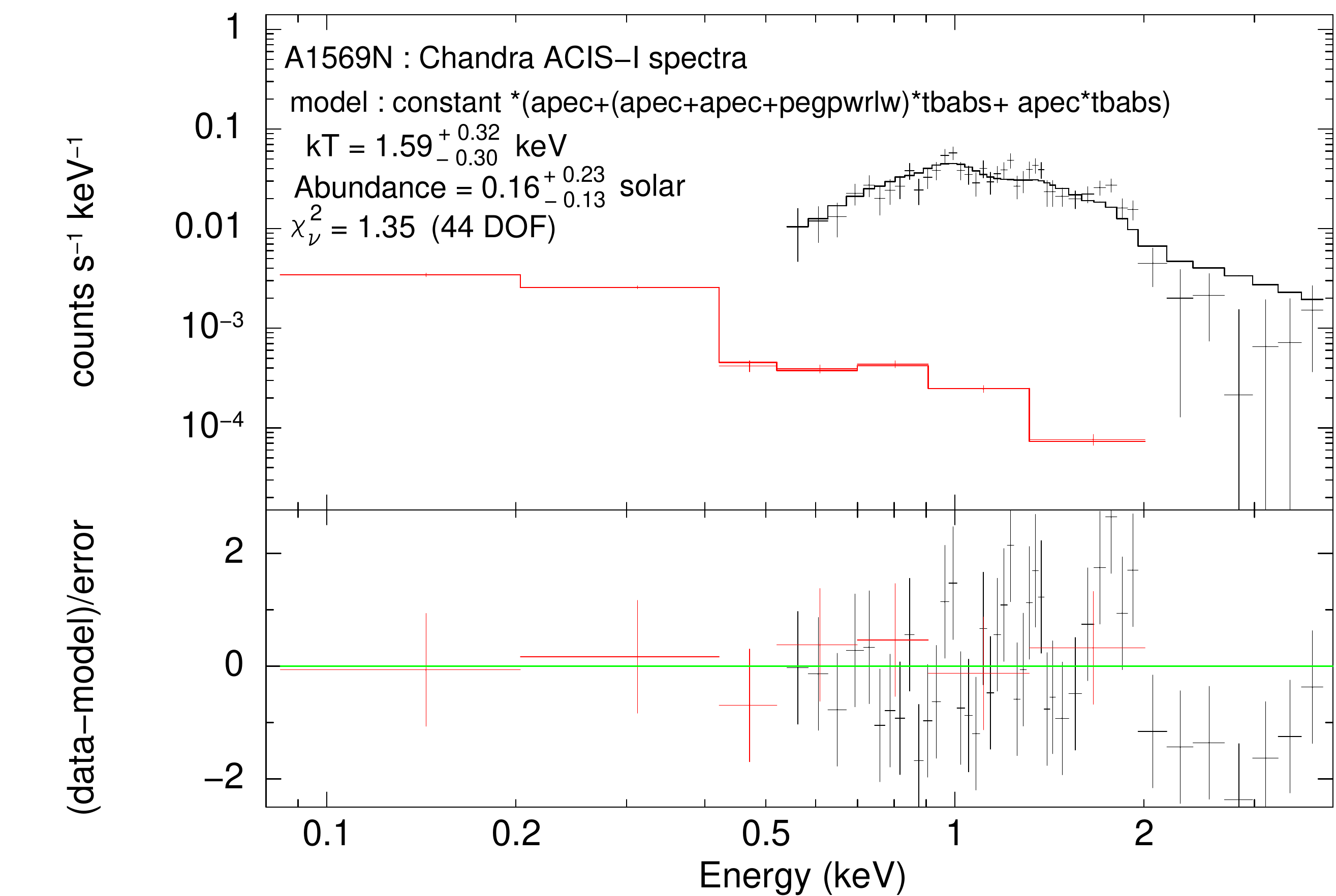}}\par
\subcaptionbox{}{\includegraphics[width=0.6\linewidth,height=8cm,angle=-90]{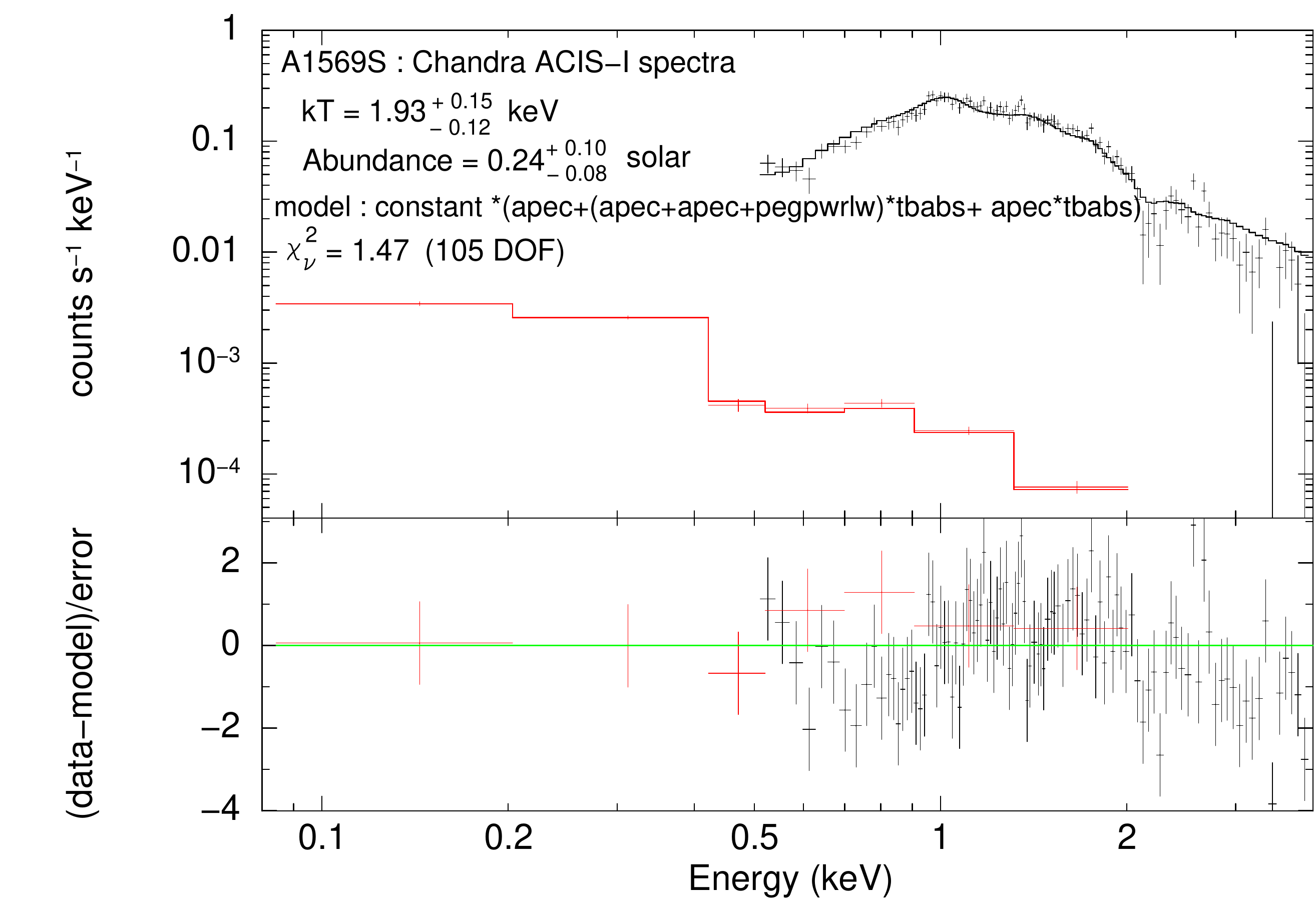}}\par
\end{multicols}
\caption{\textbf{(a)} Average spectrum of A1569N obtained from the \textit{Chandra} ACIS-I detector. The spectrum was fitted simultaneously with the \textit{RASS} diffuse background spectrum (shown in \textit{red}) using the model \textit{constant*(apec+(apec+apec+pegpwrlw)*tbabs + apec*tbabs)} in the energy range 0.5--4.0 keV. \textbf{(b)} Description same as that in panel (a) but for A1569S. The lower panels in the subplots show the residuals divided by 1$\sigma$ error bars.}
\label{fig_globalspec}
\end{figure*}

\begin{table*}
 \begin{center}
  \caption{Best-fitting parameters -- subcluster gas temperature (kT), elemental abundance (Z), and \textit{apec} normalization ($\mathcal{N}$) -- obtained from the X-ray spectral analysis of the full A1569N and A1569S regions (outlined in \textit{white} circles in Fig.\ref{morphology}a ). The minimum reduced $\chi^2$ statistic along with the degrees of freedom (DOF) is also listed. The \textit{Chandra} global spectra were fitted using the model \textit{constant*(apec+(apec+apec+pegpwrlw)*tbabs + apec*tbabs)}, simultaneously with the \textit{RASS} diffuse background spectrum. The error bars correspond to 90 per cent confidence intervals based on $\chi^2_{min}$+2.71.}
  \begin{tabular}{ccccccc}
    \hline
\hline
Region & kT & Z & apec norm. ($\mathcal{N}$) & X-ray Luminosity ($L_X$) & $(\chi_{\nu}^2)_{\text{min}}${\hskip 0.04in}(DOF)\\
&(keV)&(Z$_\odot$)& ($10^{-3}$ cm$^{-5}$)& ($10^{43}$ erg s$^{-1}$)&\\
\\
\hline
\hline
A1569N&$1.6^{+0.3}_{-0.3}$&$0.16^{+0.23}_{-0.13}$&$0.49^{+0.12}_{-0.09}$&$0.45\pm{0.04}$ \hspace{0.22cm}(0.5$-$4.0 keV) &1.35 (44)\\
A1569S&$1.9^{+0.2}_{-0.1}$&$0.24^{+0.10}_{-0.08}$&$2.90^{+0.19}_{-0.20}$&$2.31\pm{0.06}$ \hspace{0.22cm}(0.5$-$4.0 keV) &1.47 (105)\\
\hline\hline
  \end{tabular}
   \label{tab_global}
  \end{center}
 \end{table*}
 
\begin{figure*}\setlength{\columnsep}{1pt}
 {\begin{multicols}{2}
\subcaptionbox{}{\includegraphics[width=0.65\linewidth,height=4.2cm]{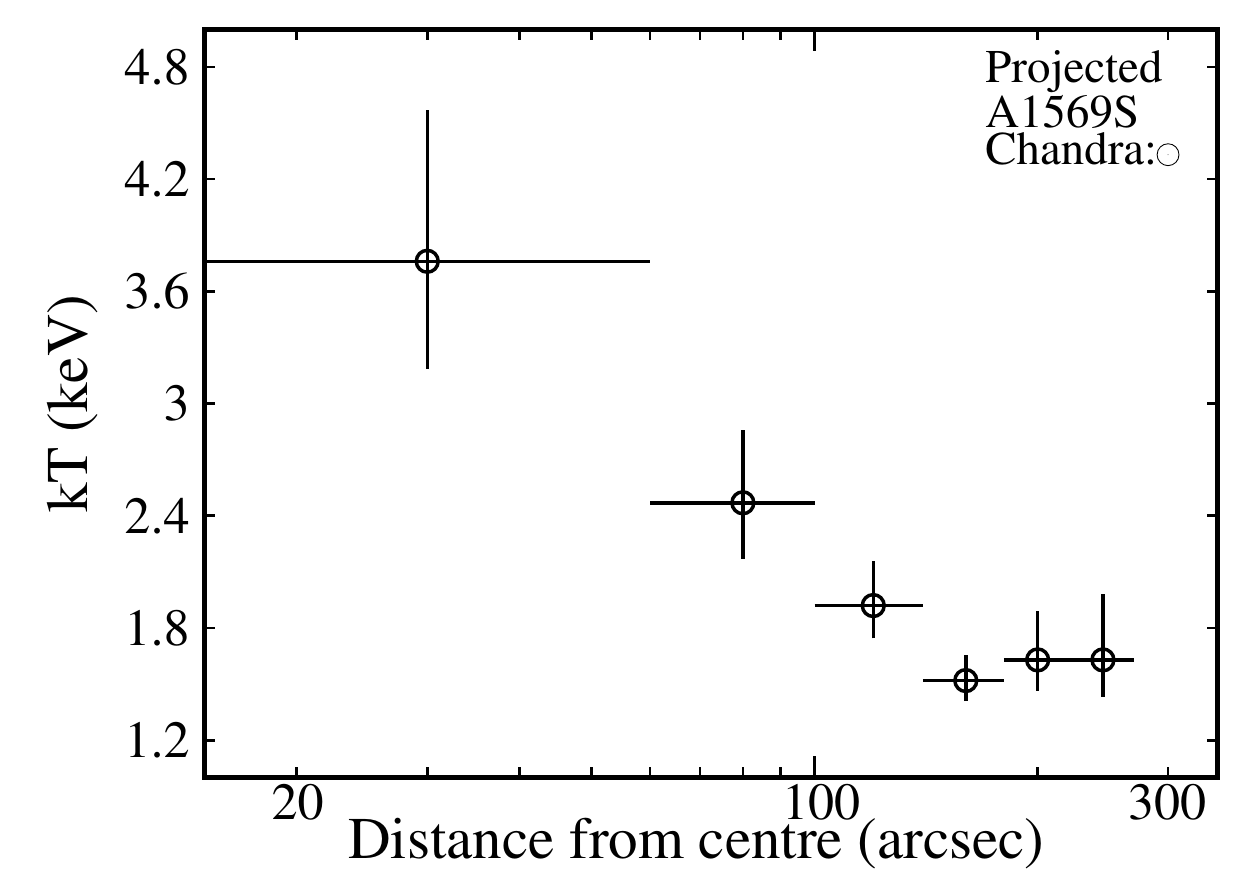}}\par
\subcaptionbox{}{\includegraphics[width=0.65\linewidth,height=4.2cm]{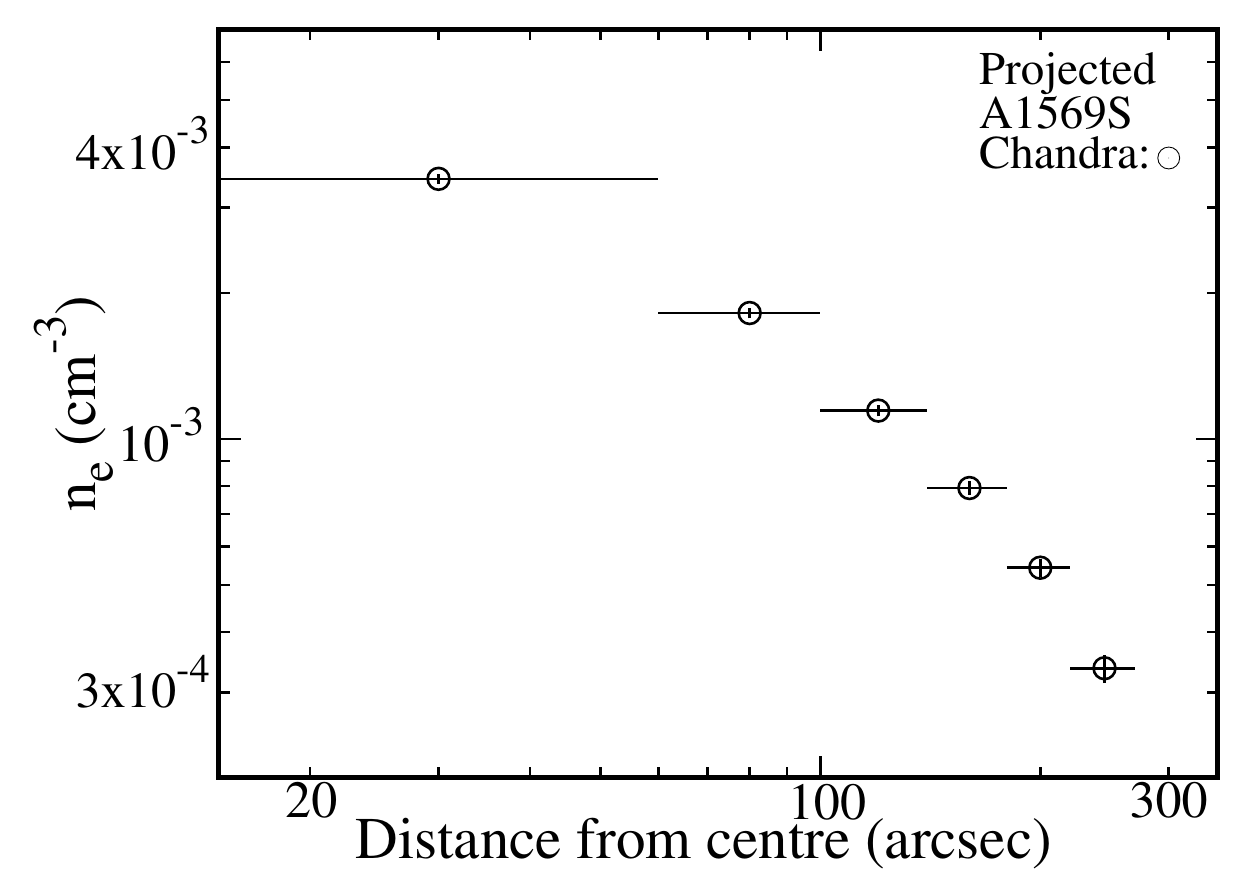}}\par
\end{multicols}
\begin{multicols}{2}
\subcaptionbox{}{\includegraphics[width=0.65\linewidth,height=4.2cm]{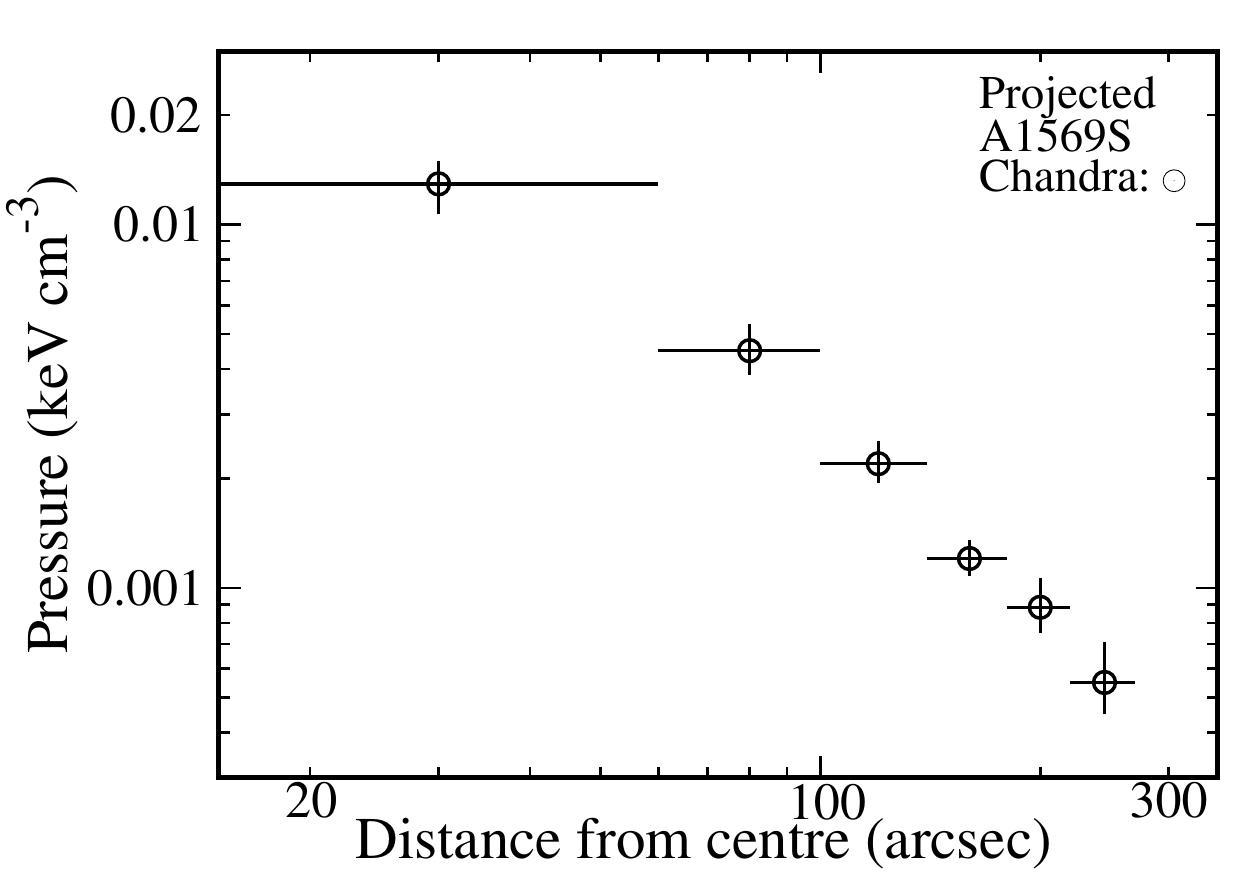}}\par
\subcaptionbox{}{\includegraphics[width=0.65\linewidth,height=4.2cm]{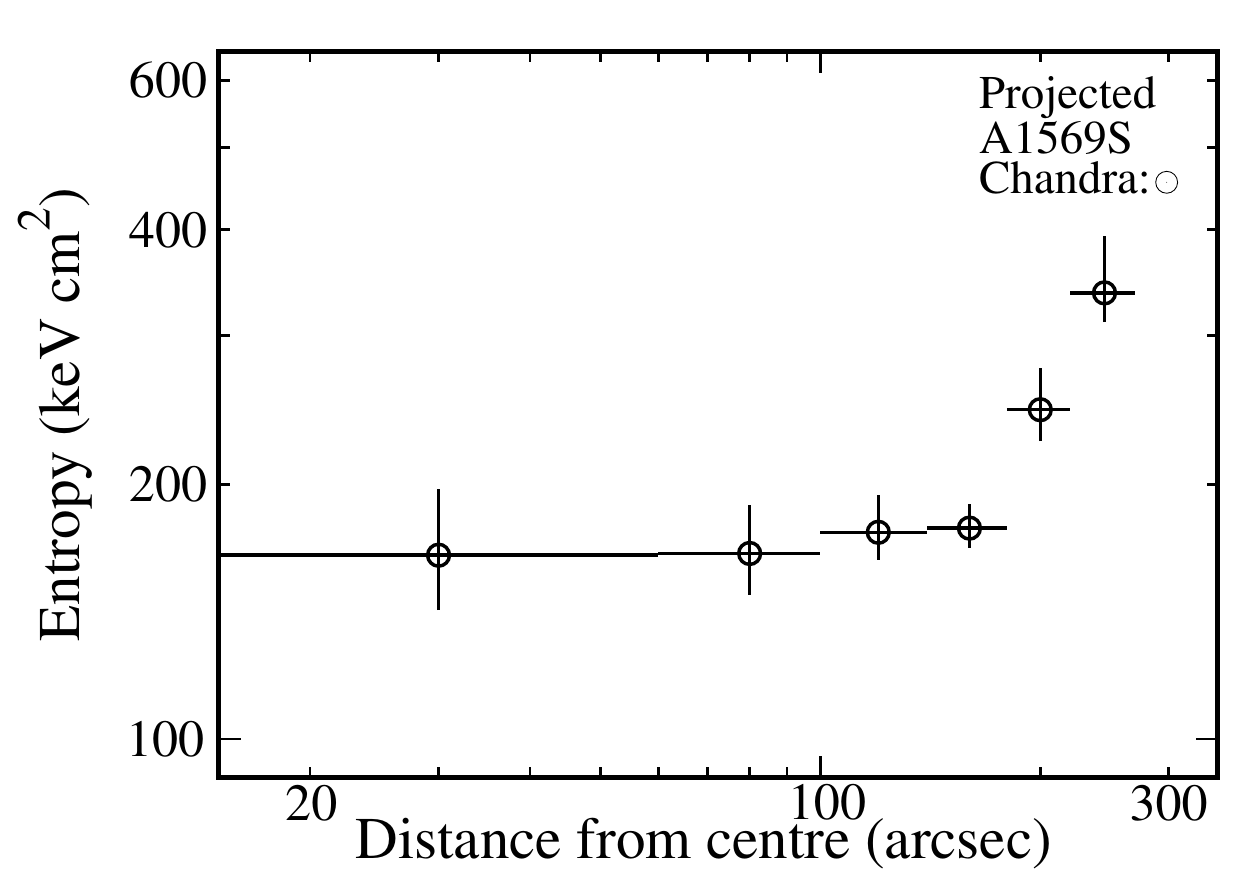}}\par
\end{multicols} }
\caption{Projected temperature (kT), electron density (n$_{e}$), pressure, and entropy profiles of A1569S obtained from the azimuthally averaged spectral analysis of \textit{Chandra} data. The abundance value in each annulus was kept frozen to the global abundance value of A1569S (0.24 Z$_{\odot}$). The error bars correspond to a 90 per cent confidence interval based on $\chi^2_{min}$+2.71.}
\label{fig_projspec}
\end{figure*}

 \begin{figure*}\setlength{\columnsep}{1pt}
{\begin{multicols}{2}
\subcaptionbox{}{\includegraphics[width=0.65\linewidth,height=4.2cm]{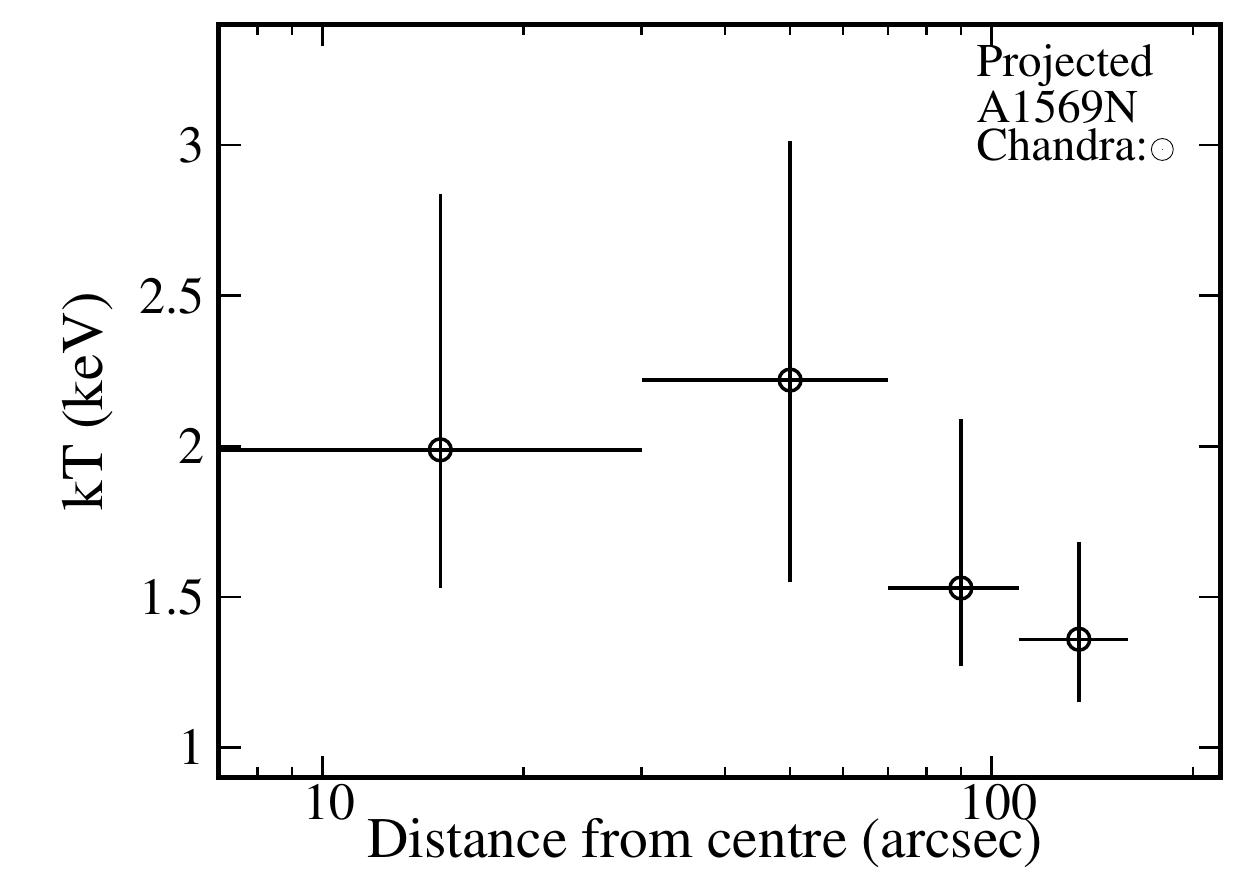}}\par
\subcaptionbox{}{\includegraphics[width=0.65\linewidth,height=4.2cm]{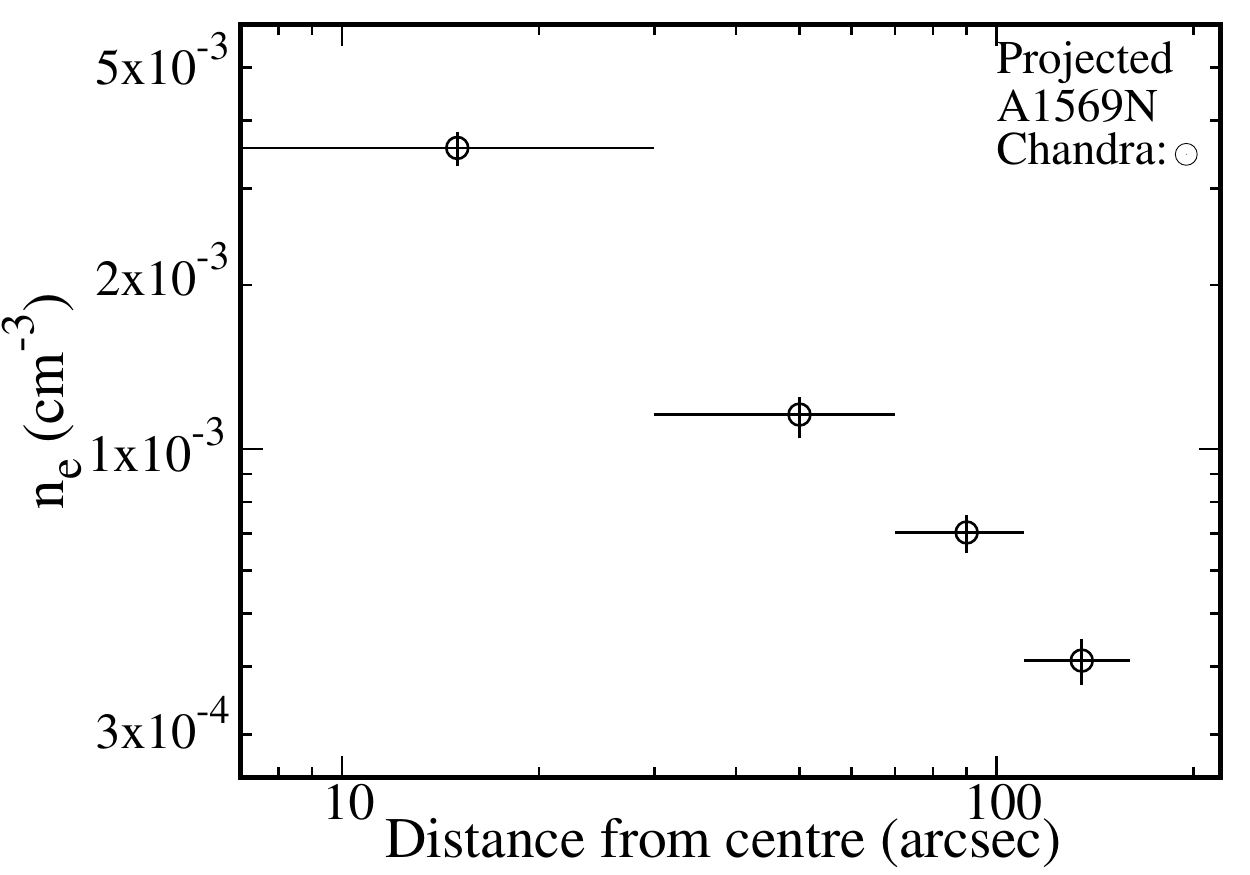}}\par
\end{multicols}
\begin{multicols}{2}
\subcaptionbox{}{\includegraphics[width=0.65\linewidth,height=4.2cm]{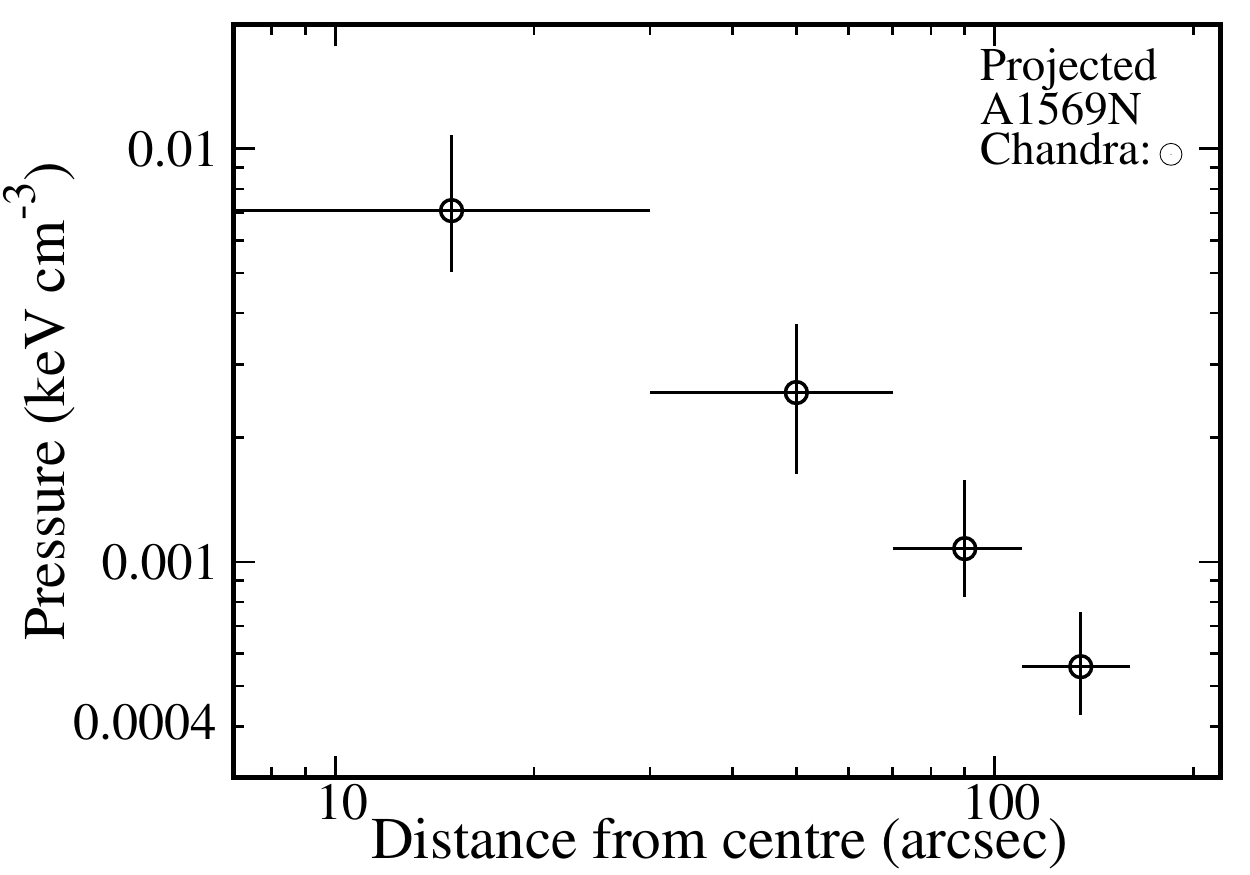}}\par
\subcaptionbox{}{\includegraphics[width=0.65\linewidth,height=4.2cm]{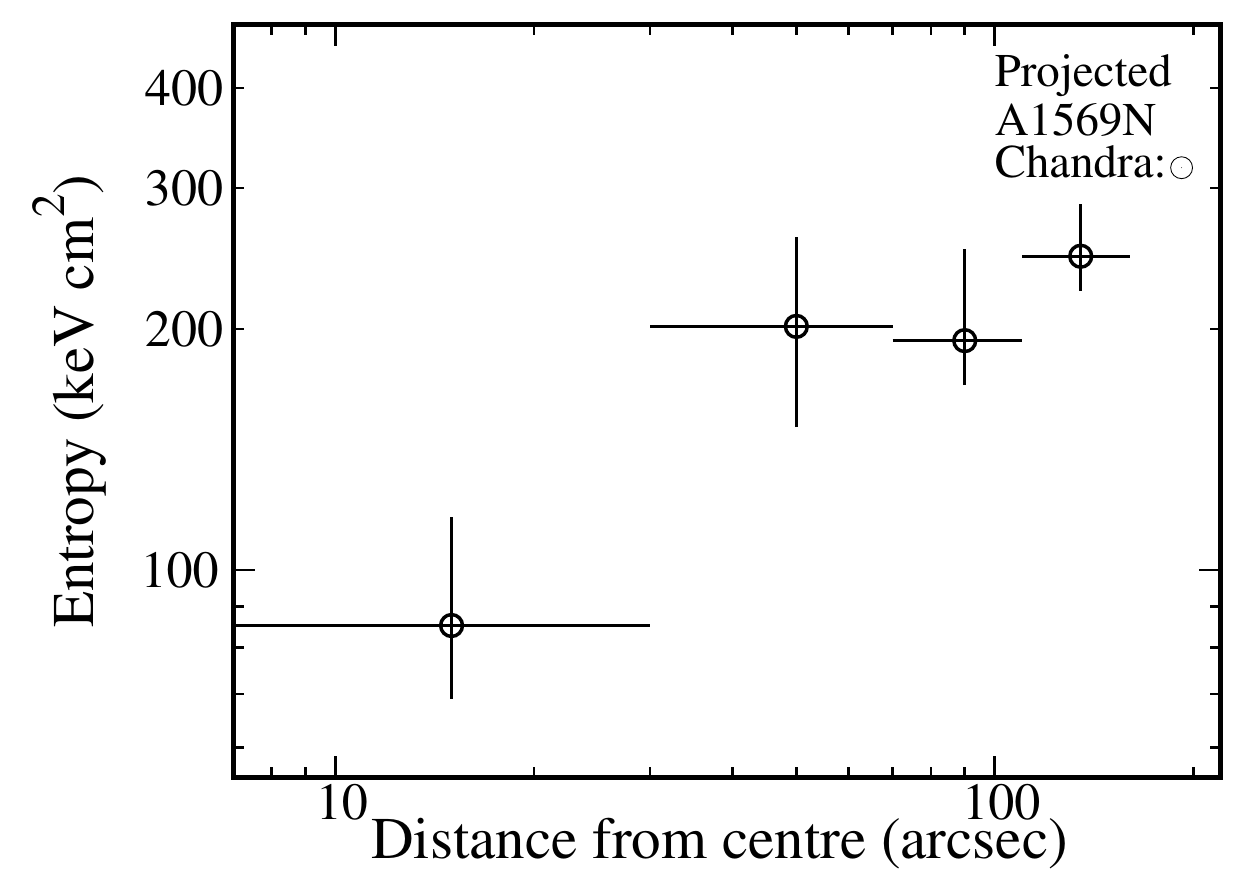}}\par
\end{multicols}}
\caption{Projected temperature (kT), electron density (n$_{e}$), pressure, and entropy profiles of A1569N obtained from the azimuthally averaged spectral analysis of \textit{Chandra} data. The abundance value in each annulus was kept frozen to the global abundance value of A1569N (0.16 Z$_{\odot}$). The error bars correspond to a 90 per cent confidence interval.}
\label{fig_projspec_A1569N}
\end{figure*}

\section{Spectral analysis}\label{spectral}
\subsection{Average X-ray spectra}\label{avgspec}
Average X-ray spectra of A1569N and A1569S were extracted from regions outlined in white color in Fig.\ref{morphology}a. The regions were centred on the BCGs of the two subclusters (\S3.1). A circular region of radius 160 arcsec was selected for the average spectral extraction of A1569N, whereas a circular region of radius 270 arcsec was chosen for A1569S. Background spectra and responses in these regions were generated as described in \textbf{\S 2.1}. The extracted spectra were simultaneously fitted with the \textit{RASS} diffuse background spectrum using the XSPEC model: \textit{constant*(\textcolor{magenta}{apec}+(\textcolor{red}{apec}+\textcolor{red}{apec}+pegpwrlw)*tbabs + \textcolor{cyan}{apec*tbabs})}. The individual model component definitions are the same as described in \S3.2 of \citet{hercules2021}. We fixed the total hydrogen column density (neutral+molecular), $N_H$, along the line of sight (LOS) to the subclusters, to values provided by the UK Swift Science Data Centre\footnote{\url{https://www.swift.ac.uk/analysis/nhtot}} \citep{willingale2013}. The $N_H$ value is $2.20 \times 10^{20}$ cm$^{-2}$ for A1569N and $2.23 \times 10^{20}$ cm$^{-2}$ for A1569S. We used a redshift value of 0.0793 for A1569N and 0.0691 for A1569S \citep{gomez1997_a1569}.

The \textit{RASS} background spectrum is in units of cts s$^{-1}$ arcmin$^{-2}$. Therefore, for consistency of units in the spectral fitting, a \textit{constant} factor was used to scale down the model normalization of the \textit{Chandra} ACIS-I data. The \textit{constant} factor used in the model is the active sky area available to the detector in arcmin$^2$ units and its value can be derived from the BACKSCAL keyword in the spectral file header. The component \textit{\textcolor{magenta}{apec}+(\textcolor{red}{apec}+\textcolor{red}{apec}+pegpwrlw)*tbabs} models the contribution of the XRB. The first \textcolor{magenta}{\textit{apec}} component here represents the unabsorbed thermal emission from the Local Hot Bubble (kT fixed at 0.1 keV). The remaining two \textcolor{red}{\textit{apec}} components represent the absorbed thermal emission from the cooler (kT fixed at 0.1 keV) and the hotter Galactic Halo (and/or emission from the Local Group) (kT fixed at 0.27 keV\footnote{This value was obtained from fitting the \textit{RASS} diffuse background spectrum with the XRB model}) \citep{snowden2008}. A1569 lies either in the direction of the North Polar Spur (NPS) outskirts or beyond them, as seen in the \textit{ROSAT} diffuse background images. The contribution of the NPS to the total soft X-ray background should therefore be very weak (if any). We added an extra absorbed \textit{apec} component to the model to account for the thermal emission from the NPS. This resulted in a very low normalization for the \textit{apec} component representing the NPS emission and left the other best-fitting parameters unchanged. This component was therefore left out of the spectral model. The \textit{pegpwrlw} component represents the cosmic X-ray background (CXB) contribution. Its photon index ($=1.4$) and normalization (2--8 keV CXB flux value $1.7\pm{0.2} \times 10^{-11} \text{erg cm}^{-2} \text{s}^{-1}\text{deg}^{-2}$) were fixed based on the study of \citet{hickox2006}. The component \textcolor{cyan}{\textit{apec*tbabs}} accounts for the absorbed X-ray emission from the cluster gas. We note that the \textit{constant} factor and the normalization value of the \textcolor{cyan}{\textit{apec}} model describing the cluster emission were set equal to 1.0 and 0.0 respectively, for the \textit{RASS} background spectrum.  The temperature, abundance and normalization of the \textit{apec} component representing the emission from the subclusters were allowed to vary freely during the fit. The spectral fitting was done in the energy range 0.5--4.0 keV. The data along with the model spectra are shown in \autoref{fig_globalspec}. The best-fitting values of gas temperature, abundance, and \textit{apec} normalization for the two subclusters are provided in \autoref{tab_global}. We note that the X-ray source visible on the eastern outskirts of A1569S (Fig. \ref{morphology}a) was removed from the spectral analysis. Additionally, the effect of including the central BCGs on the average and radial ICM thermodynamic properties (\S4.2) was tested by excluding the central 15 arcsec region from the spectral analysis. This did not produce a notable change in the derived gas properties.     

\begin{figure*}\setlength{\columnsep}{1pt}
{\begin{multicols}{2}
\subcaptionbox{}{\includegraphics[width=0.65\linewidth,height=4.2cm]{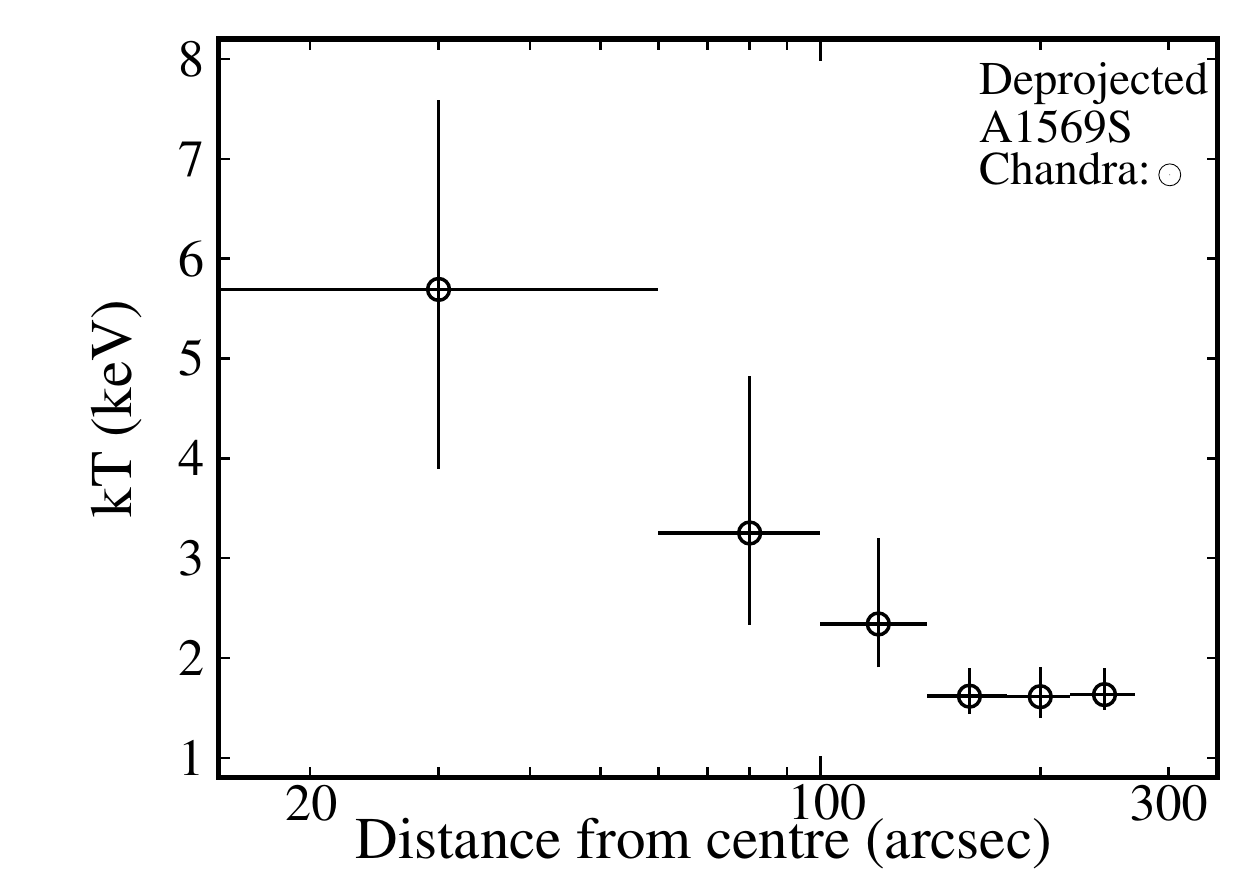}}\par
\subcaptionbox{}{\includegraphics[width=0.65\linewidth,height=4.2cm]{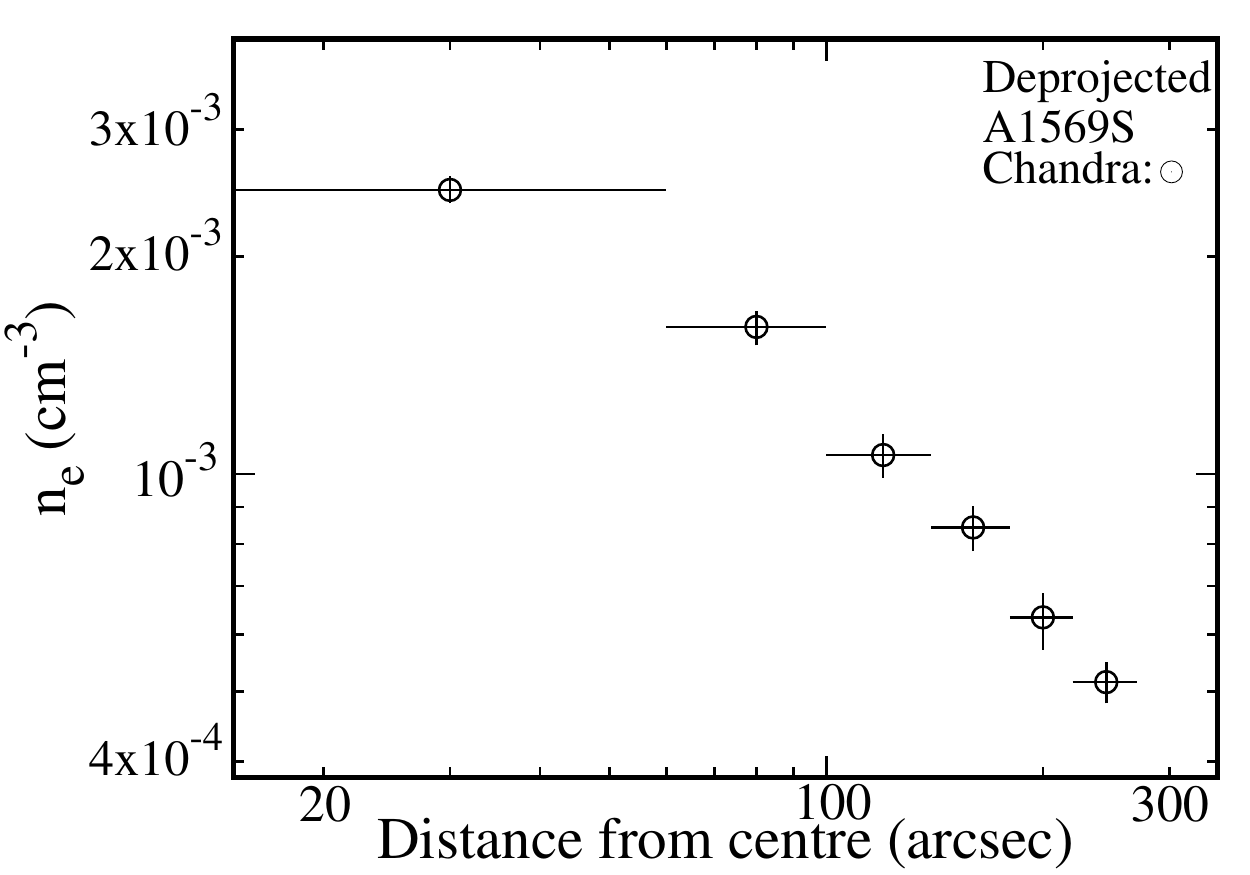}}\par
\end{multicols}
\begin{multicols}{2}
\subcaptionbox{}{\includegraphics[width=0.65\linewidth,height=4.2cm]{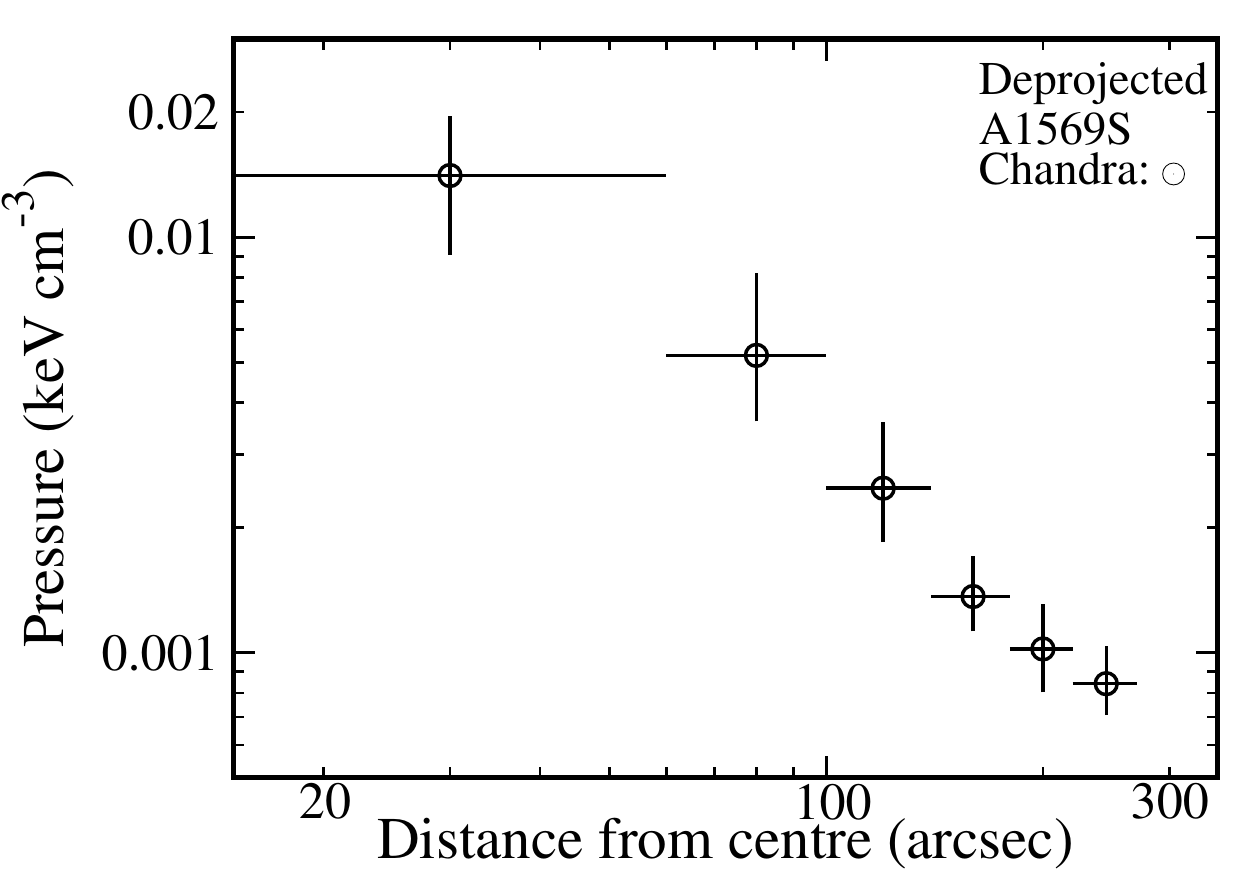}}\par
\subcaptionbox{}{\includegraphics[width=0.65\linewidth,height=4.2cm]{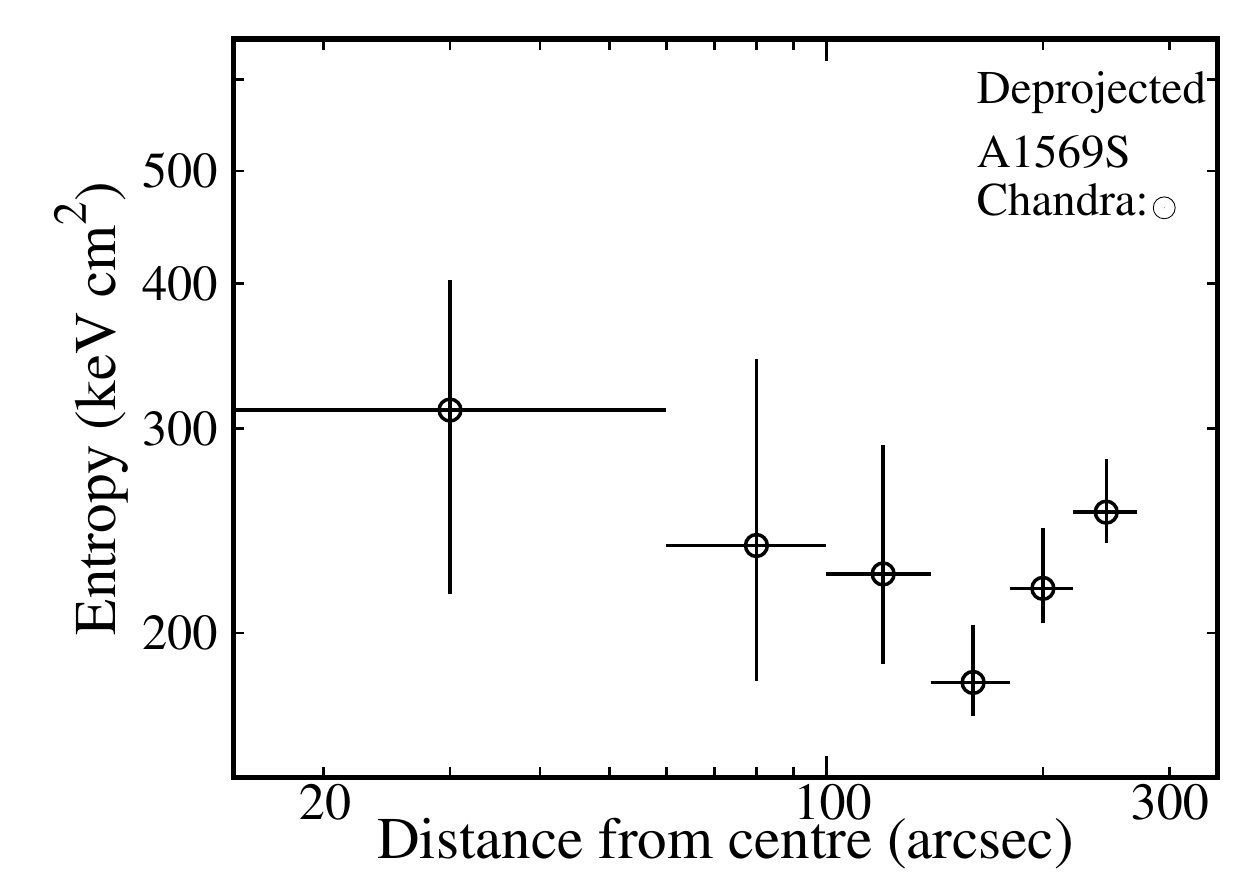}}\par
\end{multicols} }
\caption{Deprojected temperature (kT), electron density (n$_{e}$), pressure, and entropy profiles of A1569S obtained from the spectral analysis of \textit{Chandra} data. The abundance value in each annulus was kept frozen to the global abundance value of A1569S (0.24 Z$_{\odot}$). The error bars correspond to 90 per cent confidence interval.}
\label{fig_deprojspec}
\end{figure*}

\begin{table*}
 \begin{center}
 \caption{Best-fitting temperature and the derived electron density ($n_{e}$), pressure (P), and entropy (K) values obtained from the projected spectral analysis of the six annular regions in A1569S (0.5--4.0 keV) (\S4.2.1). The regions used for spectral extraction are listed. The \textit{Chandra} spectra were fitted using the model \textit{constant*(apec+(apec+apec+pegpwrlw)*tbabs + apec*tbabs)} and were simultaneously analysed with the \textit{RASS} diffuse background spectrum. The elemental abundances (Z) in the subcluster were frozen to the average value obtained from the global spectral analysis of A1569S. The errors are quoted at 90 per cent confidence level.}
  \begin{tabular}{ccccc}
    \hline
\hline
Region&kT & n$_{e}$& P &K\\
&(keV)&($10^{-3}$cm$^{-3}$)&($10^{-2}$ keV cm$^{-3}$)&(keV cm$^2$)\\
\\
\hline
\hline
0--60 arcsec&$3.8_{-0.6}^{+0.8}$&$3.44\pm{0.08}$&$1.29_{-0.22}^{+0.21}$&$165_{-23}^{+33}$\\
\\
60--100 arcsec&$2.5_{-0.3}^{+0.4}$&$1.82_{-0.05}^{+0.04}$&$0.45_{-0.06}^{+0.08}$&$166_{-18}^{+23}$\\
\\
100--140 arcsec&$1.9_{-0.2}^{+0.2}$&$1.14\pm{0.03}$&$0.22_{-0.02}^{+0.03}$&$176_{-13}^{+19}$\\
\\
140--180 arcsec&$1.5_{-0.1}^{+0.1}$&$0.79_{-0.03}^{+0.02}$&$0.12\pm{0.01}$&$178_{-9}^{+12}$\\
\\
180--220 arcsec&$1.6_{-0.2}^{+0.3}$&$0.54\pm{0.02}$&$0.09_{-0.01}^{+0.02}$&$245_{-20}^{+30}$\\
\\
220--270 arcsec&$1.6_{-0.2}^{+0.4}$&$0.34\pm{0.02}$&$0.05_{-0.01}^{+0.02}$&$337_{-26}^{+56}$\\
\hline
\hline
  \end{tabular}
  \label{protab}
  \end{center}
 \end{table*}        

 \begin{table*}
 \begin{center}
 \caption{Best-fitting temperature and the derived electron density ($n_{e}$), pressure (P), and entropy (K) values obtained from the projected spectral analysis of the four annular regions in A1569N (0.5--4.0 keV) (\S4.2.1). The regions used for spectral extraction are listed. The \textit{Chandra} spectra were fitted using the model \textit{constant*(apec+(apec+apec+pegpwrlw)*tbabs + apec*tbabs)} and were simultaneously analysed with the \textit{RASS} diffuse background spectrum. The cluster abundance (Z) was frozen to the average value obtained from the global spectral analysis of A1569N. The errors are quoted with 90 per cent confidence.}
  \begin{tabular}{ccccc}
    \hline
\hline
Region&kT & n$_{e}$& P &K\\
&(keV)&($10^{-3}$cm$^{-3}$)&($10^{-3}$ keV cm$^{-3}$)&(keV cm$^2$)\\
\\
\hline
\hline
0--30 arcsec&$2.0_{-0.5}^{+0.9}$&$3.56_{-0.27}^{+0.24}$&$7.09_{-2.04}^{+3.71}$&$85_{-16}^{+31}$\\
\\
30--70 arcsec&$2.2_{-0.7}^{+0.8}$&$1.16_{-0.11}^{+0.09}$&$2.57_{-0.94}^{+1.19}$&$201_{-51}^{+59}$\\
\\
70--110 arcsec&$1.5_{-0.3}^{+0.6}$&$0.70_{-0.06}^{+0.05}$&$1.08_{-0.26}^{+0.50}$&$193_{-23}^{+58}$\\
\\
110--160 arcsec&$1.4_{-0.2}^{+0.3}$&$0.41\pm{0.04}$&$0.56_{-0.13}^{+0.20}$&$246_{-23}^{+40}$\\
\hline
\hline
  \end{tabular}
  \label{protab_A1569N}
  \end{center}
 \end{table*}
 
 \begin{table*}
 \begin{center}
 \caption{Best-fitting temperature and the derived electron density ($n_{e}$), pressure (P), and entropy (K) values obtained from the deprojected spectral analysis of the six annular regions in A1569S (0.5--4.0 keV). The regions used for spectral extraction are listed. The spectra were fitted using the model \textit{apec+(apec+apec+pegpwrlw)*tbabs + projct*(apec*tbabs)}. The XRB model normalizations within each region were fixed to values obtained from the projected spectral analysis, and the cluster abundance (Z) was frozen to the average value obtained from the global spectral analysis of A1569S. The errors are quoted at 90 per cent confidence level.}
  \begin{tabular}{ccccc}
    \hline
\hline
Region&kT & n$_{e}$& P &K\\
&(keV)&($10^{-3}$cm$^{-3}$)&($10^{-2}$ keV cm$^{-3}$)&(keV cm$^2$)\\
\\
\hline
\hline
0--60 arcsec&$5.7_{-1.8}^{+1.9}$&$2.47\pm{0.11}$&$1.41_{-0.50}^{+0.55}$&$311_{-95}^{+92}$\\
\\
60--100 arcsec&$3.3_{-0.9}^{+1.6}$&$1.60\pm{0.09}$&$0.52_{-0.16}^{+0.30}$&$238_{-56}^{+107}$\\
\\
100--140 arcsec&$2.3_{-0.4}^{+0.9}$&$1.06\pm{0.07}$&$0.25_{-0.06}^{+0.11}$&$225_{-37}^{+66}$\\
\\
140--180 arcsec&$1.6_{-0.2}^{+0.3}$&$0.84\pm{0.06}$&$0.14_{-0.02}^{+0.03}$&$181_{-12}^{+22}$\\
\\
180--220 arcsec&$1.6_{-0.2}^{+0.3}$&$0.63_{-0.06}^{+0.05}$&$0.10_{-0.02}^{+0.03}$&$219_{-15}^{+28}$\\
\\
220--270 arcsec&$1.6_{-0.2}^{+0.3}$&$0.52\pm{0.03}$&$0.08_{-0.01}^{+0.02}$&$254_{-15}^{+28}$\\
\hline
\hline
  \end{tabular}
  \label{deprotab}
  \end{center}
 \end{table*}

\subsection{Azimuthally averaged spectral analysis: Radial profiles of gas temperature, electron density, pressure and entropy}\label{radialprof}
\subsubsection{2D projected profiles}\label{proj}
Radial profile of gas temperature was obtained for A1569S by extracting spectra in six annular regions with outer radii 60, 100, 140, 180, 220, and 270 arcsec centred on the X-ray peak mentioned in  \S3.1.  The regions are shown in \autoref{morphology}b. Each of the chosen regions had $>3000$ source counts after background subtraction. Details of spectral modelling are as described in \S4.1. The spectral fit was performed in the energy range 0.5--4.0 keV. We note that the gas abundance in each annular region was fixed at the average ICM abundance (0.24 Z$_{\odot}$) obtained for A1569S (\S4.1) since the errors in the abundance value were rather large if it was left free. The best-fitting temperature values are provided in \autoref{protab} and the radial temperature profile is shown in \autoref{fig_projspec}a. The gas electron density was derived from the cluster \textit{apec} normalization, $\mathcal{N}$, defined as: \begin{equation}\mathcal{N}=\frac{10^{-14}}{4\pi[D_{A}(1+z)]^{2}}\int n_{e}n_{H} dV \hspace{0.6cm}  (cm^{-5})\label{eq:eq1}\end{equation} where $D_{A}$ is the angular diameter distance to the source (cm), $z$ is the redshift, $V$ is the volume of the region, and $n_{e}$ and $n_{H}$ are the electron and hydrogen densities (cm$^{-3}$), respectively. Assuming that electron density does not vary within a spherical shell, and using $n_{H}/n_{e}=0.835$, we derived the projected electron density profile shown in \autoref{fig_projspec}b. The $n_{H}/n_{e}$ ratio was derived for an ICM abundance of 0.24 Z$_{\odot}$ (average abundance of A1569S obtained in \S4.1) using the method described in  \S3.3.1 of \citet{hercules2021}. The derived value of the mean molecular weight ($\mu$) was 0.608. Gas pressure (P) and entropy (K) was calculated using the following:
\begin{equation}P=n_{e}kT \hspace{0.6cm} (keV cm^{-3})\end{equation}
\begin{equation} K=n_{e}^{-2/3}kT \hspace{0.4cm} (keV cm^{2})\end{equation}
The projected pressure and entropy profiles are shown in \autoref{fig_projspec}c and \autoref{fig_projspec}d, respectively.

In order to check for any potential temperature drop towards the central $r \sim$40 kpc region of A1569S, we additionally performed a projected spectral analysis in two inner radial bins -- 0--30 and 30--60 arcsec. The best-fitting temperature value in the inner bin 0--30 arcsec (0--41 kpc) was found to be $6.4^{+5.8}_{-2.1}$ keV corresponding to a derived entropy of $201^{+193}_{-70}$ keV cm$^2$. The gas temperature and entropy in the annular bin 30--60 arcsec were found to be $3.0^{+0.7}_{-0.5}$ keV and $144^{+38}_{-27}$ keV cm$^{2}$, respectively. The clear lack of a drop in temperature towards the inner 40 kpc region together with a high central entropy value indicates that A1569S does not possess a large cool core ($\gtrsim40$ kpc) associated with the intracluster gas. We note that this analysis was done only to examine the presence of a potential ICM-associated cool core in A1569S. A deprojection analysis (described in \S4.2.2 below) including these small inner bins was not workable due to a low number of counts in these regions. For a consistent representation of the projected and deprojected radial profiles of the gas properties of A1569S, we show six data points (as indicated previously) in Fig.\ref{fig_projspec}, with the innermost data point representing the central 0--60 arcsec region.

We also performed a rather crude spectral analysis in four annular bins of A1569N with outer radii 30, 70, 110, and 160 arcsec centred on the X-ray peak mentioned in  \S3.1 (Fig.\ref{morphology}b). The regions were chosen to have a minimum of only 600 background-subtracted counts due to low photon statistics. The gas abundance in each annular region was fixed at the average ICM abundance (0.16 Z$_{\odot}$) obtained for A1569N (\S4.1). The best-fitting temperature values and the derived electron density, pressure, and entropy values are listed in \autoref{protab_A1569N}, and the radial profiles are shown in \autoref{fig_projspec_A1569N}. 

\subsubsection{Deprojected profiles}
We also carried out a deprojection analysis on the spectra extracted from the six annular regions of A1569S mentioned in \S4.2.1, using the same technique as described in \S3.3.2 of \citet{hercules2021}. This was done in order to correct for the smoothing of variation in the measured thermodynamic quantities due to projection effects. The energy range 0.5--4.0 keV was used for fitting the data. The cluster abundance value was frozen to 0.24 Z$_{\odot}$ as before. The best-fitting temperature values and the derived electron density, pressure, and entropy values are listed in \autoref{deprotab}. The deprojected profiles of these quantities are shown in \autoref{fig_deprojspec}. 

\begin{table*}
 \begin{center}
  \captionsetup{justification=centering}
\caption{Mass of the intra-group gas obtained by fitting single-$\beta$ models to the gas density profiles of A1569N and A1569S. The errors are quoted at 90 per cent confidence level. Column (1): Subcluster name; Column (2): ratio of the specific energy in galaxies to the specific energy in hot gas; Column (3): projected core radius of the subcluster; Column (4): $\rho_0=\mu n_e(0) m_p$, where $m_p$ is the mass of proton, $\mu$ is the mean molecular weight of the gas, and $n_e(0)$ is the central electron gas density; Column (5): Minimum reduced chi-square statistic and the degrees of freedom; Column (6): Radius within which the gas mass is derived; Column (7): Derived gas mass of the subcluster.}
  \begin{tabular}{ccccccc}
    \hline
\hline
Gas clump & $\beta$ & $r_{c}$  & $\rho_{0}$ &$(\chi_{\nu}^2)_{\text{min}} \text{(DOF)}$ & $r$ & $M_{gas}(r)$   \\
&&(kpc)&$(10^{4} M_{\odot} \text{ kpc}^{-3})$&&(kpc)&$(10^{12}$ M$_{\odot})$ \\\vspace{0.05cm}
(1) & (2) & (3)  & $(4)$ & (5) & (6) & (7)   \\
\hline
A1569N&$0.34^{+0.03}_{-0.02}$&$15.0_{-1.4}^{+0.6}$&$10.3\pm{0.7}$&1.27 (3) &248&$0.57_{-0.18}^{+0.15}$ \\
\\
A1569S&$0.41^{+0.10}_{-0.07}$&$87.5_{-26.5}^{+32.1}$&$4.2^{+0.5}_{-0.4}$&0.66 (1) &370&$2.3_{-1.3}^{+1.8}$ \\
\hline\hline
  \end{tabular}
  \label{tabgasmass}
  \end{center}
 \end{table*}

Deprojection was also tried on the four annular spectral bins of A1569N, resulting in $(\chi_{\nu}^2)_{\text{min}}=1.14$ (173 dof). The cluster abundance value was frozen to the average abundance value of A1569N (0.16 Z$_{\odot}$). The deprojected temperature and electron density in the central 30 arcsec region were found to be $1.82^{+0.66}_{-0.51}$ keV and $3.06^{+0.29}_{-0.33} \times 10^{-3}$ cm$^{-3}$, respectively. These are not significantly different from the projected temperature and density values. Although the deprojection analysis resulted in a high temperature (5.41 keV) in the second annular bin (30--70 arcsec region where the radio lobes of 1233+169 primarily reside), and an electron density value of $0.86\pm{0.15} \times 10^{-3}$ cm$^{-3}$, error limits on the temperature value could not be obtained. For this reason, the deprojected radial profiles of the gas properties of A1569N are not presented, here. The deprojected gas temperature and electron density were found to be $1.62^{+1.20}_{-0.43}$ keV and $0.64^{+0.11}_{-0.13} \times 10^{-3}$ cm$^{-3}$, respectively for the third annular region (70--110 arcsec), and $1.37^{+0.32}_{-0.17}$ keV and $0.54^{+0.05}_{-0.06} \times 10^{-3}$ cm$^{-3}$ respectively, for the fourth annular bin (110--160 arcsec).        

\section{X-ray luminosity, gas mass, total mass, and cooling time estimates}
The X-ray luminosities, gas mass, total mass, and cooling time values of the two subclusters -- A1569N and A1569S -- were estimated using the methods described in \S3 of \citet{hercules2021}. The 0.5--4.0 keV luminosities ($L_X$) of A1569N and A1569S are listed in \autoref{tab_global}.
  
\subsection{Gas mass estimates}
The gas density profiles of A1569N and A1569S obtained in \S4.2.1 and \S4.2.2 were used to estimate the gas mass of the two subclusters. The deprojected electron density profile was used for A1569S, whereas the projected density profile was used for A1569N owing to uncertainty in the reliability of the deprojection analysis in case of A1569N (\S4.2.2). We obtained the gas mass ($M_{gas}$) of A1569N and A1569S out to radii 160 arcsec ($\sim$248 kpc) and 270 arcsec ($\sim$370 kpc), respectively from the corresponding X-ray peaks. The results of the single-$\beta$ model fitting ($\beta$, $r_c$, and $\rho_0$) to the gas density profiles along with the gas mass estimates are listed in \autoref{tabgasmass}.\\\\The $\beta$ values obtained by fitting a single-$\beta$ model to the electron density profiles of A1569N and A1569S are consistent with those resulting from the 2D-$\beta$ modelling of the gas surface brightness assuming isotropy (\S3.2; \autoref{2dbetatable}). The core radius is, however, noticeably different for A1569N in the two cases. This is probably due to the exclusion of the central 15 arcsec region of the galaxy 1233+169 while modelling the 2D surface brightness distribution of the gas in A1569N (\S3.2), resulting in a large core radius. The electron density profile fitted with a single-$\beta$ model was, however, derived from the results of the radial spectral analysis of A1569N (\S4.2.1) which included the central 15 arcsec region in the innermost bin.

\subsection{Total mass estimates}

We estimated the total gravitational mass of A1569N within a radius of 248 kpc from the X-ray peak to be $2.0\pm{1.5}\times10^{13}$ M$_{\odot}$. The total mass of A1569S within a radius of 370 kpc was calculated to be $2.1\pm{1.6}\times 10^{13}$ M$_{\odot}$. We note that the results of the deprojection spectral analysis were used in the mass calculation of A1569S. In case of A1569N, results of the projection spectral analysis were used owing to uncertainty in the reliability of the deprojection analysis of A1569N. The errors on the mass values correspond to a 90 per cent confidence interval.

\subsection{Cooling time}

Using the deprojected temperature and electron density values obtained in \S4.2.2, the cooling time of A1569N in the central 30 arcsec region was found to be $8.6_{-4.0}^{+7.2} \times 10^{9} \text{ yr}$ and that of A1569S in the central 60 arcsec region was estimated to be $6.6_{-3.2}^{+4.4} \times 10^{10} \text{ yr}$. These values are comparable to/greater than the Hubble time ($\sim$14.5 $\times 10^{9} \text{ yr}$ in our cosmology). We note that due to the low number of counts in the observation, these estimates are averages over fairly large central regions. Nonetheless, these large cooling time estimates provide support to the observed lack of drop in gas temperature in the central $\sim$40--50 kpc subcluster regions (\S4.2.1 and \S4.2.2), thus, confirming that large cool cores associated with the intracluster gas are absent in both A1569N and A1569S.
 
\section{Radio galaxies and the surrounding gas}
\subsection{Galaxy 1233+169 in A1569N}
The apparent gas substructure (Fig. \ref{radio_overlays}b) and evident gas deficits (\S3.3) in A1569N coincident with the extended features of 1233+169, suggest the presence of an ongoing interaction between the radio galaxy 1233+169 and the surrounding ICM. These gas deficient regions appear to be cavities created in the ICM of A1569N by the radio lobes of 1233+169. Approximating the shape of these supposed cavities as spheres (circular regions marked in \textit{green} color in Fig. \ref{cavity}b), we estimate the time-averaged mechanical power output of the radio jet creating the cavities. We note that this is a rather crude estimate, since due to limited photon statistics of the X-ray observation, it is not possible to determine the exact shapes of the cavity pair. None the less, it gives an idea about the role that the central radio source can play in affecting the energetics of the surrounding ICM in A1569N. Using the standard procedure adopted in literature \citep{birzan2004}, the total energy associated with a single X-ray cavity ($E_{cav}$) was calculated as the sum of the work done by the radio jet against the surrounding gas at pressure $P$ to inflate the cavity with volume $V$ and the internal energy of the fluid within the cavity.
\begin{equation}E_{cav} = PV +\frac{PV}{\gamma-1}; \hspace{0.25cm} \gamma=C_P/C_V \hspace{0.15cm}\text{(ratio of the specific heats)}\label{eqn:ecav}\end{equation}            
Assuming that the cavity is filled with relativistic fluid ($\gamma=4/3$), \autoref{eqn:ecav} translates to:
\begin{equation}E_{cav} = 4PV\end{equation}
We derived the surrounding thermal gas pressure ($P$) at the location of the cavities (midpoint located at $\sim$43.5 arcsec (67.4 kpc) from the cluster centre) by performing azimuthally averaged spectral analysis within an annulus (inner radius 27 arcsec and outer radius 65 arcsec from the cluster centre) encompassing the two cavities. This resulted in a projected pressure value of $4.23 \times 10^{-12} \text{ erg cm}^{-3}$. The resulting cavity energy values along with the properties of the cavities are given in \autoref{cav_tab}. 
\begin{table}
 \begin{center}
 \caption{Physical properties and energetics of the two cavities carved out by 1233+169 in the ICM of A1569N.}
  \begin{tabular}{ccccc}
    \hline
\hline
Cavity parameters&Eastern cavity &Western Cavity\\\hline
Radius (kpc) &28&21\\
Volume ($V$) (m$^{3}$)&$2.7 \times 10^{63}$&$1.1 \times 10^{63}$\\
$E_{cav}$ (erg) &$4.5 \times 10^{58}$&$1.9 \times 10^{58}$\\
Distance from centre ($R$) (kpc)&67.4&67.4\\
$t_{buoy}$ (yr)&$0.95 \times 10^8$&$1.1 \times 10^8$\\
$P_{cav}$ (erg s$^{-1})$&$1.5 \times 10^{43}$&$0.56 \times 10^{43}$\\
\hline
\hline
  \end{tabular}
  \label{cav_tab}
  \end{center}
\end{table}  

The time-averaged mechanical power associated with each cavity ($P_{cav}$) is calculated by dividing $E_{cav}$ by the approximate age of the cavity. The latter is calculated by assuming that the cavity bubble has risen from the centre of the cluster to its current radius $R$ at a terminal velocity $v\sim \sqrt{2gV/SC}$, thus giving a buoyancy timescale of $t_{buoy}\sim R\sqrt{SC/2gV}$, where, $S$ is the cross-sectional area of the cavity bubble, $V$ is its volume, $R$ is the distance of the bubble from the cluster nucleus, $g$ is the gravitational acceleration at the location of the cavity, and $C$ is the drag coefficient (taken to be equal to 0.75; \citet{churazov2001}). In order to estimate $g$ at the cavity location, we derived the total gravitational mass enclosed within a sphere of radius 43.5 arcsec from the cluster centre (the midpoint of the cavities) to be $5.2 \times 10^{12} \text{M}_{\odot}$. This implies a gravitational acceleration of $\sim$$1.6 \times 10^{-10} \text{m s}^{-2}$ (using $g=GM_{(<R)}/R^2$). The resulting buoyancy timescales and mechanical power associated with the two cavities are listed in \autoref{cav_tab}. The total mechanical power of the two cavities is $\sim$$2.1 \times 10^{43} \text{erg s}^{-1}$. The $0.4-8.0 \text{ keV}$ X-ray luminosity of the ICM in A1569N measured within a radius of 65 arcsec (radius within which the extended radio features of 1233+169 and the cavities are seen) from the centre is $2.2\pm{0.2} \times 10^{42} \text{ erg s}^{-1}$. The total mechanical power associated with the cavities is an order of magnitude larger than the observed X-ray luminosity in the central region of A1569N. This indicates that the radio galaxy 1233+169 can have a substantial impact on the energetics of the ICM in the core of A1569N. We note that the circular regions chosen as cavities are based entirely on visual inspection. The errors on the cavity volume estimates can, therefore, be large due to smoothing and projection effects. In order to check how the errors in gas pressure and cavity volume affect the cavity energetics, we recalculated the total power associated with the cavity pair by assuming an upper cap of 50 per cent error in the cavity radii and 40 per cent error in the surrounding gas pressure (the actual relative error in gas pressure). The estimated total cavity power is still larger than the X-ray radiative loss by a factor of 2, indicating that accounting for errors does not significantly change the effect that the cavities have on the surrounding gas.     

\begin{figure}
\centering
\includegraphics[width=0.99\linewidth,height=6.2cm]{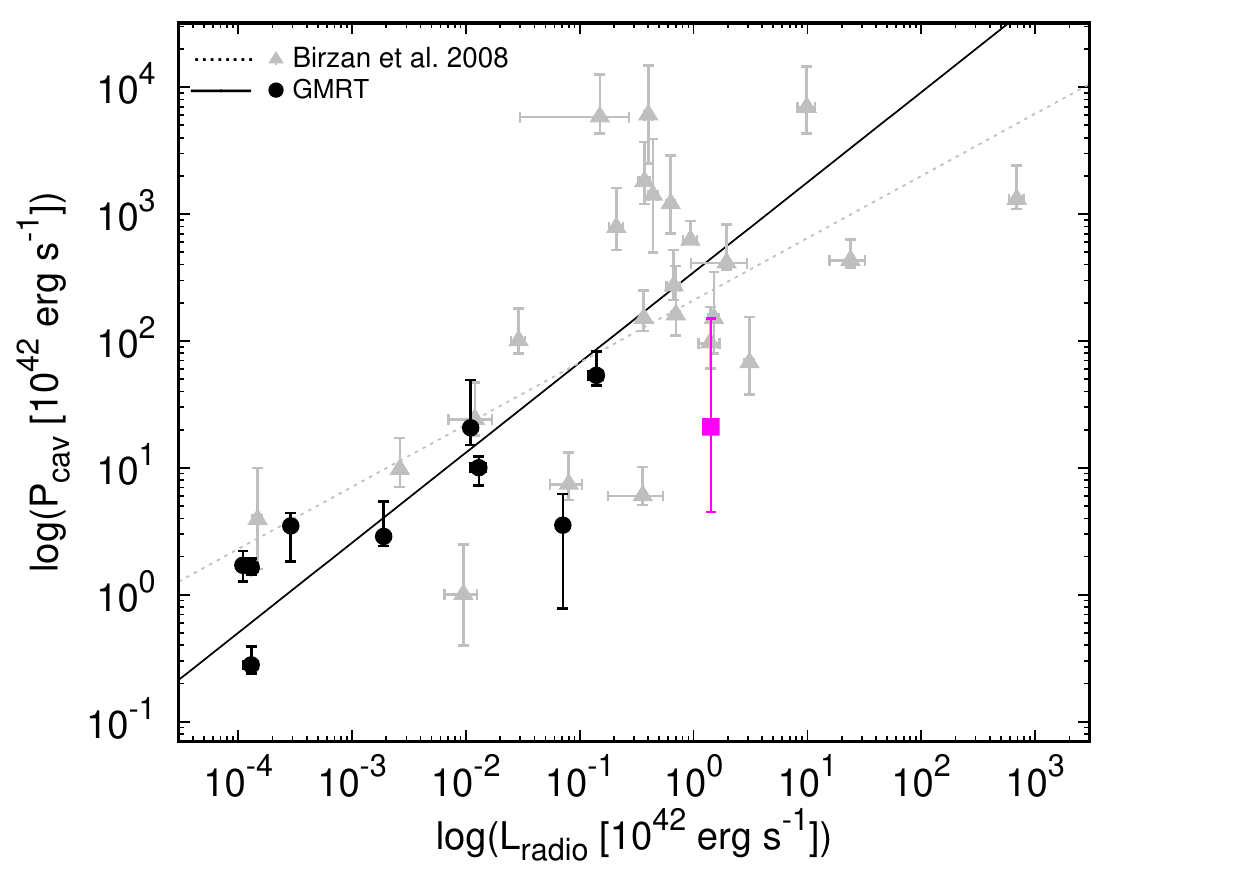}
\caption{Cavity power vs. integrated 10 MHz$-$10 GHz radio power relation plot reproduced from \citet{os11}. The \textit{solid black} line indicates the fit to all the data points. The \textit{dotted grey} line indicates the relation found by \citet{birzan2008}. The \textit{filled box} symbol in \textit{magenta} color represents the A1569N cavity system energetics.}
\label{cavity_os11}
\end{figure}
 
\begin{figure*}
\begin{multicols}{2}
\subcaptionbox{}{\includegraphics[width=0.85\linewidth,height=5.8cm]{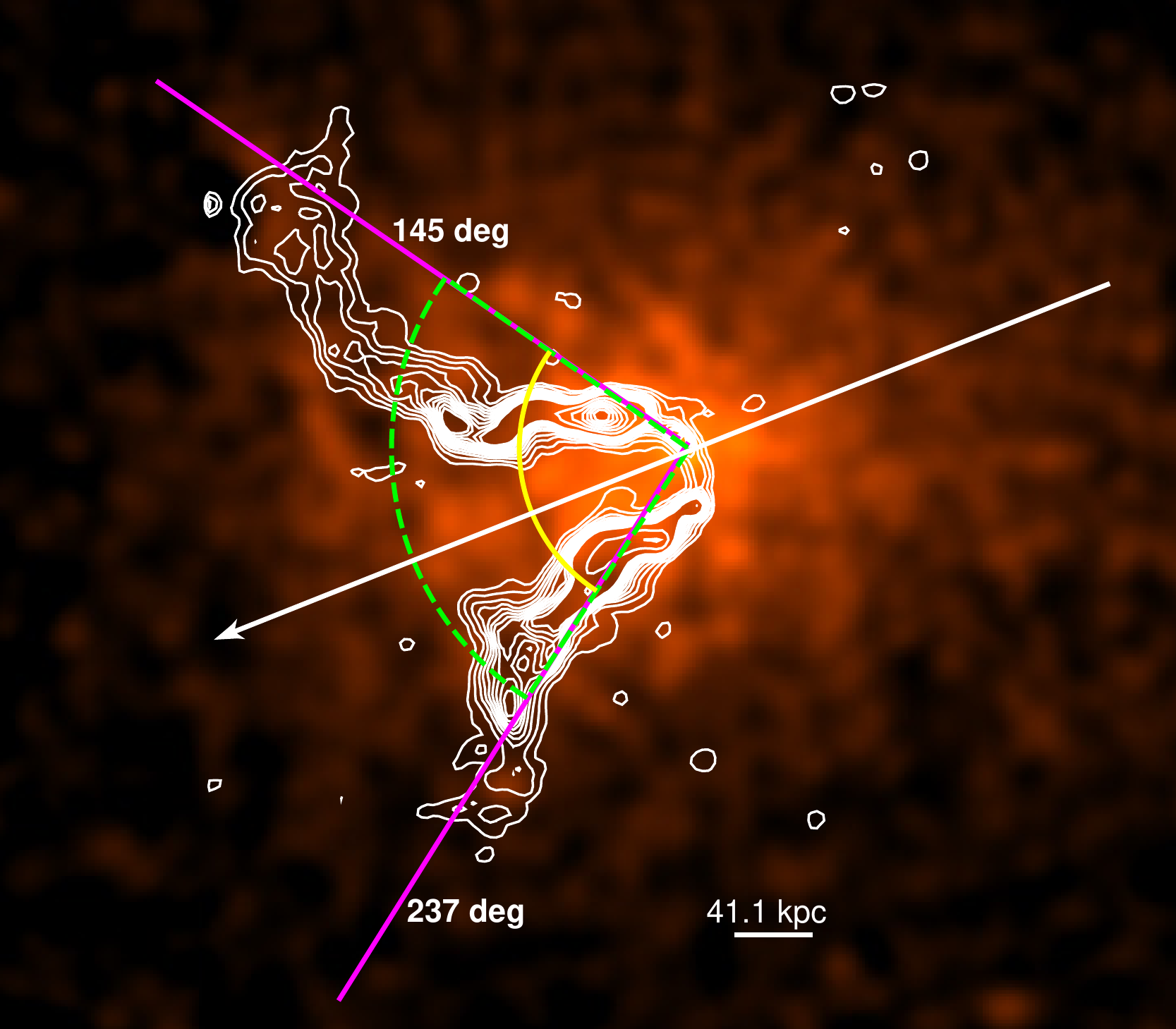}}\par
\subcaptionbox{}{\includegraphics[width=\linewidth,height=6.1cm]{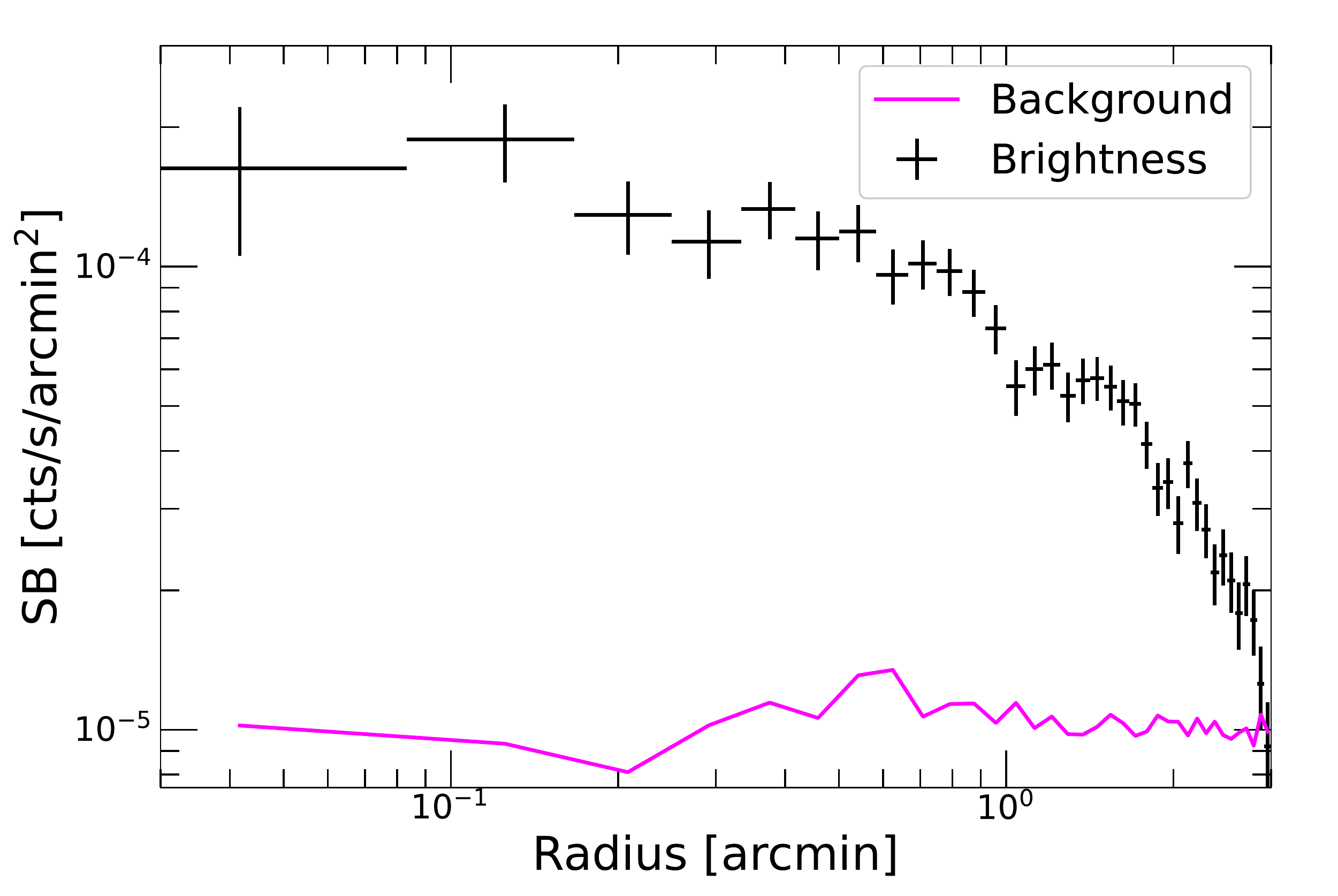}}\par
\end{multicols}
\begin{tabular}{cc}
\includegraphics[width=0.43\linewidth,height=5.5cm]{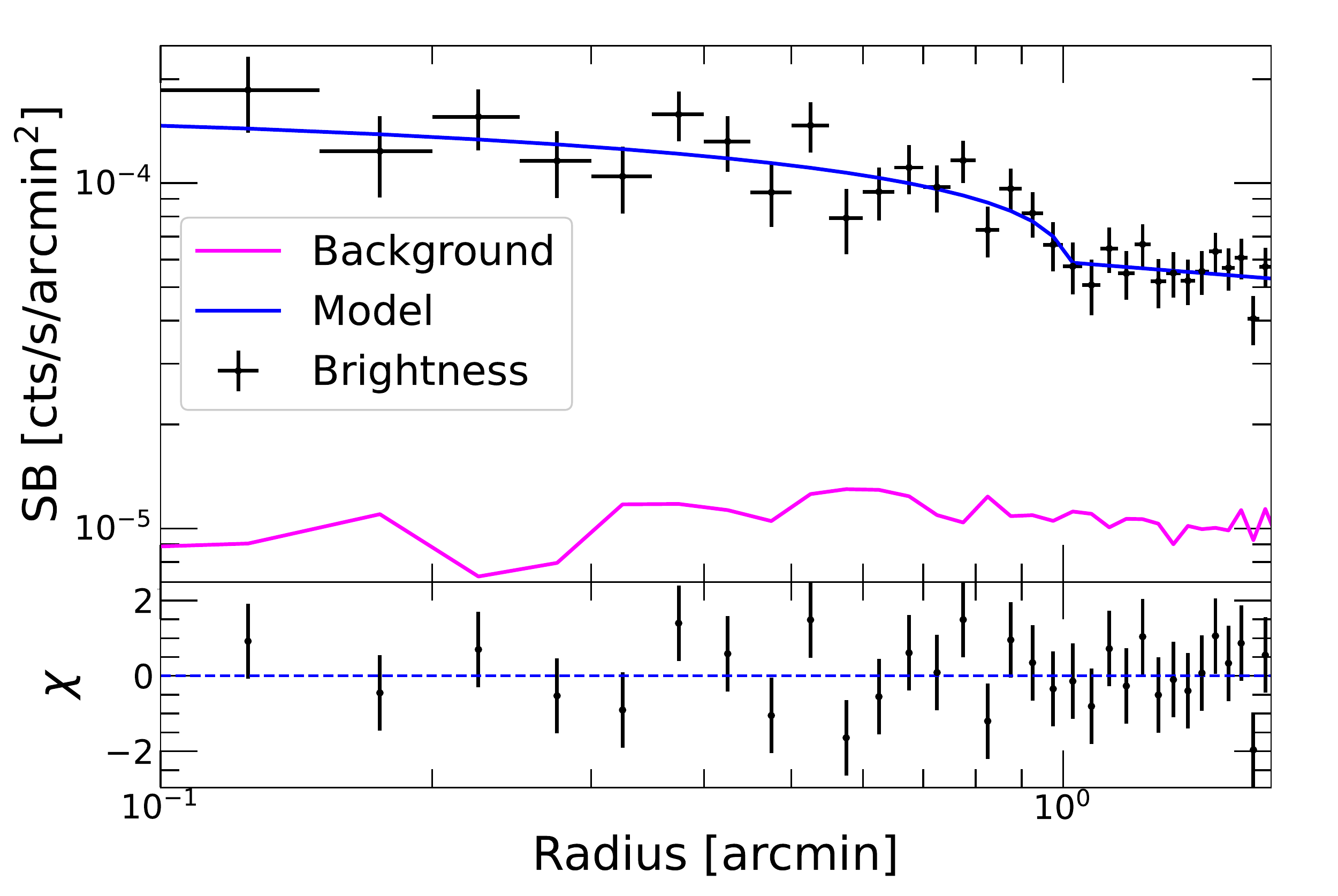}&\includegraphics[width=0.43\linewidth,height=5.5cm]{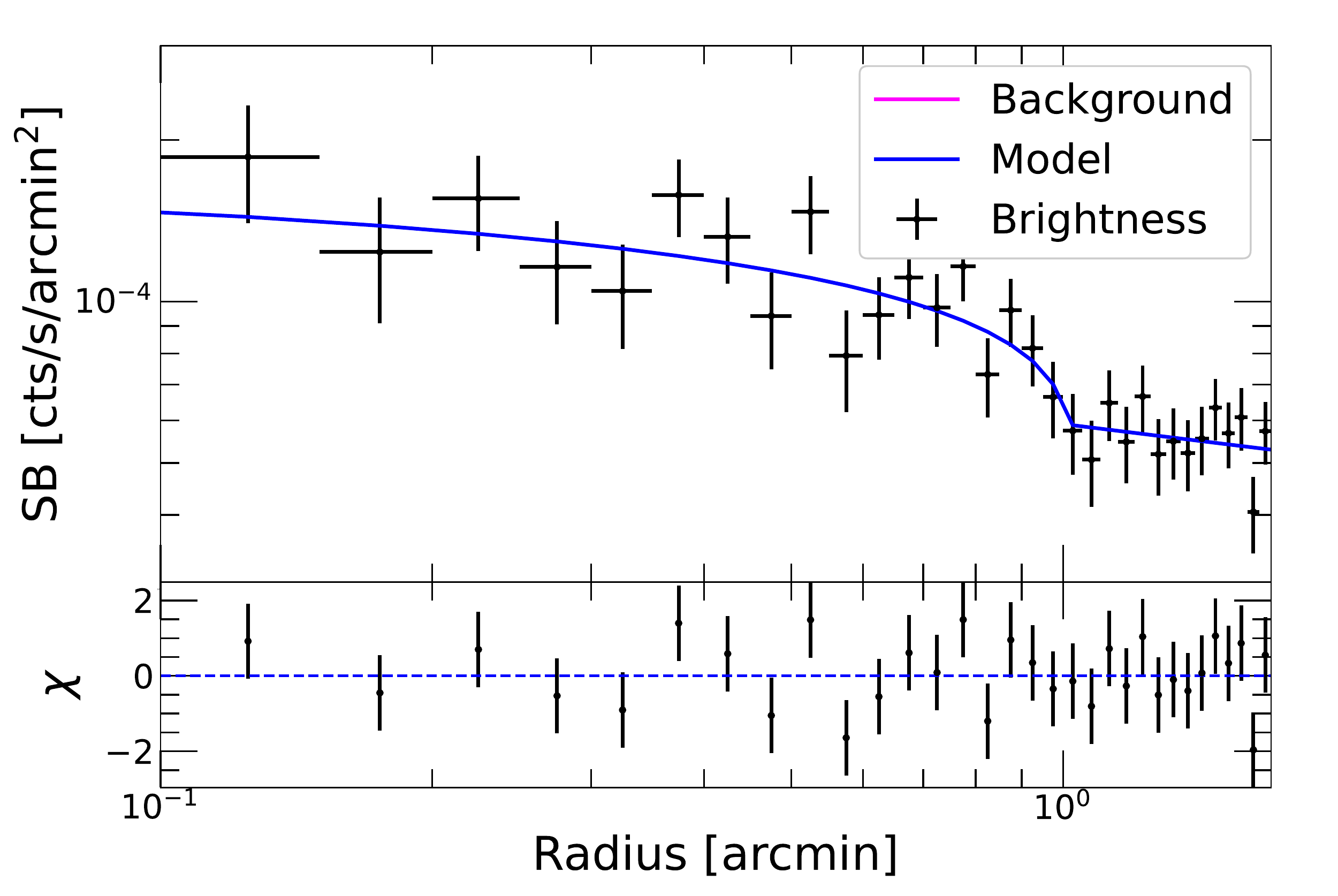}\\
\hspace{0.4cm}(c)&\hspace{1cm}(d)
\end{tabular}
\caption{\textbf{(a) }The point-source subtracted, particle-background-subtracted, and exposure-corrected image of the central region of A1569S in the energy range 0.5--4.0 keV. The extended radio features of 1233+168 are shown in \textit{white} color. The \textit{solid white} line bisects the angle between the bent radio tails of 1233+168 which is the direction along which local X-ray gas elongation is observed. The 145--237\textdegree{} eastern wedge used for extracting the X-ray surface brightness profile is highlighted using \textit{solid magenta} lines. The \textit{solid yellow} arc marks the position of the surface brightness edge at 1 arcmin. The sector highlighted in \textit{dotted green} is used for fitting the broken power-law 3D density model. \textbf{(b) }The 0.5--4.0 keV X-ray surface brightness profile extracted in the eastern sector 145--237\textdegree{}. An edge is apparent at $\sim$1 arcmin. \textbf{(c) }The best-fitting broken power-law density model (\textit{blue}) fitted in the radial range 0.01--1.7 arcmin to the X-ray surface brightness profile extracted in the sector 145--237\textdegree{}. \textbf{(d) } Same as (c) except that the background has been removed from the figure for better representation.}
\label{A1569S_centre}
\end{figure*}

\begin{figure}
\centering
\begin{multicols}{1}
\subcaptionbox{}{\hspace{0.5cm}\includegraphics[width=1.7\linewidth,height=4.8cm]{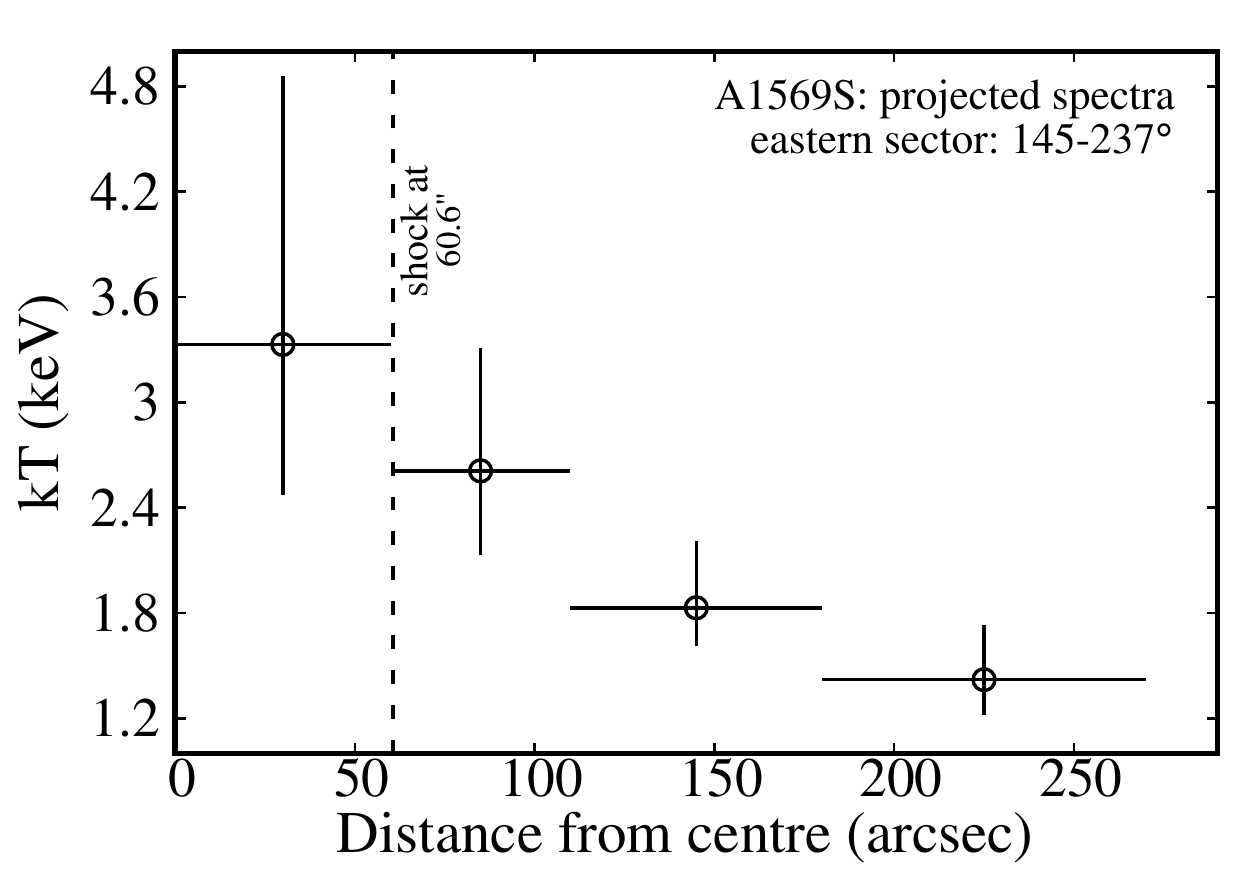}}\par
\end{multicols}
\begin{multicols}{1}
\subcaptionbox{}{\includegraphics[width=1.8\linewidth,height=4.8cm]{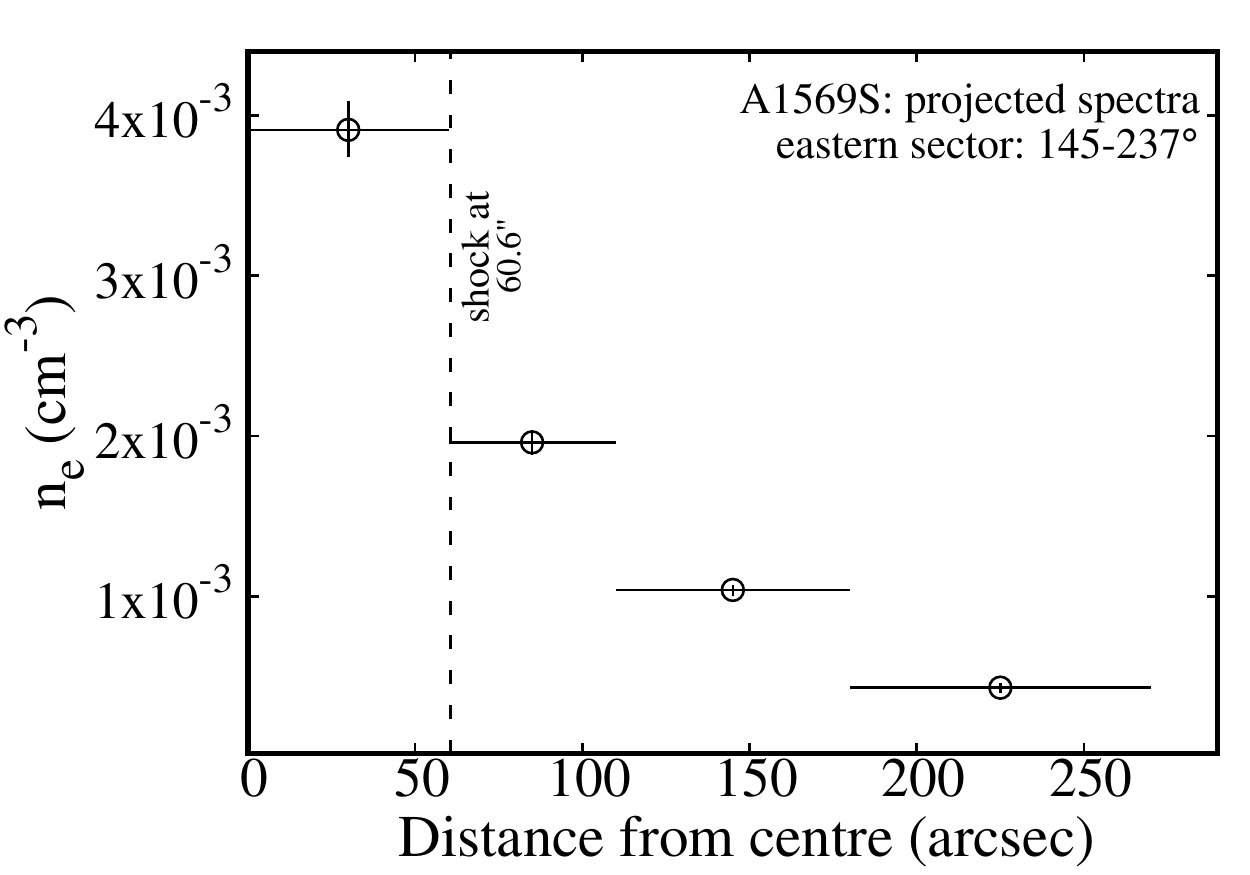}}\par
\end{multicols}
\begin{multicols}{1}
\subcaptionbox{}{\includegraphics[width=1.8\linewidth,height=4.8cm]{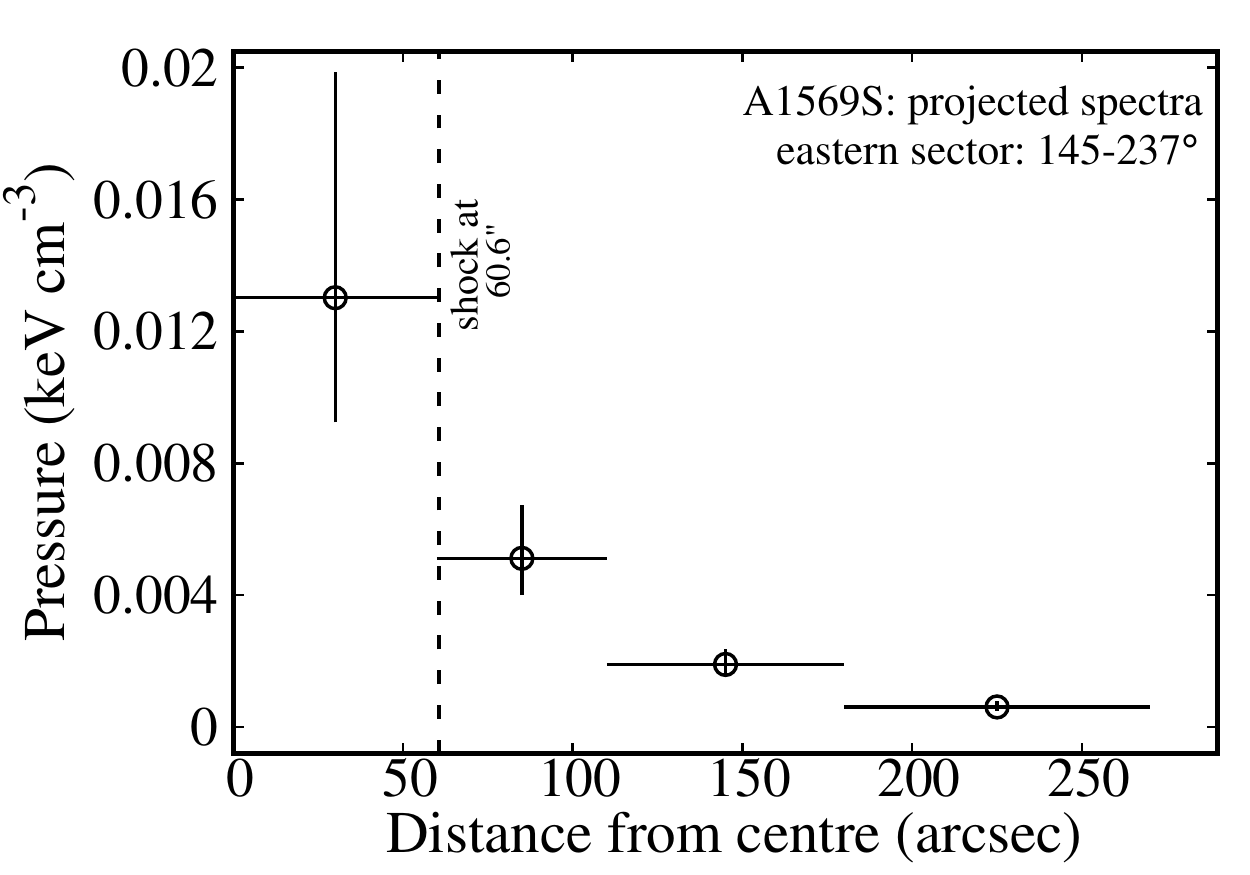}}\par
\end{multicols}
\caption{\textbf{(a) }Gas temperature ($kT$), \textbf{(b)} electron density ($n_e$), \textbf{(c)} pressure profiles obtained as a result of the projected spectral analysis performed in four radial bins along the eastern sector 145--237\textdegree{} in A1569S. The \textit{dashed} vertical line at 60.6 arcsec marks the position of the putative shock.}
\label{shockprofiles_proj}
\end{figure}

We also compared the energetics of 1233+169 with the jet (cavity) power -- radio power ($P_{cav}-L_{radio}$) relation obtained by \citet{os11} for a sample of galaxy groups and clusters harbouring radio sources, as shown in \autoref{cavity_os11}. The radio power used in this relation is the integrated 10 MHz-10 GHz power of the radio source. We calculated the total radio luminosity of 1233+169 by integrating the flux density between $\nu_{1}=10 \text{ MHz}$ and $\nu_{2}=10 \text{ GHz}$ as:
\begin{equation}L_{radio}=4\pi D_{L}^2 S_{\nu_0}\int_{\nu_{1}}^{\nu_{2}} (\nu/\nu_0)^{-\alpha} d\nu \end{equation}
where a power-law spectrum has been assumed for the radio source ($S_\nu\sim \nu^{-\alpha}$; $\alpha$ is the spectral index), $D_L=371.6 \text{ Mpc}$ is the luminosity distance to the source, and $S_{\nu_0}=620 \text{ mJy}$ is the 1.48 GHz flux density from the VLA data. We assumed a spectral index value $\alpha=1$. This results in $L_{radio}=1.42 \times 10^{42} \text{ erg s}^{-1}$. The derived $P_{cav}$ and $L_{radio}$ values place 1233+169 well within the scatter observed in the $P_{cav}-L_{radio}$ plot of \citet{os11} (Fig. \ref{cavity_os11}), indicating that the radio source contains adequate mechanical energy to create cavities in the ICM of A1569N.
\begin{table*}
 \begin{center}
  \captionsetup{justification=centering}
  \caption{Best--fitting parameters of the broken power-law 3D density model fitted to the X-ray surface brightness profile of A1569S extracted in the sector 145--237\textdegree{}. The errors are quoted at 68 per cent ($1\sigma$) confidence level.}
  \begin{tabular}{ccccccc}
    \hline
\hline
$\alpha_1$&$\alpha_2$&$r_{jump}$&SB Normalization&$C$ (density jump magnitude)&Background&$(\chi_{\nu}^2)_{\text{min}}${\hskip 0.04in}(DOF)\\
&&(arcmin)&($10^{-5}$ cts s$^{-1}$ arcmin$^{-2}$)&&($10^{-5}$ cts s$^{-1}$ arcmin$^{-2}$)&\\
\hline\hline
0.29$\pm{0.15}$&0.62$\pm{0.12}$&1.01$\pm{0.07}$&5.49$\pm{1.32}$&2.3$\pm{0.6}$&1.18&0.51 (54)\\
\hline\hline
  \end{tabular}
  \label{tabbknpow}
  \end{center}
 \end{table*}
 
\begin{table}
 \begin{center}
 \caption{Best-fitting temperature (kT) and the derived electron density ($n_{e}$), and pressure (P) values obtained from the projected spectral analysis performed along the eastern sector 145--237\textdegree{} in A1569S. The regions used for spectral extraction are listed. The errors are quoted at 90 per cent confidence level.}
  \begin{tabular}{cccc}
    \hline
\hline
Region&kT & n$_{e}$& P\\
&(keV)&($10^{-3}$cm$^{-3}$)&($10^{-2}$ keV cm$^{-3}$)\\
\\
\hline
\hline
0--60.6 arcsec&$3.3_{-0.9}^{+1.5}$&$3.91^{+0.18}_{-0.17}$&$1.30_{-0.38}^{+0.69}$\\
\\
60.6--110 arcsec&$2.6_{-0.5}^{+0.7}$&$1.96\pm{0.08}$&$0.51_{-0.11}^{+0.16}$\\
\\
110--180 arcsec&$1.8_{-0.2}^{+0.4}$&$1.04^{+0.03}_{-0.04}$&$0.19_{-0.03}^{+0.05}$\\
\\
180--270 arcsec&$1.4_{-0.2}^{+0.3}$&$0.43\pm{0.03}$&$0.06_{-0.01}^{+0.02}$\\
\hline
\hline
  \end{tabular}
  \label{shock_protab}
  \end{center}
 \end{table}  

\subsection{A subcluster--cluster merger in A1569S?} 
The gas distribution in A1569S clearly deviates from azimuthal symmetry and appears to be elongated in the central $\sim$70 kpc region (Fig. \ref{radio_overlays}e). The direction of the observed X-ray elongation is along the line that bisects the radio tails of 1233+168. A1569S clearly lacks a large cool core as indicated by the spectral analysis presented in Section \ref{radialprof}. The presence of central X-ray substructure, elongation of cluster gas distribution in the vicinity of extended radio galaxies, and absence of a cool core are often considered as common indicators of mergers in galaxy clusters \citep{pinkney1993,burns1994,burns1996,roettiger1996,gomez1997_9clusters}.  

Based on the above mentioned observational facts, we performed a systematic search for discontinuities in the surface brightness distribution of the intracluster medium of A1569S that could be indicative of any previous merger activity in the cluster. The python package \textit{Pyproffit}\footnote{\url{https://github.com/domeckert/pyproffit}} (version 0.6.0) was used for this purpose \citep{eckert2020}. A point-source-subtracted image of the cluster (0.5--4.0 keV energy range), an exposure map, and a scaled background image were supplied to \textit{Pyproffit}. Surface brightness profiles centred at R.A.(J2000) = 12$^h$36$^m$25$^s$.86 and Dec.(J2000) = +16\textdegree{}32$'$19$''$.94 were initially extracted in the eastern and western directions corresponding to sectors 90--270\textdegree{} and 270--450\textdegree{} (angles measured anticlockwise with respect to the horizontal), respectively. The azimuthal span of the sectors was then gradually decreased until a surface brightness discontinuity started to become visible in the eastern sector spanning 110--250\textdegree. No evident signs of discontinuities were noticed in the western direction. The discontinuity in the eastern direction was best visible in the sector 145--237\textdegree{} which is highlighted in \textit{magenta} color in \autoref{A1569S_centre}a. The surface brightness profile along this sector is shown in \autoref{A1569S_centre}b where a discontinuity is apparent at $\sim$1 arcmin. This is consistent with the abrupt drop in the surface brightness noticed across the \textit{yellow} arc shown in Fig. \ref{A1569S_centre}a. 

In order to check if the observed surface brightness discontinuity corresponds to a gas density jump in the ICM of A1569S, we fitted a broken power-law 3D density model projected along the line of sight to the 2D surface brightness profile extracted in the eastern sector 145--237\textdegree{}, in the vicinity of the observed discontinuity. The broken power-law 3D density model (\textit{BknPow} in \textit{Pyproffit}) is described as:
\begin{equation}\begin{split}n(r)=n_{0}\left(\frac{r}{r_{jump}}\right)^{-{\alpha}_1}\hspace{0.1cm}\text{for}\hspace{0.2cm} r < r_{jump}\\n(r)=\frac{1}{C}n_{0}\left(\frac{r}{r_{jump}}\right)^{-{\alpha}_2}\hspace{0.1cm}\text{for}\hspace{0.2cm} r \geq r_{jump}\end{split}\label{eq:bknpow}\end{equation} where $\alpha_1$ and $\alpha_2$ are the power-law indices, $n_0$ is the normalization factor, $r$ is the radius from the centre of the sector, $r_{jump}$ is the radius at which the density jump is observed, and $C$ is the density compression factor or the magnitude of the density jump. A detailed description of the 2D surface brightness projection of the broken power-law 3D density model is given in the Appendix of \citet{owers2009}. The radial range was restricted to 0.01--1.7 arcmin for fitting the broken power-law density model since the density profile beyond the apparent surface brightness edge could be approximated by a power-law only till $\sim$1.7 arcmin. A radial bin size of 3 arcsec was used. Initially, all parameters were left free during the fit. A background value of $1.18 \times 10^{-5}$ cts s$^{-1}$ arcmin$^{-2}$ was obtained. The model was fitted again with the background fixed to this value, which did not cause the other model parameters to change. The best--fitting model is shown in \autoref{A1569S_centre}c,d and its parameters are listed in \autoref{tabbknpow}. The surface brightness edge is located at 1.01 arcmin and corresponds to a gas density jump of 2.3$\pm{0.6}$.

In order to assess the nature of this density discontinuity, we performed a projected spectral analysis along the sector 145--237\textdegree{} in four radial bins ranging 0--60.6, 60.6--110, 110-180, and 180-270 arcsec (see \S4.2.1 for details of spectral modelling). Each of these regions was chosen to have a minimum of 800 counts. The best-fitting temperature, and the derived electron pressure and density values obtained for each region are provided in \autoref{shock_protab}, and the profiles shown in \autoref{shockprofiles_proj}. A deprojection analysis was also attempted but the errors on the model parameters could not be constrained due to low number statistics.

It can be seen in \autoref{shockprofiles_proj}a that the gas temperature inside (0--60.6 arcsec region) the surface brightness edge (located at 1.01 arcmin) is indicated to be slightly higher ($\sim$ $3.3^{+1.5}_{-0.9}$ keV) than the gas temperature outside the edge (60.6--110 arcsec region; $2.6^{+0.7}_{-0.5}$ keV), although the errors are large. Across the edge, the gas density jump derived from the spectral analysis is however, very significant with a value of $2.0\pm{0.2}$ (\autoref{shockprofiles_proj}b).  This is consistent with the value obtained by fitting the surface brightness profile with the broken power-law density model ($2.3\pm{0.6}$). The gas pressure across the edge drops by a factor of $2.5^{+2.4}_{-1.2}$ (\autoref{shockprofiles_proj}c). The gas density jump observed at 1.01 arcmin, along with an indication of a drop in both the gas temperature and pressure across it, is indicative of the presence of a shock front at this location (\textit{yellow} arc in \autoref{A1569S_centre}a). Additionally, we also checked for the presence of substructures resulting from any sloshing motions of the gas (e.g., spiral patterns) in the 2D-$\beta$ model subtracted image of A1569S. We did not find evidence of any such structures in the residual image, thus, supporting the argument that the observed edge is indeed a shock and not a cold front. The shock is plausibly the result of a merger between A1569S and a subcluster that has fallen in from the west along the direction indicated by the \textit{white} arrow line shown in \autoref{A1569S_centre}a, which is likely responsible for the local X-ray gas elongation observed in the inner region of A1569S. Under this scenario, the region 0--60.6 arcsec in the sector 145-237\textdegree{} is already affected by passage of the shock (post-shock) and the region 60.6--110 arcsec is yet to experience it (pre-shock). We note that the choice of the radial size of these regions is influenced by the photon statistics, and a deeper X-ray observation of A1569S will be immensely helpful in better constraining the thermodynamic properties of the gas across the discontinuity. Several other studies (\citet{owers2009, bourdin2013, ubertosi2021} to name a few) have performed a similar analysis in relatively large ($>$ 70 kpc) spectral bins to confirm whether a discontinuity is a potential shock or a cold front.

The Mach number of the shock ($M$) was calculated from the gas pressure ($P$) and density ($n$) values obtained spectrally in the post-shock (0--60.6 arcsec) and pre-shock (60.6--110 arcsec) regions using the following equations:
\begin{equation}\frac{P_{post}}{P_{pre}}=\frac{10M^2-2}{8}\label{eq:pressjump}\end{equation}\\\\resulting in $M=1.5^{+0.6}_{-0.3}$, and 
\vspace{0.3cm}\begin{equation}\frac{n_{post}}{n_{pre}}=\frac{4M^2}{3+M^2}=C\label{eq:denjump}\end{equation}\\\\resulting in $M=1.7^{+0.2}_{-0.1}$.\vspace{0.2cm}

The gas was assumed to be monoatomic with an adiabatic index  $\gamma=5/3$ in these calculations. The shock Mach number was not estimated from the gas temperature jump due to the large uncertainty in the temperature values. Using equation \eqref{eq:denjump} and a density jump $C=2.3\pm{0.6}$ obtained from fitting the surface brightness profile with a broken power-law density model, we estimated $M=2.0^{+0.8}_{-0.5}$. 

The shock Mach number was then used to calculate the shock velocity ($v$) using
\begin{equation}M=\frac{v}{c_s}\end{equation} where $c_s$ is the speed of sound in a gas of temperature 2.61 keV (temperature of the pre-shock region), which was calculated using,\begin{equation}c_s=\sqrt{\frac{\gamma kT}{\mu m_p}}\end{equation}

Using $M=1.7$ (this value, estimated using the spectrally derived gas density jump, was chosen because it has the tightest constraints), $\gamma=5/3$, $\mu=0.608$, and $m_p=1.67 \times 10^{-27}$ kg, we estimated $c_s=831.7$ km s$^{-1}$ for a 2.61 keV gas, and a shock velocity $v=1413.9$ km s$^{-1}$.
    
\section{Discussion}
Our imaging analysis of the \textit{Chandra} X-ray data (\S3.1) shows clear presence of two gas clumps in A1569 (Fig. \ref{morphology}a) as noticed earlier in the \textit{ROSAT} PSPC image of the cluster by \citet{gomez1997_a1569}. The northern clump -- A1569N -- is smaller (radial extent $\sim$248 kpc), and appears to be extended in the NW-SE direction (Fig. \ref{morphology}a) which is confirmed by the 2D-$\beta$ modelling of its gas surface brightness presented in \S3.2 (\autoref{2dbetatable}). The southern gas clump -- A1569S -- is larger (radial extent $\sim$370 kpc) (Fig. \ref{morphology}a) and is elongated in the E-W direction with an ellipticity of $0.27\pm{0.02}$ (\S3.2; \autoref{2dbetatable}) consistent with the value obtained by \citet{gomez1997_9clusters}. A global spectral analysis of A1569N results in a gas temperature of $1.6^{+0.3}_{-0.3}$ keV and an average elemental abundance of $0.16^{+0.23}_{-0.13}$ Z$_{\odot}$ (\autoref{tab_global}). The spectral analysis of A1569N has been performed for the first time in this study. We find that the gas in A1569S has an average temperature of $1.9^{+0.2}_{-0.1}$ keV which is in agreement with the \textit{ROSAT} result of \citet{gomez1997_9clusters} but has much tighter constraints. The average abundance of metals in A1569S is $0.24^{+0.10}_{-0.08}$ Z$_{\odot}$ and is reported for the first time here. The X-ray luminosity of A1569N is $4.5\pm{0.4} \times 10^{42}$ erg s$^{-1}$ (0.5--4.0 keV) while that of A1569S is $2.31\pm{0.06} \times 10^{43}$ erg s$^{-1}$ (0.5--4.0 keV), making it $\sim$five times more luminous than A1569N. The temperature and luminosity values obtained for the two gas clumps are representative of groups of galaxies \citep{bahcall1999,helsdon2000,rasmussen2007,lovisari2015,zou2016}. The derived gas mass (\S5.1) of A1569N within a radius of 248 kpc is $5.7^{+1.5}_{-1.8} \times 10^{11}$ $M_{\odot}$ and that of A1569S (within 370 kpc) is $2.3^{+1.8}_{-1.3} \times 10^{12}$ M$_{\odot}$ (\autoref{tabgasmass}). The total gravitational mass of A1569N within a radius of 248 kpc is $2.0\pm{1.5} \times 10^{13}$ M$_{\odot}$ and that of A1569S within a radius of 370 kpc is $2.1\pm{1.6} \times 10^{13}$ M$_{\odot}$ (\S5.2). These values are also typical of galaxy groups \citep{mulchaey2000,lovisari2015} in contrast to galaxy clusters which have a greater extent and possess higher mass, temperature and luminosity \citep{bohringer2002,reiprich2002,sanderson2006}.    

\begin{figure}
\centering
\includegraphics[width=0.92\linewidth,height=6cm]{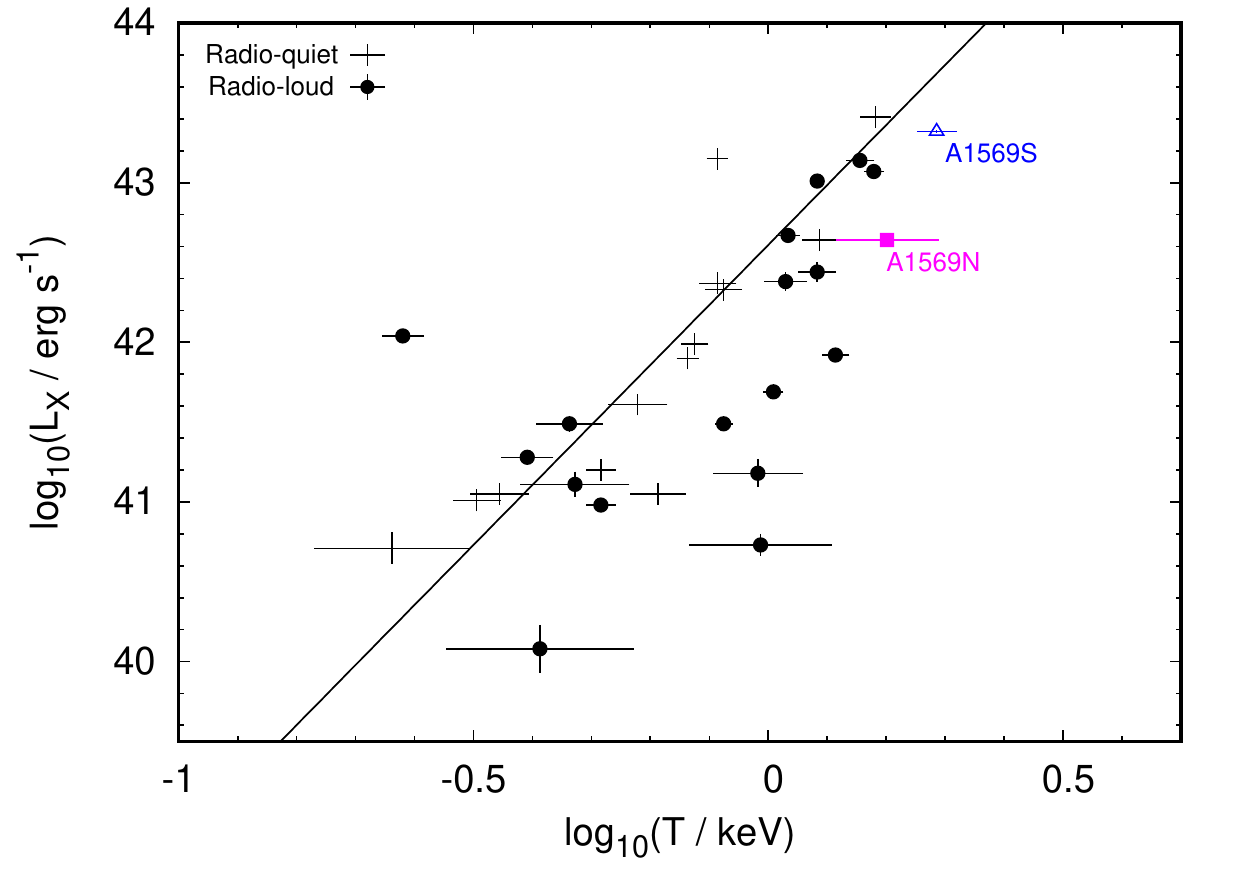}
\caption{$L_{X}-T_{X}$ plot for a sample of radio-loud (\textit{filled circles}) and radio-quiet (\textit{plus} symbols) groups adopted from \citet{croston2005}. The overplotted \textit{black} line is the best-fitting radio-quiet relation. The \textit{filled magenta box} and \textit{filled blue triangle} symbols represent A1569N and A1569S, respectively. There is a clear tendency for radio-loud groups to be on the hotter side of the radio-quiet relation.}
\label{croston05_repro}
\end{figure}

A pair of X-ray cavities have been discovered within the central $\sim$100 kpc ($\sim$65 arcsec radius) region of A1569N, consistent with a clear deviation found in the X-ray isophotes from azimuthal symmetry in the same region of A1569N (Fig. \ref{morphology}a, c). The presence of the cavity pair is confirmed by the unsharp-masked image (Fig. \ref{cavity}a) and the elliptical 2D-$\beta$ model subtracted residual image (Fig. \ref{cavity}b) presented in \S3.3. These X-ray cavities are very likely to have been excavated in the ICM  by the displacement of hot gas by the radio lobes of the central elliptical galaxy 1233+169, as indicated by the spatial coincidence of the X-ray deficits with the presence of radio emission from 1233+169 (Fig. \ref{radio_overlays}b). The eastern and western deficits have about 3$\sigma$ and 2.5$\sigma$ significance respectively.
Calculation of the energetics of the cavities (\S6.1) shows that the total mechanical power associated with the cavity pair is $\sim$$2.1 \times 10^{43} \text{erg s}^{-1}$, which is an order of magnitude larger than the X-ray luminosity of ($2.2\pm{0.2} \times 10^{42} \text{ erg s}^{-1}$) observed in the central region of A1569N. These values indicate that the radio galaxy 1233+169 can play a significant role in affecting the properties of the intragroup gas in A1569N.

We find evidence for heating of the intragroup gas in A1569N by the central radio source 1233+169. This is apparent from the temperature and entropy rise observed in the second annular spectral bin (30-70 arcsec) where the radio lobes of 1233+169 lie (Fig.\ref{fig_projspec_A1569N}a, d)  and also from the location of A1569N with respect to the $L_{X}-T_{X}$ relation obtained for a sample of 13 radio-quiet galaxy groups by \citet{croston2005} (top right panel of fig. 2 of their paper) as displayed in \autoref{croston05_repro}. For a sample of 30 $ROSAT$-observed galaxy groups, \citet{croston2005} obtained the X-ray luminosity (0.1--2.4 keV) and temperature by performing a spectral analysis within a circular region of radius $r_{cut}$, which is the extent up to which X-ray emission from the group was detected. Galaxy groups were classified as radio-loud and radio-quiet based on a radio-luminosity cutoff value ($L_{1.4\hspace{0.1cm}GHz}=c1=1.2 \times 10^{21}$ W Hz$^{-1}$ used in Fig. \ref{croston05_repro}). The authors found that the radio-loud groups are likely to be hotter at a given X-ray luminosity than the radio-quiet groups. They tested different models and attributed the observed effect to the heating of the host group gas by the central radio source. The location of A1569N, a radio-loud group, clearly on the hotter side of the radio-quiet relation (\textit{black} line) in Fig. \ref{croston05_repro} indicates that the central radio source 1233+169 is responsible for heating the gas in A1569N. Analogous results suggesting that radio-loud groups are likely to have a higher temperature than radio-quiet groups of the same luminosity have been reported by \citet{croston2008} in an \textit{XMM-Newton} study of galaxy groups hosting low-power radio galaxies. Another similar study by \citet{maglio2007} finds that the $L_{X}-T_{X}$ relation steepens for low-temperature clusters ($\lesssim$ 3 keV) hosting central radio objects with extended jets and/or lobe structures, and attribute this effect to overheating of the intracluster gas, likely caused by the interplay between the extended radio structures and the intracluster medium. There is a marginal indication that the gas temperature in A1569N is on the higher side of the correlation shown (Fig. \ref{croston05_repro}), pointing towards the possibility of some heating by the central radio source.

The lack of signatures of large-scale ($\gtrsim$ 46 kpc) cooling in the central region of A1569N further corroborates the radio-source heating proposition. The projected spectral analysis presented in \S4.2.1 does not show a central drop in temperature (Fig. \ref{fig_projspec_A1569N}a) or a sharply peaked electron density profile (Fig. \ref{fig_projspec_A1569N}b). Low temperature in the central region of A1569N is not observed even after taking projection effects into account (\S4.2.2). A single $\beta-$model fit to the projected electron density radial profile of A1569N (\S5.1) gives a central density value of $6.85\pm{0.44} \times 10^{-3}$ cm$^{-3}$ unlike the high densities found in the central regions of cool-core galaxy groups/clusters. Absence of observed cooling is also supported by a cooling time of $8.6^{+7.2}_{-4.0}$ Gyr within the central 46.5 kpc (30 arcsec) region of A1569N which is comparable to the Hubble time (\S5.3). Radiative losses from the thermal ICM are believed to be balanced by feedback from the central radio source in massive galaxy clusters. This picture has not been explored as much in the case of galaxy groups. However, due to shallower gravitational potential of galaxy groups, and hence, lower binding energy per particle, the same non-gravitational heating energy (supplied by the central radio source) per particle has a greater effect in groups than clusters \citep{mcnamara2007,gitti2012}. \citet{sun2012} report that almost all galaxy groups with a large-scale cool core do not host a central radio source. For groups which have a cool core along with a central radio source, the cool core is a coronal cool core corresponding to that of the central galaxy and not the group. \citet{best2007} point out that radio-heating probably overcompensates the radiative cooling losses in groups of galaxies. The increased importance of overheating of the intracluster gas by central radio sources in low-mass clusters (kT $\lesssim$ 3 keV) has also been highlighted by \citet{maglio2007}. \citet{jetha2007} suggest that energy injection from central radio sources is sufficient to balance cooling in the core of the galaxy groups. Based on the facts that the lobes of 1233+169 are confined to the very central region of A1569N and that the jet power is an order of magnitude larger than the X-ray radiative loss in the region, together with evidence for heating of the intragroup gas highlighted previously, we attribute the lack of a large, ICM-associated cool core in A1569N likely to cavity heating caused by 1233+169. A1569N may, however, possess a small-scale ($r\sim$few kpc) coronal cool core associated with the central BCG. This is indicated by the two-dimensional image fitting of A1569N presented in \S3.2, where a single 2D-$\beta$ model could well describe the cluster emission, only if the central 15 arcsec ($\sim$23 kpc) region corresponding to the central galaxy was excluded from the modelling. We note that it has not been possible to confirm the presence of a coronal cool core in A1569N in this work via a spectral analysis in smaller radial bins within the central 30 arcsec ($<$ 46 kpc) region or through double-$\beta$ image modelling due to a low number of counts.

The gas distribution within the southern subcluster, A1569S, is not azimuthally symmetric (Fig. \ref{morphology}a). Local elongation of the intracluster gas is observed in between the radio tails of the central source 1233+168 (Fig. \ref{radio_overlays}e). These properties are typical of dynamically disturbed systems which have undergone mergers or are in the process of formation \citep{burns1996,roettiger1996,schuecker2001,sarazin2002}. We detect an X-ray surface brightness discontinuity at 1.01 arcmin ($\sim$83 kpc) to the east (sector spanning 145--237\textdegree{}) from the subcluster centre (Fig. \ref{A1569S_centre}b). The discontinuity corresponds to a gas density jump of $2.3\pm{0.6}$ (Table \ref{tabbknpow}, Fig. \ref{A1569S_centre}c, d) and lies perpendicular to the line bisecting the angle between the radio tails of 1233+168 (Fig. \ref{A1569S_centre}a). Projected spectral analysis carried out along the eastern sector containing the SB edge indicates a drop in both the gas temperature and pressure across the discontinuity, indicative of the presence of a shock in the subcluster (\S6.2; Fig. \ref{shockprofiles_proj}). The observed pressure drop across the edge is $2.5^{+2.4}_{-1.2}$ and the estimated shock Mach value is $\sim$1.7, suggesting that the putative shock is a weak shock resulting from a small-scale merger between A1569S and another subcluster. The proposed merger scenario is also supported by the absence of a cool core in A1569S as indicated by the spectral analysis presented in \S4.2.1 and \S4.2.2, and a cooling time of $6.6_{-3.2}^{+4.4} \times 10^{10} \text{ yr}$ within the central 60 arcsec ($\sim$82 kpc) region of A1569S (\S5.3), since merger activity in galaxy clusters tends to disrupt and inhibit the formation of cool cores \citep{ritchie2002,zuhone2011}, particularly if the merger takes place in the early evolutionary stage of the cluster \citep{burns2008,henning2009}. Shocks associated with mergers increase the entropy of the gas \citep{markevitch1996,sarazin2004}. Merger activity can disperse the low-entropy gas in the cluster centre and mix it with the high-entropy ICM at larger radii, which leads to high central entropy and flattening of the entropy profile in the inner regions of galaxy clusters \citep{susana2009,zuhone2011}. The entropy profile of A1569S is flat (Fig. \ref{fig_projspec}d) that is consistent with the absence of a large-scale ($\gtrsim$ 40 kpc) cool core. Deprojection spectral analysis shows high entropy ($311^{+92}_{-95}$ keV cm$^2$) in the central 60 arcsec ($\sim$82 kpc) region of A1569S (Fig. \ref{fig_deprojspec}d), lending further support for the merger scenario. Similar high central entropy values ($\gtrsim$300 keV cm$^2$) have been found in other low surface brightness clusters undergoing mergers (e.g., Abell 1631 -- \citet{babazaki2018}; Abell 2399 -- \citet{mitsuishi2018}).

Relative motions between extended radio galaxies and the dense host ICM, result in significant ram pressure on the radio-emitting material of the jets, which leads to the observed jet bending. \citet{gomez1997_a1569} estimated $400-2500$ km s$^{-1}$ as the range of relative velocities of radio sources with respect to the ICM that is responsible for the observed WAT radio morphology. The central WAT source 1233+168 in A1569S, however, has a peculiar velocity of 215 km s$^{-1}$ with respect to the subcluster dynamical centre. This velocity is insufficient to account for the bending of the jets \citep{gomez1997_a1569}. The relative motion between the radio galaxy and the host ICM required for ram pressure exertion can very well arise from the gas itself moving across the galaxy. Simulations show that bulk flow motions of gas with high velocities of $\gtrsim$1000 km s$^{-1}$ are generated in cluster mergers \citep{roettiger1993,loken1995,roettiger1996}. Within the central regions of galaxy clusters ($r$$\sim$200 kpc), these motions may remain above 1000 km s$^{-1}$ for timescales of $\sim$2 Gyr which are much longer than the typical lifetimes of radio galaxies ($\sim$$10^7-10^8$ yr) \citep{loken1995,roettiger1996}. This bulk flow of the ICM (relative to the central extended radio galaxy) arising in cluster mergers has been invoked as the primary reason for the observed bending of central WAT sources \citep{burns1994,pinkney1994,loken1995,gomez1997_a1569,gomez1997_9clusters,douglass2011}. The overall E-W morphology of A1569S, the direction of local elongation of the ICM in between the radio tails of 1233+168, and the presence of gas density discontinuity perpendicular to the line bisecting the angle between the radio tails of 1233+168, suggest that the most plausible geometry of the ongoing interaction is a head-on merger between A1569S and a subcluster falling in from the west along the line bisecting the WAT tails in the plane of the sky. Using a pre-shock gas temperature of 2.6 keV and a merger shock Mach value equal to 1.7, we estimate the ICM bulk flow velocity to have an upper limit of $\sim$1414 km s$^{-1}$ (equal to the shock velocity) (\S6.2). The proposed merger geometry and the estimated flow velocity is the most likely explanation for the observed bending of the WAT source 1233+168, thus supporting the merger hypothesis of \citet{gomez1997_a1569}.  

We do not detect cavities in the substructure maps of A1569S. This is in support of the calculation of \citet{gomez1997_a1569} who found the radio tails of 1233+168 to be underpressured by more than a factor of 10 with respect to the surrounding ICM, likely due to the entrainment of the ICM into the radio tails. As a result, we attribute the temperature excess observed in A1569S (Fig. \ref{croston05_repro}), a radio-loud group, to be the heating effect of the above-mentioned merger. Deeper X-ray observations are, however, required to confirm the presence/absence of cavities in A1569S, and the presence of any AGN driven shocks in the ICM of both A1569N and A1569S. These observations will also enable us to produce high resolution thermodynamic maps of the two subclusters, detect any metallicity substructure around the radio lobes and visually identify SB edges in the X-ray images. The study of X-ray cavities and mergers in small-scale galaxy clusters and groups has been sparse due to the low surface brightness of these objects, and should benefit greatly from deeper exposures or larger X-ray telescopes. Future radio observations of A1569S will prove helpful in confirming the observed merger activity in A1569S.

\section{Conclusions}
In this work, we have presented a detailed study of the thermodynamic properties of the intracluster gas in the two subclusters of A1569 -- A1569N and A1569S. 
The superior spatial and spectral resolution of \textit{Chandra} compared to the \textit{ROSAT}, has allowed us to detect substructure, and provide better constraints on the thermodynamic properties of the hot gas in the two subclusters. We have used 1.4 GHz \textit{Very Large Array} archival observations of the central radio galaxies in the two subclusters -- 1233+169 in A1569N and 1233+168 in A1569S -- to investigate their interaction with the surrounding hot gas. The key findings of this study are as follows.
\begin{itemize}\setlength\itemsep{0.3em}
\item The northern subcluster of Abell 1569 -- A1569N -- has $L_{X\text{(0.5--4.0 keV)}}=4.5\pm{0.4} \times 10^{42}$ erg s$^{-1}$, and is $\sim$5 times fainter than the southern subcluster -- A1569S -- that has $L_{X\text{(0.5--4.0 keV)}}=2.31\pm{0.06} \times 10^{43}$ erg s$^{-1}$.
\item The average temperature of A1569N is $1.6^{+0.3}_{-0.3}$ keV while that of A1569S is $1.9^{+0.2}_{-0.1}$ keV.
\item X-ray emission from A1569N and A1569S extends to radius  $r\sim$248 kpc and $r\sim$370 kpc respectively, which along with the low temperature and luminosity values confirms that the two subclusters are indeed galaxy groups. 
\item A1569N has an average elemental abundance of $0.16^{+0.23}_{-0.13}$ Z$_{\odot}$ while this value is $0.24^{+0.10}_{-0.08}$ Z$_{\odot}$ for A1569S.
\item Mass of hot gas is estimated to be $5.7^{+1.5}_{-1.8} \times 10^{11}$ M$_{\odot}$ (within a radius of 248 kpc) for A1569N and $2.3^{+1.8}_{-1.3} \times 10^{12}$ M$_{\odot}$  for A1569S (within a radius of 370 kpc).
\item The total gravitational mass of A1569N within a radius of 248 kpc from the X-ray peak is $2.0\pm{1.5}\times10^{13}$ M$_{\odot}$ while that of A1569S within a radius of 370 kpc is $2.1\pm{1.6}\times 10^{13}$ M$_{\odot}$.
\item Both A1569N and A1569S lack the presence of a large ($\gtrsim$ 40--50 kpc) cool core associated with the intracluster gas.
\item A pair of cavities coincident with extended radio emission from the central galaxy 1233+168 in A1569N is detected.
\item The mechanical power associated with the cavity pair ($\sim$$2.1 \times 10^{43}$ erg s$^{-1}$) is an order of magnitude larger than the X-ray radiative loss in the central region of A1569N ($2.2\pm{0.2} \times 10^{42}$ erg s$^{-1}$), pointing towards cavity-induced heating of the gas in A1569N.
\item A surface brightness edge is detected at 1.01 arcmin along the eastern sector 145--237\textdegree{} in A1569S. The edge corresponds to a gas density jump of $2.3\pm{0.6}$ and a pressure drop of $2.5^{+2.4}_{-1.2}$.
\item The detected density discontinuity is perpendicular to the line bisecting the angle between the tails of the central WAT galaxy, 1233+168, in A1569S, which is indicative of a head-on merger occurring between A1569S and a subcluster falling in from the west along the line bisecting the WAT tails.
\item The shock associated with the merger is a weak shock with Mach number $\sim$1.7 and velocity equal to 1414 km s$^{-1}$.
\item The proposed merger scenario and the accompanying ICM bulk flow is likely responsible for the bending of the WAT source 1233+168.
\end{itemize}
\textbf{ACKNOWLEDGEMENTS}\vspace{1.2mm}
\\ We have used data from X-ray observations carried out with the Chandra X-ray observatory, managed by NASA's Marshall Center. The X-ray data were downloaded from the High Energy Astrophysics Science Archive Research Center (HEASARC), maintained by NASA's Goddard Space Flight Center. We have used radio observations obtained with the Very Large Array telescope of the National Radio Astronomy Observatory (NRAO), and optical imaging data from data release 12 of the Sloan Digital Sky Survey-III (\url{http://www.sdss3.org/}). Funding for SDSS-III has been provided by the Alfred P. Sloan Foundation, the Participating Institutions, the National Science Foundation, and the U.S. Department of Energy Office of Science. This research has made use of SAOImageDS9, developed by Smithsonian Astrophysical Observatory, and the HEASoft FTOOLS (\url{http://heasarc.gsfc.nasa.gov/ftools}). This research has also made use of the NASA/IPAC Extragalactic Database (NED), operated by the Jet Propulsion Laboratory, California Institute of Technology, under contract with the National Aeronautics and Space Administration, and the SIMBAD database, operated at CDS, Strasbourg, France. We are grateful to Eric Greisen for assistance with radio data reduction. We thank the reviewer for helpful suggestions that improved the paper.
\\\textbf{\\\\Data availability}\vspace{2.2mm}
\\This article has made use of archival data from X-ray observations with the Advanced CCD Imaging Spectrometer (ACIS) onboard the \textit{Chandra} Observatory and radio observations with the \textit{Very Large Array} telescope. The X-ray data were obtained from the HEASARC archive and are publicly accessible via \url{https://heasarc.gsfc.nasa.gov/cgi-bin/W3Browse/w3browse.pl}. The radio data were obtained from the NRAO Science Data Archive and can be retrieved from \url{https://archive.nrao.edu/archive/advquery.jsp}.
\bibliographystyle{mnras}
\bibliography{a1569} 
\bsp
\label{lastpage}
\end{document}